\documentclass[onecolumn,superscriptaddress,nofootinbib]{revtex4}

\pdfoutput=1
\usepackage{amssymb,amsmath,amsthm,graphicx}
\usepackage{graphics, color}
\usepackage{subfigure} 
\usepackage{latexsym}
\usepackage{bm}
\usepackage{epsfig}
\usepackage[pdftex]{hyperref}
\usepackage[utf8]{inputenc} 

\def\({\left(}
\def\){\right)}
\def\[{\left[}
\def\]{\right]}
\def\<{\langle}
\def\>{\rangle}

\newcommand{\be}{\begin{equation}}
\newcommand{\ee}{\end{equation}}
\newcommand{\bea}{\begin{eqnarray}}
\newcommand{\eea}{\end{eqnarray}}
\newcommand{\bwt}{\begin{widetext}}
\newcommand{\ewt}{\end{widetext}}

\newcommand{\bi}{\begin{itemize}}
\newcommand{\ei}{\end{itemize}}
\newcommand{\ben}{\begin{enumerate}}
\newcommand{\een}{\end{enumerate}}
\newcommand{\bca}{\begin{cases}}
\newcommand{\eca}{\end{cases}}
\newcommand{\bln}{\begin{align}}
\newcommand{\eln}{\end{align}}
\newcommand{\bst}{\begin{split}}
\newcommand{\est}{\end{split}}

\newcommand{\lp}{\left(}
\newcommand{\rp}{\right)}

\def\scalar{{\mathbf S}} 
\def\vector{{\mathbf V}}

\begin{document}
\title{AdS nonlinear instability: breaking spherical and axial symmetries}
\author{\'Oscar J.~C.~Dias}
\email{ojcd1r13@soton.ac.uk}
\affiliation{STAG research centre and Mathematical Sciences, University of Southampton, UK}
\author{Jorge E.~Santos}
\email{J.E.Santos@damtp.cam.ac.uk}
\affiliation{DAMTP, Centre for Mathematical Sciences, University of Cambridge, \\ Wilberforce Road, Cambridge CB3 0WA, UK}
\begin{abstract}\noindent{
Considerable effort has been dedicated to study the nonlinear instability of Anti-de Sitter (AdS) within spherical symmetry, but little is known about this nonlinear instability in the purely gravitational sector, where spherical symmetry is necessarily broken. In \cite{Bizon:2011gg} the onset of such nonlinear instability was associated with the existence of irremovable secular resonances at third order in perturbation theory. Furthermore, it was also conjectured in \cite{Bizon:2011gg} that certain very fine tuned initial data would not collapse. Such solutions, upon linearisation, correspond to individual normal modes of AdS, which can be consistently backreacted to all orders in perturbation theory. However, the analysis of \cite{Bizon:2011gg} was restricted to spherical symmetry. The perturbative arguments of \cite{Bizon:2011gg} were then generalised to gravitational perturbations in \cite{Dias:2011ss}, and in particular certain time-periodic solutions were also conjectured to exist - these were coined geons. However, in \cite{Dias:2011ss}, only a certain class of perturbations was considered, for which the perturbative analysis considerably simplifies. In this manuscript we present details of the systematic computational formalism and an exhaustive and complementary analysis of physical properties of the geons and gravitational AdS instability that were absent in our companion Letter \cite{Dias:2016ewl}. In particular, we find that, unlike in spherical symmetry, a (single) gravitational normal mode of AdS can be backreacted to generate a nonlinear solution only in very exceptional circumstances. We also show that weak turbulent perturbative theory predicts the existence of direct and inverse cascades, and give evidence suggesting that the former dominates the latter for equal energy two-mode seeds.
}
\end{abstract}
\maketitle

\tableofcontents
\newpage


\section{Introduction}

The nonlinear instability of Anti-de Sitter (AdS) spacetime has been an active topic of research in recent years \cite{Dafermos2006,DafermosHolzegel2006,Bizon:2011gg,Dias:2011ss,Dias:2012tq,Buchel:2012uh,Buchel:2013uba,Maliborski:2013jca,Bizon:2013xha,Maliborski:2012gx,Maliborski:2013ula,Baier:2013gsa,Jalmuzna:2013rwa,Basu:2012gg,2012arXiv1212.1907G,Fodor:2013lza,Friedrich:2014raa,Maliborski:2014rma,Abajo-Arrastia:2014fma,Balasubramanian:2014cja,Bizon:2014bya,Balasubramanian:2015uua,daSilva:2014zva,Craps:2014vaa,Basu:2014sia,Okawa:2014nea,Deppe:2014oua,Dimitrakopoulos:2014ada,Horowitz:2014hja,Buchel:2014xwa,Craps:2014jwa,Basu:2015efa,Yang:2015jha,Fodor:2015eia,Okawa:2015xma,Bizon:2015pfa,Dimitrakopoulos:2015pwa,Green:2015dsa,Deppe:2015qsa,Craps:2015iia,Craps:2015xya,Evnin:2015gma,Menon:2015oda,Jalmuzna:2015hoa,Evnin:2015wyi,Freivogel:2015wib,Dias:2016ewl,Evnin:2016mjx,Deppe:2016gur,Dimitrakopoulos:2016tss,Dimitrakopoulos:2016euh,Rostworowski:2016isb,Rostworowski:2017tcx,Martinon:2017uyo,Moschidis:2017lcr,Moschidis:2017llu}. Its existence was first conjectured by Dafermos and Holzegel in \cite{Dafermos2006,DafermosHolzegel2006} and its first numerical evidence was put forward in the seminal work of Bizon and Rostworowski \cite{Bizon:2011gg}\footnote{Note that this is a dynamical instability and thus does not contradict the positivity of energy proved in the seminal work of Abbott and Deser \cite{Abbott:1981ff}.}. Rather remarkably, this nonlinear instability was recently proved for the spherically symmetric Einstein-massless Vlasov system in \cite{Moschidis:2017lcr,Moschidis:2017llu}.

In simple terms, this instability is triggered by nonlinear interactions of waves in certain confining backgrounds (notably, in global AdS with energy preserving boundary conditions or in a spherically symmetry box in Minkowski with Dirichlet boundary conditions at the walls of the cavity). The analysis performed in \cite{Dias:2012tq}, strongly suggests that the commensurability of the linear spectrum of normal modes around such backgounds is a necessary condition for triggering this nonlinear instability. That is to say, that the linear spectrum is fully resonant in the sense that the sum or difference of any two frequencies still belongs to the linear spectrum (see also \cite{Maliborski:2014rma}). An open question is wether  {\it any} solution of the Einstein equation with a fully resonant spectrum is unstable.

One of the simplest (and initial) setups where we can find and analyse this nonlinear instability is within AdS-Einstein's theory of gravity. To keep the discussion technically simpler it is further desirable to assume spherical symmetry. To circumvent the fact that there are no gravitational waves in such a setup, one can add a massless scalar field that controls the dynamics of the system through the associated Klein-Gordon equation of motion \cite{Bizon:2011gg}. In this scenario, the numerical analysis of \cite{Bizon:2011gg}, and subsequent works, give strong evidence that AdS is nonlinearly unstable to the formation of {\it arbitrarily} small global AdS Schwarzschild black hole, whose mass is constrained by the energy ejected in the initial data. This is in sharp contrast with Minkowski and de Sitter spacetimes, for which  nonlinear stability has long been established \cite{Christodoulou:1993uv,friedrich86}. 

Ref. \cite{Bizon:2011gg} discussed the nonlinear problem not only from a numerical time evolution perspective but also proposed an analytical approach to identify the onset of the instability. In short, one can consider standard perturbation theory whereby the expansion parameter is the amplitude $\varepsilon$ of a linear seed, and this seed is a linear combination of normal modes of global AdS which have a quantized spectrum of frequencies. Extending the perturbative expansion to third order, one generically finds  secular term(s) of the form $t\,\varepsilon^3$. Such a linear growth in time of the amplitude at third order invalidates the perturbation scheme for timescales $t\geq \varepsilon^{-2}$. The fact that the black hole formation timescale observed in the numerical simulations seems to scale with the inverse square of the amplitude of the initial data suggests that the perturbative analysis $-$ known as a weakly perturbative turbulent analysis $-$ is appropriately describing the mechanism responsible for the onset of the instability  \cite{Bizon:2011gg,Dias:2012tq}. Such secular resonant modes do appear generically (and not just accidentally) in global AdS because its linear spectrum is fully commensurable and, as such, nonlinearities can easily create resonances \cite{Bizon:2011gg,Dias:2012tq}. 

This standard perturbative analysis is furthermore invaluable to get further physical insights on the nature of the nonlinear instability. For progressively small amplitudes of the initial data, black hole formation occurs only after an increasingly large number of bounces at the outer boundary. Of course, this multiple reflection process cannot itself explain the horizon formation. For this explanation, it is important to observe that the time simulation indicates that as time evolves there is a direct/reverse cascade of energy from low/high frequency modes to high/low frequency modes  \cite{Bizon:2011gg}. When the black hole forms, the net competition is won by the direct cascade. This net transfer of energy to Fourier modes with higher frequency suggests that the system is focusing the available energy till the collision of waves meet some sort of `local hoop conjecture'. Further supporting this explanation, and the weak perturbative turbulent origin of the instability, the perturbative analysis summarized in the previous paragraph also indicates that there are direct and reverse cascades of frequency as we climb the perturbation ladder, as pointed out in our Letter \cite{Dias:2016ewl}. In the present manuscript, we will expand on this discussion: we will find that in general the irremovable secular resonances that appear at third order have modes with both higher and lower frequencies than those of the linear seed. We will further propose an argument that seems to give perturbative evidence that the direct cascade should dominate the net competition.

Meanwhile, improved modifications of standard perturbation theory were developed to describe the nonlinear system up to timescales  $\mathcal{O}\left(\varepsilon^{2}\right)$. These schemes are essentially equivalent and dubbed as two time scale formalism \cite{Balasubramanian:2014cja}, renormalisation group perturbation methods \cite{Craps:2014vaa,Craps:2014jwa} and resonant approximation \cite{Bizon:2015pfa}. In addition to extend the perturbative description of the nonlinear system to larger times, these improved  schemes have the added value of suggesting that there is a scaling symmetry present in the problem. This scaling maps a solution with unit amplitude at time $t$ into a solution with amplitude $\varepsilon$ at time $t/\varepsilon^2$. To appreciate the consequences of this scaling, note that as the amplitude of the initial data becomes arbitrarily small, it becomes increasingly harder (and forbiddingly costly) to observe the formation of a black hole numerically. Hence, one could argue that it is conceivable that black hole formation occurs only for amplitudes above a critical amplitude $\varepsilon_C$, with the latter being small but finite. In this hypothetical scenario, the collapse process in AdS would have a critical amplitude much like in the original Choptuik collapse in Minkowski space \cite{Choptuik:1992jv}. This scenario is not ruled out, but it is highly unlikely, specially in light of \cite{Moschidis:2017lcr,Moschidis:2017llu}.

The AdS nonlinear instability seems to be present for a wide variety of initial data. There are however `islands of stability' \cite{Dias:2012tq}, \emph{i.e.} perturbations which do not necessarily lead to an instability \cite{Dias:2012tq,Buchel:2012uh,Maliborski:2013jca,Dimitrakopoulos:2015pwa,Green:2015dsa,Craps:2015xya}. An open question in the field is wether we can understand how large are these islands as a function of the properties of the initial data, in particular what happens when the energy of the system becomes arbitrarily small.

Much of what has been discussed in the above paragraphs was built from studies considering the spherical symmetric collapse of a scalar field in global AdS. A pertinent question is wether the instability is still present away from these fine-tuned spherical conditions. 
{\it \`A priori} there is a strong argument suggesting that the nonlinear instability and associated gravitational collapse to a black hole should shut-down once we break spherical symmetry. Recall that perturbations in global AdS can be expanded in spherical harmonics which are described by the usual quantum numbers $\ell$ and $m$. The first is essentially related to the number of polar zeros, while the latter is the azimuthal number which is non-zero for non-axisymmetric perturbations. Breaking spherical symmetry amounts to consider $\ell\neq 0$. Once we break spherical symmetry we can also break axial symmetry, \emph{i.e}. we can consider  $m\neq 0$. This amounts to give angular momentum to the system. In such case, it seems natural to expect that the associated centrifugal effects could balance the gravitational collapse and halt the black formation. Surprisingly, the perturbative analysis carried on in \cite{Dias:2011ss,Dias:2016ewl} and further expanded in the present manuscript, indicates that this intuition is too naive and actually incorrect. Even more remarkably, as a matter of fact the addition of angular momentum can further contribute to the nonlinear instability and associated black hole formation. For even initial data with a single non-spherical normal mode of AdS can develop resonances at third order, unlike the spherical symmetric scalar field case. Moreover, quite often the collision of two normal modes can generate more than a pair of irremovable resonances, as oppose to the case of a spherically symmetric scalar field. These observations are one of the main conclusions of the present study. 

The AdS nonlinear instability is closely related with another topic, namely with the spectrum of solutions in the phase diagram of asymptotically global AdS spacetimes. Indeed, \cite{Dias:2011ss} observed that AdS-Einstein theory allows for a one-parameter family of {\it time-periodic} solutions that were coined AdS geons (following the nomenclature proposed by Wheeler \cite{Wheeler1955}\footnote{The asymptotically flat geons proposed by Wheeler do not to exist because of dispersion of energy at asymptotic infinity. However, in AdS they can exist because gravitational radiation emitted by the geon can be balanced by absorption of waves reflected at the asymptotic boundary.}). Geons are solitonic lumps of gravitational energy which are held together by their own self gravity. Typically, geons are nor time-independent neither axisymmetric but they are time-periodic since they have a single helical Killing vector field that is a linear combination of the stationary ($\partial_t$) and axial  ($\partial_\phi$) vector fields. In the simplest case, they are the gravitational back-reaction of a normal mode of global AdS \cite{Dias:2011ss,Dias:2016ewl}. The geons of \cite{Dias:2011ss,Dias:2016ewl} were constructed perturbatively and their fully nonlinear existence was then confirmed in the numerical studies of \cite{Horowitz:2014hja} and \cite{Martinon:2017uyo}. Spherically symmetric, time-periodic solitonic solutions with a complex or real scalar field have also been constructed. They are known as boson stars \cite{Basu:2010uz,Dias:2011tj,Buchel:2013uba}, in the case of a spherically symmetric complex scalar field, or oscillons (or AdS breathers) \cite{Maliborski:2013jca,Fodor:2013lza,Fodor:2015eia}, in the case of a real spherically symmetric scalar field.  Rotating time-periodic boson stars were also constructed in \cite{Dias:2011at}. Given a geon, we can add a small black hole at its core with the same angular rotation to get a black resonator, \emph{i.e.} a time-periodic black hole with a single Killing vector field that is also its horizon generator \cite{Dias:2015rxy}. Similarly, given a rotating boson star, we can add a small rotating \cite{Dias:2011at} or a small charged black hole at its centre (with the same thermodynamic potential, \emph{i.e.} with the same angular rotation or chemical potential) to get an hairy black hole \cite{Basu:2010uz,Dias:2011tj,Dias:2016pma,Markeviciute:2016ivy}. Interestingly, these black holes establish a connection between the nonlinear instability of AdS and the superradiant instability of AdS black holes. This is because on one hand their zero-size limit corresponds to geons or boson stars and, on the other hand, they also merge with the Kerr or Reissner-Nordstr\"om black holes of the theory at the onset of the superradiant instability \cite{Basu:2010uz,Dias:2011tj,Dias:2011ss,Dias:2011at,Dias:2015rxy,Markeviciute:2016ivy}.   

Meanwhile, \cite{Dias:2016ewl} found it is not true that we can always back-react a {\it single} normal mode of AdS to get a geon. Actually, in general, given a gravitational normal mode of AdS, its back-reaction does not yield a regular soliton, \emph{i.e.} it does not yield a geon. This is an interesting result given that in the case of a spherically symmetric scalar field  {\it any} normal mode can be back-reacted to give a boson star or oscillon. Ref. \cite{Dias:2016ewl} pointed out that this feature is due to the fact that the linear frequency spectrum of gravitational normal modes is degenerate as oppose to what happens in the case of a spherically symmetric scalar field. In this manuscript, we will expand on this discussion. The perturbative method presented in \cite{Dias:2011ss,Dias:2016ewl} is a systematic method that allows to construct any geon with broken spherical and, if that is the case, with broken axial symmetry up to any perturbation order. It is also a systematic method that allows to identify all the irremovable secular resonances that appear at third order in perturbation theory when we start with a single or linear combination of normal modes of AdS that drives the system nonlinearly unstable. In the present manuscript, we will give details of this systematic method that were missing in the Letters \cite{Dias:2011ss,Dias:2016ewl}.  

Recently, \cite{Rostworowski:2016isb,Rostworowski:2017tcx,Martinon:2017uyo} have reproduced some of the axially symmetric ($m=0$) results of \cite{Dias:2016ewl} and \cite{Martinon:2017uyo} also confirmed independently the existence of the non-axisymmetric geons of \cite{Dias:2011ss,Dias:2016ewl,Horowitz:2014hja}, namely the geons with $\ell=m=2$ and $\ell=m=4$. More importantly, these references have also found that we can construct wider families of regular solitons if we start not with a single normal mode, but with a {\it linear combination} of gravitational normal modes with the same frequency as a seed. Although these solitons are not a back-reaction of a \emph{single} normal mode of AdS, they can still be called geons since they are asymptotically global AdS solitons that are regular everywhere. A central result that emerges from \cite{Dias:2011ss,Horowitz:2014hja,Dias:2016ewl,Rostworowski:2016isb,Rostworowski:2017tcx,Martinon:2017uyo} can be summarized as follows. Choose a frequency that is present in the linear spectrum of gravitational normal modes of AdS. Given the degeneracy of the spectrum, this frequency can be associated with different eigenmodes. Within perturbation theory, at linear order start with a seed that is a linear combination of all normal modes that have the chosen frequency. In these conditions, it seems that we can always back-react this linear seed at all orders to get geons. For that, at each order we need to correct the frequency using a Poincar\'e-Lindstedt  procedure and tune the several amplitudes of the normal modes that constitute the linear seed to avoid secular resonant terms. For a linear seed with $k$ normal modes with the same eigenfrequency, there are $k$ different combinations of amplitudes that yield regular geons.
 
The plan of this manuscript is the following. Firstly, note that this manuscript is a planned long version of the companion Letter  \cite{Dias:2016ewl} with a detailed description of the systematic perturbation procedure and an expanded discussion of the physical results. Section \ref{sec:General} states the perturbation problem and describes the general steps (to be complemented by more specific analyses in later sections) required to solve it to any higher order in a systematic way that is valid for any linear seed. Section \ref{sec:geons} applies this systematic procedure to cases where the linear seed is a single gravitational normal mode of AdS. In some cases, such an individual normal mode can be back-reacted to third and higher order (and at full nonlinear level) yielding a geon. We describe in detail the physical properties of some of these geons in subsection \ref{sec:Grav4geon}. However, generically, individual gravitational normal modes of AdS cannot be back-reacted up to third order without finding resonances that are irremovable and that signal the breakdown of the perturbation theory. Section  \ref{sec:Grav} shows that the weakly perturbative turbulent analysis indicates that the time evolution of generic initial data should have a direct frequency cascade but also of a reverse frequency cascade. Previously, it was thought that the reverse cascade could not be anticipated by the perturbation theory. We also argue that there is a perturbative criterion to suspect that the direct cascade should win the overall competition in a time evolution simulation. Finally, in Section \ref{sec:Grav4} we point out the perturbative theory analysis suggests that, in a sense, AdS should have a stronger nonlinear instability in the non-spherical gravitational sector than in the spherically symmetric scalar field sector. The systematic procedure that we use to find the perturbative solutions and our main results relies heavily on the Kodama-Ishibashi decomposition of the perturbations \cite{Kodama:2003jz,Kodama:2003kk} (which is an extension of the  Regge-Wheeler$-$Zerilli decomposition \cite{Regge:1957td,Zerilli:1970se} to backgrounds with a cosmological constant). Therefore,  to complement the analysis in the main text, we describe this decomposition with the necessary detail in Appendix \ref{appendix:KI}.

\section{Statement of the problem. Systematic procedure to solve it \label{sec:General}}

In this section we describe in more detail the systematic procedure used in the companion  Letters  \cite{Dias:2011ss,Dias:2016ewl}  to solve the perturbation theory required to discuss geons and the weakly perturbative turbulent instability of AdS$_4$ in the gravitational sector. This approach is general, \emph{i.e.} it applies to any  seed that is a single or a linear combination of gravitational normal modes of AdS$_4$ that can break spherical and axial symmetry and solves the problem to any order in perturbation theory (although at an increasingly higher computational cost). This systematic procedure is then applied to several linear seeds in the following sections to extract the main physical results.

\subsection{Perturbation theory problem. Systematic approach to find its solution at any order \label{sec:Grav0}}

Consider Einstein-AdS theory in four dimensions with action and equation of motion 
\begin{equation}
S=\int \mathrm{d}^4x\;\sqrt{-g}\left(R+\frac{6}{L^2}\right), \qquad R_{\mu\nu}+\frac{3}{L^2}g_{\mu\nu}=0\,,
\label{eq:action}
\end{equation}
where $L$ is the AdS length scale. 

We are interested in the simplest solution of the theory, namely the global AdS solution described by the line element
\begin{equation}\label{globalAdS}
\bar{g} = -f(r)dt^2+\frac{dr^2}{f(r)}+r^2(d\theta^2+\sin^2\theta d\phi^2), \qquad f(r)=1+\frac{r^2}{L^2}\,,
\end{equation} 
where $\{t,r,\theta,\phi \}$ are the standard global coordinates.

We will consider generic perturbations around global AdS in a perturbation scheme to higher order.
Therefore we take the expansion 
\begin{equation}\label{gexpansion}
g = \bar{g}+\sum_k h^{(k)}\varepsilon^k
\end{equation}
where  $\varepsilon$ is  a perturbation parameter (whose physical meaning will be understood later), and $\bar{g}$ is the global AdS metric \eqref{globalAdS}.

At each order in perturbation theory, the Einstein equations yield
\begin{equation}
\Delta_L h_{\mu\nu}^{(k)} = T^{(k)}_{\mu\nu},
\label{eq:perturb}
\end{equation}
where $\Delta_L$ is a second order operator constructed uniquely from the background $\bar{g}$, namely
\begin{equation}\label{linearGab}
2\Delta_L h_{\mu\nu}^{(k)} \equiv -\bar \nabla^2 h_{\mu\nu}^{(k)}-2 \bar{R}_{\mu\phantom{\alpha}\nu\phantom{\beta}}^{\phantom{\mu}\alpha\phantom{\nu}\beta}h_{\alpha\beta}^{(k)}-\bar{\nabla}_{\mu}\bar{\nabla}_\nu h^{(k)}+2\bar{\nabla}_{(\mu} \bar{\nabla}^\alpha h^{(k)}_{\nu)\alpha}.
\end{equation}
Here, $h^{(k)}= \bar{g}^{ab}h_{ab}^{(k)}$ is the trace of the perturbation, and $\bar{R}_{abcd}$ is the AdS Riemann tensor (if we decided to work in the traceless-transverse gauge, \eqref{linearGab} would reduce to the standard Lichnerowicz operator).  
On the other hand the source term $T^{(k)}$ at the $k^{\rm th}$ order  is a function of the lower order perturbations $\{h^{(j\leq k-1)}\}$ and their derivatives.
It follows from the Bianchi identities that $\bar \nabla^a T^{(k)}_{ab} =0$ for each $k$. At leading order, $k=1$, there is no source term $T^{(1)}_{ab} =0$. The source term for  $k=2$ and $k=3$ will be explicitly constructed in \eqref{O2:TLL} and  \eqref{O3:TLL}, respectively.

We will be interested on constructing {\it regular} finite energy and angular momentum solutions of (\ref{eq:perturb}) for $k \geq 1$. The metric perturbations $h^{(k)}$ and the source $T^{(k)}$ are a function of $\{t,r,\theta,\phi \}$.
However, we can use the Kodama-Ishibashi decomposition of the perturbations \cite{Kodama:2003jz,Kodama:2003kk} to simplify considerably the treatment, as described next\footnote{\label{footKI} In four dimensions and when $L\to\infty$, the Kodama-Ishibashi (KI) gauge invariant formalism  reduces exactly to the analysis firstly done by Regge and Wheeler \cite{Regge:1957td} and Zerilli  \cite{Zerilli:1970se}. Indeed, the KI vector master equation is the Regge-Wheeler master equation for odd (also called axial) perturbations \cite{Regge:1957td}, and the KI scalar master equation is the Zerilli master equation for even (also called polar) perturbations \cite{Zerilli:1970se}. For the relation between the KI gauge invariant formalism and the other well-known Teukolsky gauge invariant formalism see \cite{Dias:2013sdc}.
}. 

Global AdS is locally the product of a 2-dimensional orbit spacetime (parametrized by the time and radial coordinates) and a base space $S^{2}$. Thus, we can do a harmonic decomposition of any regular two-tensor (including the perturbations and source terms) according to how they transform under coordinate transformations on $S^{2}$. This introduces a spherical harmonic decomposition of gravitational perturbations whereby the angular dependence of the perturbations and its sources is encoded in the angular harmonics that are well known. There are two classes of harmonics: the scalar $\scalar(\theta,\phi)$ and vector $\vector_j(\theta,\phi)$ harmonics which will be distinguished by the subscripts $({\rm \bf s})$ and ({\rm \bf v}), respectively. Global AdS has the Killing isometry generated by  $\partial_\phi$. This allows a Fourier decomposition $e^{i m \phi}$ of any tensor along $\phi$. This is explicitly encoded in the Kodama-Ishibashi harmonics which are of the form $ \scalar(\theta,\phi)\sim e^{i m_{\rm \bf s} \phi}\scalar(\theta)$ and $\vector_j(\theta,\phi)\sim e^{i m_{\rm \bf v} \phi}\vector_j(\theta)$. Each of these harmonics thus depends on an azimuthal quantum number $m_{\rm \bf j}$ but also on a polar quantum number $\ell_{\rm \bf j}$, both of which are quantized by regularity of the harmonics. This scalar and vector decompositions are reviewed in detail in Appendix \ref{appendix:KI}.

In addition, the global AdS background is also time-independent. This permits to Fourier expand the harmonic coefficients of the linear perturbation $-$ collectively denote them as $F(t,r)$ $-$ along the time direction as $e^{-i \bar{\omega} t}$, where $\bar{\omega}$ is the normal mode frequency which is quantized as discussed later in \eqref{spectrumS} and \eqref{spectrumV}. 

Since we will go beyond the linear order ($k=1$) in perturbation theory, we need to use a real representation for the spherical harmonics. This means that we have to decompose each harmonic component of the linear perturbation as a sum of a $\cos\left( \bar{\omega} t - m\phi \right)= {\rm Re}[e^{-i \bar{\omega}}e^{i m \phi}]$ and a $\sin\left( \bar{\omega} t - m\phi \right)= -{\rm Im}[e^{-i \bar{\omega}}e^{i m \phi}]$ contributions. 

Also important, we want the perturbations (and thus the associated sources) to keep the $t-\phi$ symmetry of the global AdS background, {\it i.e.} the invariance of $h_{\mu\nu} dx^\mu dx^\nu$ under $\{t,\phi\} \to \{-t,-\phi\}$. To guarantee this the case we perform the operations described in Appendix \ref{appendix:KI}.

We can now summarise the systematic procedure to back-react any linear seed to any perturbation order. We will expand the perturbation and source two-tensors in a basis of eigenfunctions of the operator $\Delta_L$ defined in global AdS in \eqref{eq:perturb} and \eqref{linearGab}. Any regular two-tensor, say $h^{(k)}$, can be expressed as an infinite sum of a scalar and vector building blocks,
\begin{equation}
h^{(k)} = \sum_{\ell_{\rm \bf s},m_{\rm \bf s}} h^{({\rm \bf s})}_{\ell_{\rm \bf s},m_{\rm \bf s}}+\sum_{\ell_{\rm \bf v},m_{\rm \bf v}} h^{({\rm \bf v})}_{\ell_{\rm \bf v},m_{\rm \bf v}}.
\label{eq:linearsol1}
\end{equation}
Here, the scalar eigenfunction $h^{({\rm \bf s})}_{\ell_{\rm \bf s},m_{\rm \bf s}}$ with quantum numbers $\ell_{\rm \bf s},m_{\rm \bf s}$ and $\omega_{\ell_{\rm \bf s},p_{\rm \bf s}}$ preserves the $t-\phi$ symmetry and can be written in its real representation as
  \begin{eqnarray}\label{hmetricS}
 \left(h^{({\rm \bf s})}_{\ell_{\rm \bf s},m_{\rm \bf s}}\right)_{\mu\nu} dx^\mu dx^\nu&=&   \hat{h}^{\rm \bf s}_{ab} dx^a dx^b +\, \hat{h}^{\rm \bf s}_{ai} dx^a dx^i +  h^{\rm \bf s}_{ij} dx^i dx^j \nonumber\\
 &&+  \cos\left( \omega_{\ell_{\rm \bf s},p_{\rm \bf s}} t - m_{\rm \bf s} \phi \right) \leftrightarrow \sin\left( \omega_{\ell_{\rm \bf s},p_{\rm \bf s}} t - m_{\rm \bf s}\phi \right),
\end{eqnarray}
where $\{a,b\}=\{t,r\}$ and $\{i,j\}=\{\theta,\phi\}$. So a scalar mode is in general parametrized by $7 \times 2$ arbitrary functions of $r$.
Similarly, in \eqref{eq:linearsol1} the vector eigenfunction $h^{({\rm \bf v})}_{\ell_{\rm \bf v},m_{\rm \bf v}}$ with quantum numbers $\ell_{\rm \bf v},m_{\rm \bf v},\omega_{\ell_{\rm \bf v},p_{\rm \bf v}}$ can be expanded as
  \begin{equation}\label{hmetricV}
    \left(h^{({\rm \bf v})}_{\ell_{\rm \bf v},m_{\rm \bf v}}\right)_{\mu\nu} dx^\mu dx^\nu= \hat{h}^{\rm \bf v}_{ai} dx^a dx^i +  \hat{h}^{\rm \bf v}_{ij} dx^i dx^j +  \cos\left( \omega_{\ell_{\rm \bf v},p_{\rm \bf v}} t - m_{\rm \bf v} \phi \right) \leftrightarrow \sin\left( \omega_{\ell_{\rm \bf v},p_{\rm \bf v}} t - m_{\rm \bf v} \phi \right), 
\end{equation}
and thus a vector mode is parametrized by $3\times 2$ functions of $r$. 
In these expressions, the components $\hat{h}^{\rm \bf s}_{ab}, \hat{h}^{\rm \bf s}_{ai}, h^{\rm \bf s}_{ij}$ and $\hat{h}^{\rm \bf v}_{ai}, \hat{h}^{\rm \bf v}_{ij}$ are functions that can be read from \eqref{KI:scalar} and \eqref{KI:vector} in Appendix  \ref{appendix:KI}, respectively. Moreover, the hat in some of the metric components reminds us the replacement rule (mentioned above and discussed in detail in Appendix \ref{appendix:KI}) that guarantees that the $t-\phi$ symmetry is preserved. To easy our higher order analysis, while keeping the fundamental ingredients, we will start with a linear seed where the sine contribution in  \eqref{hmetricS} or \eqref{hmetricV} is absent.

Note that it straightforwardly follows from the previous discussion that at, each order $k$, the source term $T^{(k)}$ can be expressed as a sum of fundamental scalar blocks $\mathcal{T}^{(k)}_{({\rm \bf s})\ell_{\rm \bf s},m_{\rm \bf s}}$ and fundamental vector blocks $\mathcal{T}^{(k)}_{({\rm \bf v})\ell_{\rm \bf v},m_{\rm \bf v}}$, with each one of them specified by the harmonic quantum numbers $\ell_{\rm \bf j}$ and $m_{\rm \bf j}$ and by the frequency $\omega_{\ell_{\rm \bf j}}$,
\begin{equation}
T^{(k)} = \sum_{\ell_{\rm \bf s},m_{\rm \bf s}} \mathcal{T}^{(k)}_{({\rm \bf s})\ell_{\rm \bf s},m_{\rm \bf s}}+\sum_{\ell_{\rm \bf v},m_{\rm \bf v}} \mathcal{T}^{(k)}_{({\rm \bf v})\ell_{\rm \bf v},m_{\rm \bf v}}.
\label{decomposeT}
\end{equation}

Decomposing  $h^{(k)}$ as in (\ref{eq:linearsol1}) and $T^{(k)}$ as in \eqref{decomposeT} and introducing this into \eqref{eq:perturb}, the system decouples. That is, (\ref{eq:perturb}) is obeyed if {\it each} mode described by $h^{(k)}_{({\rm \bf j})\ell_{\rm \bf j},m_{\rm \bf j}}$ and  $\mathcal{T}^{(k)}_{({\rm \bf j})\ell_{\rm \bf j},m_{\rm \bf j}}$ independently obeys \eqref{eq:perturb}. Moreover, it follows from an analysis {\it \`a la} Kodama-Ishibashi \cite{Kodama:2003kk} (see Appendix \ref{appendix:KI}) that each mode solution associated to the quantum numbers ${\ell_{\rm \bf j},m_{\rm \bf j}}$ and $\omega_{\ell_{\rm \bf j}}$ (after a Fourier transform in time) of \eqref{eq:perturb} can be described by a ODE of the form
\begin{equation}
\Box_2 \Phi^{(k)}_{({\rm \bf j})\ell_{\rm \bf j},m_{\rm \bf j}}(t,r)-\frac{1}{f}V^{(k)}_{({\rm \bf j})\ell_{\rm \bf j}}(r) \Phi^{(k)}_{({\rm \bf j})\ell_{\rm \bf j},m_{\rm \bf j}}(t,r)=\frac{1}{f}\widetilde{\mathcal{T}}^{(k)}_{({\rm \bf j})\ell_{\rm \bf j},m_{\rm \bf j}}(t,r),
\label{eq:master}
\end{equation}
where ${\rm \bf j}=\{{\rm \bf s}, {\rm \bf v}\}$ is an index selecting wether we are considering scalar or vector modes, {\it i.e.} one of the equations governs scalar-type modes, while the other describes the vector-type modes. $\Phi^{(k)}_{({\rm \bf j})\ell_{\rm \bf j},m_{\rm \bf j}}(t,r)$ is a Kodama-Ishibashi gauge invariant variable from which  $h^{(k)}_{({\rm \bf j})\ell_{\rm \bf j},m_{\rm \bf j}}$ can be recovered (in a particular gauge)  through a linear differential map \cite{Kodama:2003jz}. 
For scalar perturbations this gauge and differential map are described by 
\eqref{KI:gaugeS} and \eqref{KI:metricS} in Appendix \ref{appendix:KI}.\footnote{For the vector modes we find that, at least for the several cases that we considered and that are quite generic, we can solve the original Einstein equation without needing to resort to the Kodama-Ishibashi formalism.} $\Box_2$ is the d'Alambertian associated with the auxiliary orbit space $ds^2 = -f(r) dt^2+dr^2/f(r)$ and $V^{(k)}_{({\rm \bf j})\ell_{\rm \bf j}}(r)$ is a potential given by 
\begin{equation}
 V^{(k)}_{({\rm \bf j})\ell_{\rm \bf j}}(r)= \ell_{\rm \bf j}  (\ell_{\rm \bf j} +1)\frac{f}{r^2}, \quad \hbox{for} \:\:{\rm \bf j}=\{ {\rm \bf s},{\rm \bf v} \} .
\label{eq:masterPotV}
\end{equation}

Also in \eqref{eq:master}, $\widetilde{\mathcal{T}}^{(k)}_{({\rm \bf j})\ell_{\rm \bf j},m_{\rm \bf j}}(r)$ is the Kodama-Ishibashi source term, that can be expressed as a function of the components of the original source $\mathcal{T}^{(k)}_{({\rm \bf j})\ell_{\rm \bf j},m_{\rm \bf j}}$ and its derivatives; see \eqref{eq:perturb}  and  \eqref{decomposeT}. For example, for scalar modes\footnote{\label{foot:vector}For the vector modes we are able to solve the original Einstein equations for $h^{(k)}_{({\rm \bf j})\ell_{\rm \bf j},m_{\rm \bf j}}$ without resorting to the Kodama-Ishibashi formulation. Therefore, we do not present the associated map for the source here.}, after using the Bianchi identity, this relation can be written as
\begin{eqnarray}\label{sourceKI}
\hspace{-0.5cm}\widetilde{\mathcal{T}}^{(k)}_{({\rm \bf s})\ell_{\rm \bf s},m_{\rm \bf s}} &=& \frac{2}{\ell_{\rm \bf s} ^2+\ell_{\rm \bf s} -2} {\biggl \{} -r \mathcal{T}_{tt}+f {\biggl (}2 (f-1)+r f \partial_r{\biggr )} \int \mathcal{T}_{tr} \, \mathrm{d}t  +2 r f \,\mathcal{T}_L \nonumber\\
&& \hspace{1cm}+\frac{r f}{\sqrt{\ell_{\rm \bf s} (\ell_{\rm \bf s} +1)}}{\biggl (}(7 f-2)+r f \partial _r {\biggr )} \mathcal{T}_r  \\
&& \hspace{1cm}+ \frac{1}{\sqrt{\ell_{\rm \bf s} (\ell_{\rm \bf s} +1)}}\left[f {\biggl (}\ell_{\rm \bf s} (\ell_{\rm \bf s} +1) + 4 -6 f-2 r f \partial _r{\biggr )}\int \mathcal{T}_t \, \mathrm{d}t -r^2 \partial_t  \mathcal{T}_t\right] \nonumber\\ 
&& \hspace{1cm}+\frac{r}{\ell_{\rm \bf s} (\ell_{\rm \bf s} +1)}  {\biggl (} 4f[ \ell  (\ell +1)-3 f ] -r^2f^2 \partial_r^2+r^2 \partial_t^2-2 r f(4 f-1)\partial_r  {\biggr )}  \mathcal{T}_T {\biggr \}},\nonumber
\end{eqnarray}
where we use the short-hand notation $\{\mathcal{T}_{ab},\mathcal{T}_a,\mathcal{T}_T,\mathcal{T}_L\} \equiv \{\mathcal{T}_{ab},\mathcal{T}_a,\mathcal{T}_T,\mathcal{T}_L\}^{(k)}_{({\rm \bf s})\ell_{\rm \bf s},m_{\rm \bf s}}$ for the Kodama-Ishibashi coefficients defined in \eqref{KI:scalar} of Appendix \ref{appendix:KIscalar}.
Note that this map and the master equation \eqref{eq:master} are valid for a {\it generic} time dependence of the source $\mathcal{T}(t,r)$. Typically, these fields will be periodic in which case we naturally do a Fourier decomposition in time. However, we will also find a few cases in which  the source and the associated master variable will have a  time dependence $t \sin(\omega t)$ or $t \cos(\omega t)$. These are the secular resonant cases to be discussed below. Equations \eqref{eq:master} and \eqref{sourceKI} also hold in this special case.  

Regular solutions of (\ref{eq:master}) are in one to one correspondence with smooth solutions of \eqref{eq:perturb}. Asymptotically, we impose Dirichlet boundary conditions on the metric so that  the perturbed system \eqref{gexpansion} remains conformal to the static Einstein universe $R_t \times S^2$:  $
\lim_{R\to\infty} \frac{L^2}{r^2}\, ds^2=-dt^2+d\theta^2+\sin^2\theta\, d\phi^2$.
The homogeneous solution ($\widetilde{T}^{(k)}_{({\rm \bf j})\ell_{\rm \bf j},m_{\rm \bf j}}=0$) of the master equation (\ref{eq:master}) behaves asymptotically as
 \begin{equation}\label{KIasymp}
\Phi^{(k)}_{({\rm \bf j})\ell_{\rm \bf j},m_{\rm \bf j}} {\bigl |}_{r\to \infty}  \sim A^{(k)}_{({\rm \bf j})\ell_{\rm \bf j},m_{\rm \bf j}}(t)+B^{(k)}_{({\rm \bf j})\ell_{\rm \bf j},m_{\rm \bf j}}(t) \,\frac{L}{r} +\cdots  \qquad  \hbox{for} \quad  j=\{ {\rm \bf s},{\rm \bf v}\}\,,
\end{equation}
for arbitrary functions $A^{(k)}_{({\rm \bf j})\ell_{\rm \bf j},m_{\rm \bf j}}$ and $B^{(k)}_{({\rm \bf j})\ell_{\rm \bf j},m_{\rm \bf j}}$. The requirement that the metric perturbations preserves the conformal boundary of the background translates to impose the boundary condition on the homogeneous master variables \cite{Michalogiorgakis:2006jc,Dias:2011ss,Dias:2013sdc}:
   \begin{eqnarray} 
&&  \qquad B^{(k)}_{({\rm \bf s})\ell_{\rm \bf s},m_{\rm \bf s}}=0,  \label{KI:BCs} \\ 
&&  \qquad  A^{(k)}_{({\rm \bf v})\ell_{\rm \bf v},m_{\rm \bf v}}=0.   \label{KI:BCv} 
\end{eqnarray}
When a source is present, imposing the boundary condition is less straightforward and typically we need to impose the asymptotically globally AdS boundary condition {\it directly} on $h^{(k)}$  and \emph{not} on $\Phi^{(k)}$. Concretely, we have to ensure that the asymptotic behaviour of the particular solution of (\ref{eq:master}) is such that the metric it generates is still conformal to the static Einstein universe. 
It turns out that the structure of the vector harmonics that are excited at $k^{\rm th}$ order by the linear seeds we consider is always such that imposing the desired asymptotic boundary condition still amounts to simply impose the boundary condition \eqref{KI:BCv} on the general solution $\Phi^{(k)}_{({\rm \bf v})}$.
However, this is not the case for scalar modes. In practice, we find that the simplest systematic way of getting the required asymptotics is to still start by imposing \eqref{KI:BCs} on $\Phi^{(k)}_{({\rm \bf s}}$. Then, we reconstruct the associated  $h^{(k)}$ using the linear differential map aforementioned and described in Appendix \ref{appendix:KI}. Finally, we add the necessary gauge transformations $-$ discussed in \eqref{KI:gaugeS0} of Appendix \ref{appendix:KI} $-$ that yield an asymptotically globally AdS spacetime. 
In more detail, we find that the gauge transformation generated by the gauge parameter \eqref{KI:gaugeS0} with
\begin{equation}  \label{gaugetransf} 
\xi_t^{(k)}=0,\quad \xi_r^{(k)}=\mathcal{C}_{\ell_{\rm \bf s},m_{\rm \bf s}} r^{\ell_{\rm \bf s} +2} f^{-\frac{1}{2} (\ell_{\rm \bf s} +4)},\quad L^{(k)}=0,
\end{equation} 
and constant coefficient $\mathcal{C}_{\ell_{\rm \bf s},m_{\rm \bf s}}$ yields an asymptotically global AdS metric. $\mathcal{C}_{\ell_{\rm \bf s},m_{\rm \bf s}}$ depends on the particular mode and for some of the modes these coefficients vanish. Further note that $\xi_r^{(k)}$  is chosen to get the desired asymptotic decay while ensuring that the solution is regular at the origin, as discussed next.

In addition to the asymptotic boundary condition, at each order in perturbation theory, we impose regularity of $h^{(k)}$, seen as a two-tensor on a fixed global AdS background \cite{Dias:2010maa,Dias:2015nua}. In particular, since the background also as a fictitious boundary at the origin of the spherical coordinates, for $h^{(k)}_{\ell,m}$ to have a regular centre we must require that 
  \begin{equation} 
\Phi^{(k)}_{\ell_{\rm \bf j},m_{\rm \bf j}}\sim \mathcal{O}\left(r^{\ell_{\rm \bf j}}\right)\quad  \hbox{for} \quad  j=\{ {\rm \bf s},{\rm \bf v}\}\,.  \label{KI:BCorigin}
\end{equation}
We will find that this regularity requirement is at the heart of the weak perturbatively turbulent mechanism.

\subsection{Leading order analysis: normal mode basis for the perturbative expansion\label{sec:Grav1}}

Generic linear seeds can be Fourier decomposed as a sum of normal modes of AdS. Therefore, in this subsection we study linear perturbations ($k=1$) around global AdS$_4$ and compute the frequency spectrum of all the normal modes of AdS$_4$.

Consider first the scalar sector of linear perturbations. At first order, $\widetilde{T}^{(1)}_{\ell,m}\equiv 0$. In this case, \eqref{eq:master} is a single ODE  
of the homogeneous Sturm-Liouville type.  
Its most general solution is a linear combination of two hypergeometric functions $\, _2F_1$,
  \begin{eqnarray} \label{linearS}
\Phi^{(1)}_{({\rm \bf s})\ell_{\rm \bf s}} &=&\frac{r^{\ell_{\rm s} +1}} {\left(r^2+L^2\right)^{\frac{\ell_{\rm s} }{2}+1}} 
{\biggl [}\mathcal{A}_{\ell_{\rm s}} \sqrt{L^2+r^2} \, _2F_1\left(\frac{1}{2} (\ell_{\rm s} -\bar{\omega}_{\ell_{\rm \bf s}} L +1),\frac{1}{2} (\ell_{\rm s} +\bar{\omega}_{\ell_{\rm \bf s}} L +1);\frac{1}{2};\frac{1}{\frac{r^2}{L^2}+1}\right) \nonumber \\
&& +\mathcal{B}_{\ell_{\rm s}} \, _2F_1\left(\frac{1}{2} (\ell_{\rm s} -\bar{\omega}_{\ell_{\rm \bf s}} L +2),\frac{1}{2} (\ell_{\rm s} +\bar{\omega}_{\ell_{\rm \bf s}} L +2);\frac{3}{2};\frac{1}{\frac{r^2}{L^2}+1}\right){\biggr ]},
  \end{eqnarray}
with arbitrary constant amplitudes $\mathcal{A}_{\ell_{\rm s}}$ and $\mathcal{B}_{\ell_{\rm s}}$.  
Asymptotically, this solutions behaves as  $\Phi^{(1)}_{({\rm \bf s})\ell_{\rm \bf s}}\sim \mathcal{A}_{\ell_{\rm s}}+\mathcal{B}_{\ell_{\rm s}}/r$. Requiring the asymptotic boundary condition \eqref{KI:BCs}  demands that we set $\mathcal{B}_{\ell_{\rm s}}= 0$. In these conditions, a Taylor expansion of \eqref{linearS} about the origin $r=0$ diverges as $\mathcal{A}_{\ell_{\rm s}} \left[\Gamma \left(\frac{1}{2} (\ell_{\rm s} -L \bar{\omega} +1)\right) \right]^{-1} r^{-\ell_{\rm \bf s}}$. However, we can have a regular centre if we use the Gamma function property $\Gamma [-p]=\infty$ for $p\in \{0,1,2,\cdots\}$ to quantize the linear spectrum of frequencies as 
  \begin{equation}  \label{spectrumS}
\bar{\omega}_{\ell_{\rm \bf s},p_{\rm \bf s}}L= 1+\ell_{\rm \bf s} +2\,p_{\rm \bf s}, \quad \hbox{where}\quad p_{\rm \bf s}\in\{0,1,2\ldots \}
 \end{equation}
 is the radial overtone (number of nodes of $\Phi$ along the radial direction). The positivity of $\bar{\omega}^2_{\ell_{\rm \bf s}}$ indicates that AdS is linearly stable. The bar over $\omega$ emphasizes that this is an intrinsic property of the global AdS background.\footnote{A WKB analysis reproduces exactly this result \cite{Dias:2012tq}.} 

The regular orthogonal basis of eigenfunctions associated to the eigenvalues \eqref{spectrumS} is then 
\begin{equation}
\left\{ e_{({\rm \bf s})\ell_{\rm \bf s},p_{\rm \bf s}} \right\}= \left\{  \frac{r^{\ell_{\rm s}+1}} {\left(r^2+L^2\right)^{\frac{\ell_{\rm s}+1 }{2}}}  \, _2F_1\left(\frac{1}{2} (\ell_{\rm s} -\bar{\omega}_{\ell_{\rm \bf s}} L +1),\frac{1}{2} (\ell_{\rm s} +\bar{\omega}_{\ell_{\rm \bf s}} L +1);\frac{1}{2};\frac{L^2}{r^2+L^2}\right) \right\}, 
\label{basisS}
\end{equation}
with $\ell_{\rm \bf s} \ge 0$. Modes with $\ell_{\rm \bf s}=0$ just shift the mass of the background and modes with $\ell_{\rm \bf s}=1$ are pure gauge \cite{Kodama:2003jz}. We want to consider gravitational waves which requires $\ell_{\rm \bf s} \ge 2$. Note that if we restrict to the ground state, $p=0$, this hypergeometric function simplifies and we simply get $\left\{ e_{({\rm \bf s})\ell_{\rm \bf s},0} \right\}= \left\{ r^{\ell_{\rm \bf s}+1}(r^2+L^2)^{-\frac{\ell_{\rm \bf s}+1}{2}} \right\} $. 

Consider now the vector sector of linear perturbations.  These modes obey exactly the same equation \eqref{eq:master}-\eqref{eq:masterPotV} as the scalar modes, so the most most general solution of $\Phi^{(1)}_{({\rm \bf v})\ell_{\rm \bf v}}$ is also given by \eqref{linearS} with ${\rm \bf s}\to {\rm \bf v}$. However, the asymptotic boundary condition for $\Phi^{(1)}_{({\rm \bf v})\ell_{\rm \bf v}}$ is different and given by \eqref{KI:BCv}. So this time we have to set  $\mathcal{A}_{\ell_{\rm s}}= 0$. Regularity at the radial origin then requires that the vector normal mode spectrum is quantized as  
  \begin{equation}  \label{spectrumV}
\bar{\omega}_{\ell_{\rm \bf v},p_{\rm \bf v}}L= 2+\ell_{\rm \bf v} +2 \,p_{\rm \bf v}, \quad \hbox{with}\quad p_{\rm \bf v}\in\{0,1,2\ldots \}
  \end{equation}
and the regular orthogonal basis of eigenfunctions associated to these eigenvalues is  
\begin{equation}
\left\{ e_{({\rm \bf v})\ell_{\rm \bf v},p_{\rm \bf v}} \right\}= \left\{  \frac{r^{\ell_{\rm v}+1}} {\left(r^2+L^2\right)^{\frac{\ell_{\rm v}}{2}+1}}  \, _2F_1\left(\frac{1}{2} (\ell_{\rm v} -\bar{\omega}_{\ell_{\rm \bf v}} L +2),\frac{1}{2} (\ell_{\rm v} +\bar{\omega}_{\ell_{\rm \bf v}} L +2);\frac{3}{2};\frac{L^2}{r^2+L^2}\right) \right\},
\label{basisV}
\end{equation}
with $\ell_{\rm \bf v} \ge 1$. Modes with $\ell_{\rm \bf v}=1$ just shift the angular momentum \cite{Kodama:2003jz}. We will consider only modes with $\ell_{\rm \bf s} \ge 2$ which generate gravitational waves.

At any perturbation order $k$, we will expand the gravitational field perturbation and source in the basis of eigenfunctions \eqref{basisS} and \eqref{basisV} of the operator $\Delta_L$ defined with respect to the global AdS background.

We emphasize three key properties of the linearized spectrum of frequencies \eqref{spectrumS} and \eqref{spectrumV}:
\begin{description}
\item 1) the spectrum is purely real and thus the solution is stable at linear order;
\item 2) the spectrum is commensurable, \emph{i.e.} the sum or difference of any two frequencies yields a frequency that still belongs to the spectrum. 
\item 3) the spectrum is degenerate.
 \end{description}

Ref. \cite{Dafermos2006,DafermosHolzegel2006} used property 1) to conjecture that global AdS is nonlinearly unstable. In \cite{Dias:2012tq} it was argued that a commensurable linear spectrum of frequencies is a necessary condition to have generically (and not just accidently) the irremovable secular resonances that we will find later. To date, as far as we are aware, all gravitational systems that were found to be nonlinearly unstable indeed have this property. The degeneracy of the spectra \eqref{spectrumS} and \eqref{spectrumV} will be responsible for some of our main findings.

\section{Back-reaction of  single normal modes and the associated geons\label{sec:geons}}

Some {\it individual} gravitational normal modes of AdS can be back-reacted to higher order in perturbation theory while keeping the solution regular at the origin at each order. If this is the case we say that we have a geon, \emph{i.e.} an asymptotically AdS solution that is horizonless and regular everywhere. 
In this section we want to find under which conditions this is the case. To address this question we have to do a tour de force and solve the linearized Einstein equation \eqref{eq:perturb} for each normal mode, albeit in a systematic way. We will find good evidence that in a few cases  geons in the above conditions do indeed exist. However for most of the cases the back-reaction of a single normal mode of AdS develops a irremovable secular resonance at third order. 

As pointed out in \cite{Rostworowski:2016isb,Rostworowski:2017tcx,Martinon:2017uyo}, due to the degeneracy of the linear spectrum of gravitational normal modes of AdS, there are also families of geons that can be constructed back-reacting a linear seed that is a {\it linear combination} of normal modes that have the same frequency but different quantum numbers. We will not discuss such cases here.

\subsection{Leading order analysis. Single-mode seed.  \label{sec:Grav1geon}}

Given the many examples that we will consider, we opt to summarise the main conclusions in the three Tables \ref{Table:geonsS1}, \ref{Table:geonsS2} and \ref{Table:geonsV}. 
In these tables, the first column describes our linear seed, \emph{i.e.} the particular (single) normal mode with which we start the expansion around AdS$_4$. Tables \ref{Table:geonsS1} and \ref{Table:geonsS2} consider the back-reaction of scalar normal modes, while  Table \ref{Table:geonsV} describes the back-reaction of vector normal modes. The linear seed is uniquely characterized by the quantum numbers and frequencies $\{ \ell,m, p, \bar{\omega}\}_{\rm \bf j}$ discussed in section \ref{sec:Grav1}.  

\begin{table}[t]
\begin{eqnarray}
\nonumber
\begin{array}{||  c  ||||  l  ||||  l  |  c  | c ||}\hline\hline\hline\hline
\hbox{Normal mode}  & \hspace{0.7cm}\hbox{Excited modes} &  \hspace{1.1cm}\hbox{Excited modes}&  \hbox{Removable} & \hbox{Secular} \\
\{\ell,m,p,\bar{\omega}\}  & \hspace{1.2cm}\{\ell,m,\omega\}  & \hspace{1.6cm}\{\ell,m,\omega\} &  \hbox{resonance} & \hbox{resonances } \\
\hbox{ at } \mathcal{O} \lp \varepsilon \rp  & \hspace{1.1cm} \hbox{ at }  \mathcal{O} \lp \varepsilon^2 \rp &   \hspace{1.5cm}  \hbox{ at }  \mathcal{O} \lp \varepsilon^3 \rp  &  \hbox{$\lp \: -L\,\omega^{(2)} \: \rp$} & \{\ell,m,p,\omega\} \\
\hline \hline
{\bf \{2,0,0,\frac{3}{L}\}_{\rm \bf s} } & \{\ell,0,0\}_{\rm \bf s},\: \ell=0,2,4  &  \{\ell,0,\frac{3}{L}\}_{\rm \bf s},\: \ell=0,2,4,6 &   \{2,0,0,\frac{3}{L}\}_{\rm \bf s} & \hbox{None} \\
     & \{\ell,0,\frac{6}{L}\}_{\rm \bf s}, \: \ell=0,2,4 & \{\ell,0,\frac{9}{L}\}_{\rm \bf s},\: \ell=0,2,4,6 &   \lp \: \frac{3663}{8960} \: \rp & \hbox{\bf (Geon ?)}   \\
 \hline
\{2,0,1,\frac{5}{L}\}_{\rm \bf s}   & \{\ell,0,0\}_{\rm \bf s},\: \ell=0,2,4  &  \{\ell,0,\frac{5}{L}\}_{\rm \bf s},\: \ell=0,2,4,6 &   \{2,0,1,\frac{5}{L}\}_{\rm \bf s} & \{4,0,0,\frac{5}{L}\}_{\rm \bf s} \\
                              & \{\ell,0,\frac{10}{L}\}_{\rm \bf s}, \: \ell=0,2,4 & \{\ell,0,\frac{15}{L}\}_{\rm \bf s},\: \ell=0,2,4,6 & \lp \: \frac{34397}{5376}\:  \rp &   \\
\hline
 \{4,0,0,\frac{5}{L}\}_{\rm \bf s} & \{\ell,0,0\}_{\rm \bf s},\: \ell=0,2,4,6,8  &  \{\ell,0,\frac{5}{L}\}_{\rm \bf s},&   \{4,0,0,\frac{5}{L}\}_{\rm \bf s} &  \{2,0,1,\frac{5}{L}\}_{\rm \bf s} \\
     &  & \qquad \ell=0,2,4,6,8,10,12  &   \lp \: \frac{52311625}{21446656} \: \rp &      \\
     & \{\ell,0,\frac{10}{L}\}_{\rm \bf s}, \: \ell=0,2,4,6,8 & \{\ell,0,\frac{15}{L}\}_{\rm \bf s},  &    &  \\
     &  & \qquad \ell=0,2,4,6,8,10,12  &    & \\
 \hline
{\bf \{2,1,0,\frac{3}{L}\}_{\rm \bf s} }     & \{\ell,0,0\}_{\rm \bf s},\: \ell=0,2,4 &  \{\ell,1,\frac{3}{L}\}_{\rm \bf s},\: \ell=2,4,6  &   \{2,1,0,\frac{3}{L}\}_{\rm \bf s} & \hbox{None} \\
                               & \{\ell,2,\frac{6}{L}\}_{\rm \bf s},\: \ell=2,4&  \{\ell,3,\frac{9}{L}\}_{\rm \bf s},\: \ell=4,6 & \lp\: \frac{123}{64} \:\rp  &  \hbox{\bf (Geon ?)}  \\
                               & \{\ell,0,0\}_{\rm \bf v},\: \ell=1,3 &  \{\ell,1,\frac{3}{L}\}_{\rm \bf v},\: \ell=1,3,5 &  &  \\
                               &   &  \{3,3,\frac{9}{L}\}_{\rm \bf v} &  &   \\                   
 \hline
{\bf \{2,2,0,\frac{3}{L}\}_{\rm \bf s} }  & \{\ell,0,0\}_{\rm \bf s},\: \ell=0,2,4 &  \{\ell,2,\frac{3}{L}\}_{\rm \bf s},\: \ell=2,4,6  &   \{2,2,0,\frac{3}{L}\}_{\rm \bf s} & \hbox{None} \\
                               & \{4,4,\frac{6}{L}\}_{\rm \bf s}  \phantom{,\: \ell=x,x,}&  \{6,6,\frac{9}{L}\}_{\rm \bf s} & \lp\: \frac{14703}{1120} \:\rp  &   \hbox{\bf (Geon)}  \\
                               & \{\ell,0,0\}_{\rm \bf v},\: \ell=1,3 &  \{\ell,2,\frac{3}{L}\}_{\rm \bf v},\: \ell=3,5 &  &   \\
 \hline
\{2,2,1,\frac{5}{L}\}_{\rm \bf s}    & \{\ell,0,0\}_{\rm \bf s},\: \ell=0,2,4 & \{\ell,2,\frac{5}{L}\}_{\rm \bf s},\: \ell=2,4,6 &   \{2,2,1,\frac{5}{L}\}_{\rm \bf s} &  \{4,2,0,\frac{5}{L}\}_{\rm \bf s} \\
                               & \{4,4,\frac{10}{L}\}_{\rm \bf s} \phantom{,\: \ell=x,x,} &  \{6,6,\frac{15}{L}\}_{\rm \bf s} \phantom{,\: \ell=x,x,} & \lp\:  \frac{9409723}{70560} \: \rp &  \\
                               & \{\ell,0,0\}_{\rm \bf v},\: \ell=1,3 & \{\ell,2,\frac{5}{L}\}_{\rm \bf v},\: \ell=3,5  &    & \{3,2,0,\frac{5}{L}\}_{\rm \bf v} \\
\hline\hline\hline\hline
\end{array}
\end{eqnarray}
\caption{Back-reaction of linear seeds with a single {\it scalar} normal mode. The first column describes the quantum numbers and frequency $\{ \ell,m, p, \bar{\omega}\}_{\rm \bf j}$ of the normal mode we start with at linear order. Each row describes a distinct case. The second and third column collects  all the scalar $\{ \ell,m, \omega\}_{\rm \bf s}$  and vector $\{ \ell,m, \omega\}_{\rm \bf v}$ harmonics that are excited at second and third order, respectively. In the fourth column we identify the quantum numbers $\{ \ell,m, p, \omega\}_{\rm \bf j}$ (it coincides with the normal mode we start with) of the third order removable resonance. In this column it is also given the frequency correction $-L\,\omega^{(2)}$ that removes the resonance. Finally, the last column identifies, when they are present, the secular resonances  $\{ \ell,m, p, \omega\}_{\rm \bf j}$ of the system.} \label{Table:geonsS1}
\end{table}

\begin{table}[ht]
\begin{eqnarray}
\nonumber
\begin{array}{||  c  ||||  l  ||||  l  |  c  | c ||}\hline\hline\hline\hline
\hbox{Normal mode}  & \hspace{0.7cm}\hbox{Excited modes} &  \hspace{1.1cm}\hbox{Excited modes}&  \hbox{Removable} & \hbox{Secular} \\
\{\ell,m,p,\bar{\omega}\}  & \hspace{1.2cm}\{\ell,m,\omega\}  & \hspace{1.6cm}\{\ell,m,\omega\} &  \hbox{resonance} & \hbox{resonances } \\
\hbox{ at } \mathcal{O} \lp \varepsilon \rp  & \hspace{1.1cm} \hbox{ at }  \mathcal{O} \lp \varepsilon^2 \rp &   \hspace{1.5cm}  \hbox{ at }  \mathcal{O} \lp \varepsilon^3 \rp  &  \hbox{$\lp \: -L\,\omega^{(2)} \: \rp$} &  \{\ell,m,p,\omega\} \\
\hline \hline
{\bf \{3,3,0,\frac{4}{L}\}_{\rm \bf s} }  & \{\ell,0,0\}_{\rm \bf s},\: \ell=0,2,4,6 &  \{\ell,3,\frac{4}{L}\}_{\rm \bf s},\: \ell=3,5,7,9  &   \{3,3,0,\frac{4}{L}\}_{\rm \bf s} & \hbox{None} \\
                               & \{6,6,\frac{8}{L}\}_{\rm \bf s}  \phantom{,\: \ell=x,x,}&  \{9,9,\frac{12}{L}\}_{\rm \bf s} & \lp\: \frac{27881625}{32032} \:\rp  &   \hbox{\bf (Geon)}  \\
                               & \{\ell,0,0\}_{\rm \bf v},\: \ell=1,3,5 &  \{\ell,3,\frac{4}{L}\}_{\rm \bf v},\: \ell=4,6,8 &  &   \\                              
 \hline
\{3,2,0,\frac{4}{L}\}_{\rm \bf s}   & \{\ell,0,0\}_{\rm \bf s},\:\ell=0,4,6 & \{\ell,2,\frac{4}{L}\}_{\rm \bf s},\:\ell=3,5,7,9 &  \{3,2,0,\frac{4}{L}\}_{\rm \bf s}  &   \\
                               & \{\ell,4,\frac{8}{L}\}_{\rm \bf s},\:\ell=4,6 & \{\ell,6,\frac{12}{L}\}_{\rm \bf s},\:\ell=7,9 &  \lp\:  \frac{8081875}{72072} \:\rp &   \\
                               & \{\ell,0,0\}_{\rm \bf v},\: \ell=1,3 & \{\ell,2,\frac{4}{L}\}_{\rm \bf v},\:\ell=2,4,6,8  &    &  \{2,2,0,\frac{4}{L}\}_{\rm \bf v} \\
                                  &  & \{6,6,\frac{12}{L}\}_{\rm \bf v}  &    &   \\
\hline
{\bf \{4,4,0,\frac{5}{L}\}_{\rm \bf s} }   &  \{\ell,0,0\}_{\rm \bf s},  &  \{\ell,4,\frac{5}{L}\}_{\rm \bf s}, &  \{4,4,0,\frac{5}{L}\}_{\rm \bf s} & \hbox{None} \\
                               & \quad \ell=0,2,4,6,8 &  \quad  \ell=4,6,8,10,12 &  \lp\: \frac{7010569125}{77792} \:\rp & \hbox{\bf (Geon)}  \\
                               & \{8,8,\frac{10}{L}\}_{\rm \bf s}  & \{12,12,\frac{15}{L}\}_{\rm \bf s}  &    &   \\
                               & \{\ell,0,0\}_{\rm \bf v},\:\ell=1,3,5,7 &  \{\ell,4,\frac{5}{L}\}_{\rm \bf v},\:  &    &   \\
                               &     &  \quad \ell=5,7,9,11 &   &   \\
 \hline
\{4,2,0,\frac{5}{L}\}_{\rm \bf s}   &  \{\ell,0,0\}_{\rm \bf s} & \{\ell,2,\frac{5}{L}\}_{\rm \bf s}, &  \{4,2,0,\frac{5}{L}\}_{\rm \bf s}  & \{2,2,1,\frac{5}{L}\}_{\rm \bf s} \\
& \quad \ell=0,2,4,6,8 & \quad \ell=2,4,6,8,10,12 &  \lp\: \frac{163492329375}{243955712} \:\rp  &  \\
                               &  \{\ell,4,\frac{10}{L}\}_{\rm \bf s},\: \ell=4,6,8 & \{\ell,6,\frac{15}{L}\}_{\rm \bf s},\: \ell=6,8,10,12 &   &  \\
                                & \{\ell,0,0\}_{\rm \bf v},\: \ell=1,3,5,7 & \{\ell,2,\frac{5}{L}\}_{\rm \bf v},\: \ell=3,5,7,9,11 &    & \{3,2,0,\frac{5}{L}\}_{\rm \bf v} \\
                                &   & \{\ell,6,\frac{15}{L}\}_{\rm \bf v},\: \ell=7,9 &    &   \\
 \hline
{\bf \{6,6,0,\frac{7}{L}\}_{\rm \bf s} }  &  \{\ell,0,0\}_{\rm \bf s},  &  \{\ell,6,\frac{7}{L}\}_{\rm \bf s}, &   \{6,6,0,\frac{7}{L}\}_{\rm \bf s} & \hbox{None} \\
          &  \quad \ell=0,2,4,6,8,10,12  &  \quad  \ell=6,8,10,12,14,16,18  &   \lp \: \frac{8231910851500875}{3090464}\:  \rp  &  \hbox{\bf (Geon)}  \\
&  \{12,12,\frac{14}{L}\}_{\rm \bf s} &  \{18,18,\frac{21}{L}\}_{\rm \bf s} &  &   \\
&  \{\ell,0,0\}_{\rm \bf v}, &  \{\ell,6,\frac{7}{L}\}_{\rm \bf v}, &   &   \\
&  \quad \ell=1,3,5,7,9,11 &  \quad \ell=7,9,11,13,15,17 &   &   \\
\hline\hline\hline\hline
\end{array}
\end{eqnarray}
\caption{Back-reaction of linear seeds with a single {\it scalar} normal mode. The information is displayed as in Table \ref{Table:geonsS1}.} \label{Table:geonsS2}
\end{table}
\begin{table}[ht]
\begin{eqnarray}
\nonumber
\begin{array}{||  c  ||||  l  ||||  l  |  c  | c ||}\hline\hline\hline\hline
\hbox{Normal mode}  & \hspace{0.7cm}\hbox{Excited modes} &  \hspace{1.1cm}\hbox{Excited modes}&  \hbox{Removable} & \hbox{Secular} \\
\{\ell,m,p,\bar{\omega}\}  & \hspace{1.2cm}\{\ell,m,\omega\}  & \hspace{1.6cm}\{\ell,m,\omega\} &  \hbox{resonance} & \hbox{resonances } \\
\hbox{ at } \mathcal{O} \lp \varepsilon \rp  & \hspace{1.1cm} \hbox{ at }  \mathcal{O} \lp \varepsilon^2 \rp &   \hspace{1.5cm}  \hbox{ at }  \mathcal{O} \lp \varepsilon^3 \rp  &  \hbox{$\lp \: -L\,\omega^{(2)} \: \rp$} & \{\ell,m,p,\omega\} \\
\hline \hline
{\bf \{2,0,0,\frac{4}{L}\}_{\rm \bf v} } & \{\ell,0,0\}_{\rm \bf s},\: \ell=0,2,4   &   \{\ell,0,\frac{4}{L}\}_{\rm \bf v},\: \ell=2,4,6  &   \{2,0,0,\frac{4}{L}\}_{\rm \bf v} & \hbox{None} \\
                    &  \{\ell,0,\frac{8}{L}\}_{\rm \bf s},\: \ell=0,2,4 &  \{\ell,0,\frac{12}{L}\}_{\rm \bf v},\: \ell=2,4,6   &  \lp\: \frac{1469}{26880}  \:\rp &  \hbox{\bf (Geon ?)}  \\
 \hline
\{2,0,1,\frac{6}{L}\}_{\rm \bf v}  & \{\ell,0,0\}_{\rm \bf s},\: \ell=0,2,4   &   \{\ell,0,\frac{6}{L}\}_{\rm \bf v},\: \ell=2,4,6  &   \{2,0,1,\frac{6}{L}\}_{\rm \bf v} &  \{4,0,0,\frac{6}{L}\}_{\rm \bf v} \\
                    &  \{\ell,0,\frac{12}{L}\}_{\rm \bf s},\: \ell=0,2,4 &  \{\ell,0,\frac{18}{L}\}_{\rm \bf v},\: \ell=2,4,6   &  \lp\: \frac{19081}{376320} \:\rp &   \\
 \hline
\{2,1,0,\frac{4}{L}\}_{\rm \bf v}  &  \{\ell,0,0\}_{\rm \bf s},\: \ell=0,2,4 &  \{\ell,1,\frac{4}{L}\}_{\rm \bf s},\: \ell=1,3,5 &  &  \{3,1,0,\frac{4}{L}\}_{\rm \bf s} \\
               &  \{\ell,2,\frac{8}{L}\}_{\rm \bf s},\: \ell=2,4 & \{3,3,\frac{12}{L}\}_{\rm \bf s}  &  \   &   \\
               &   \{\ell,0,0\}_{\rm \bf v},\: \ell=1,3  &\{\ell,1,\frac{4}{L}\}_{\rm \bf v},\: \ell=2,4,6    &  \{2,1,0,\frac{4}{L}\}_{\rm \bf v}  &   \\
      &    & \{\ell,3,\frac{12}{L}\}_{\rm \bf v},\: \ell=4,6 &  \lp\: \frac{72361}{322560}\:\rp  &   \\             
 \hline
\{2,2,0,\frac{4}{L}\}_{\rm \bf v}  &  \{\ell,0,0\}_{\rm \bf s},\: \ell=0,2,4 &  \{\ell,2,\frac{4}{L}\}_{\rm \bf s},\: \ell=3,5 &  &  \{3,2,0,\frac{4}{L}\}_{\rm \bf s} \\
               &  \{4,4,\frac{8}{L}\}_{\rm \bf s} & \{\ell,2,\frac{4}{L}\}_{\rm \bf v},\: \ell=2,4,6  &  \{2,2,0,\frac{4}{L}\}_{\rm \bf v}  &   \\
               &   \{\ell,0,0\}_{\rm \bf v},\: \ell=1,3  & \{6,6,\frac{12}{L}\}_{\rm \bf v} &  \lp\: \frac{1247}{1008} \:\rp  &   \\
 \hline
\{3,2,0,\frac{5}{L}\}_{\rm \bf v}  & \{\ell,0,0\}_{\rm \bf s},\: \ell=0,4,6 & \{\ell,2,\frac{5}{L}\}_{\rm \bf s},\: \ell=2,4,6,8 &  & \{2,2,1,\frac{5}{L}\}_{\rm \bf s} \\
           & \{\ell,4,\frac{10}{L}\}_{\rm \bf s},\: \ell=4,6 &  \{6,6,\frac{15}{L}\}_{\rm \bf s} &   &   \\
           &  \{\ell,0,0\}_{\rm \bf v},\: \ell=1,3,5 &   \{\ell,2,\frac{5}{L}\}_{\rm \bf v},\: \ell=3,5,7,9 &   \{3,2,0,\frac{5}{L}\}_{\rm \bf v}   & \{4,2,0,\frac{5}{L}\}_{\rm \bf s}  \\
            &     &  \{\ell,6,\frac{15}{L}\}_{\rm \bf v},\: \ell=7,9 &  \lp\: \frac{31995875}{4612608} \:\rp  &   \\
 \hline
\{7,6,0,\frac{9}{L}\}_{\rm \bf v}  &  \{\ell,0,0\}_{\rm \bf s},  & \{\ell,6,\frac{9}{L}\}_{\rm \bf s}, \:\ell=6,8,10, &  & \{6,6,1,\frac{9}{L}\}_{\rm \bf s} \\
&  \quad \ell=0,2,4,6,8,10,12,14  & \quad \qquad 12,14,16,18,20 &  & \{8,6,0,\frac{9}{L}\}_{\rm \bf s} \\
&  \{\ell,12,\frac{18}{L}\}_{\rm \bf s}, \:\ell=12,14  & \{18,18,\frac{27}{L}\}_{\rm \bf s} &    &   \\
&   \{\ell,0,0\}_{\rm \bf v},  &  \{\ell,6,\frac{9}{L}\}_{\rm \bf v}, \: \ell=7,9,11, &  \{7,6,0,\frac{9}{L}\}_{\rm \bf v}   &   \\
&  \quad \ell=1,3,5,7,9,11,13  & \quad \qquad 13,15,17,19,21 &   \lp\:  \frac{8548214990390361}{19124224} \:\rp  &   \\
\hline\hline\hline\hline
\end{array}
\end{eqnarray}
\caption{Back-reaction of linear seeds with a single {\it vector} normal mode. The information is displayed as in Table \ref{Table:geonsS1}.} \label{Table:geonsV}
\end{table}
%

\subsection{Second order analysis. Frequency corrections.  \label{sec:Grav2geon}}

Consider now perturbation theory at second order, $k=2$. The second order metric $ h^{(2)}$ obeys the linearized Einstein equation \eqref{eq:perturb}, $\Delta_L h_{\mu\nu}^{(2)} = T^{(2)}_{\mu\nu}$, with $\Delta_L$ given by \eqref{linearGab} and source tensor $T^{(2)}$ that is quadratic in the first order metric $h^{(1)}$ and its derivatives. This energy-momentum tensor is explicitly given  by\footnote{Note that \eqref{O2:TLL}-\eqref{O2:TLLaux} reduces to the familiar Landau-Lifshitz pseudotensor if we choose the traceless transverse gauge, $h=\bar{\nabla}_\mu h^{\mu\nu}=0$ \cite{Landau:1951}.} 
\begin{equation}
T^{(2)}_{\mu\nu}(h^{(1)}) =\frac{1}{8\pi G_N} \left(R_{\mu\nu}^{(2)}(h^{(1)})-\frac{1}{2} \bar{g}_{\mu\nu} \bar{g}^{\alpha\beta}  R^{(2)}_{\alpha\beta} \left(h^{(1)}\right)\right) ,
\label{O2:TLL}
\end{equation}
with (in the right hand side of this expression we use the short-hand notation $h_{\mu\nu}\equiv h^{(1)}_{\mu\nu}$)
\begin{eqnarray}
R_{\mu\nu}^{(2)}(h^{(1)})&=& -\frac{1}{2}{\biggl[}\frac{1}{2}\lp \bar{\nabla}_\mu h_{\alpha\beta}\rp \bar{\nabla}_\nu
 h^{\alpha\beta}
 + h^{\alpha\beta}\lp \bar{\nabla}_\nu \bar{\nabla}_\mu h_{\alpha\beta}+\bar{\nabla}_\alpha \bar{\nabla}_\beta h_{\mu\nu}
   - \bar{\nabla}_\alpha \bar{\nabla}_\mu h_{\nu\beta} -\bar{\nabla}_\alpha \bar{\nabla}_\nu
   h_{\mu\beta}\rp \nonumber\\
&& \hspace{0.7cm}+  \bar{\nabla}_\alpha h^{\beta}_{\:\:\mu}\lp
\bar{\nabla}^\alpha h_{\beta\nu}-\bar{\nabla}_\beta h^{\alpha}_{\:\:\nu} \rp -
\bar{\nabla}_\alpha h^{\alpha\beta}\lp \bar{\nabla}_\mu h_{\beta\nu}+\bar{\nabla}_\nu
h_{\mu\beta}-\bar{\nabla}_\beta h_{\mu\nu} \rp\nonumber\\
&& \hspace{0.7cm}+ \frac{1}{2}\bar{\nabla}_\gamma\left(\bar{\nabla}_\nu h^\gamma_{\:\:\mu}+ \bar{\nabla}_\mu h^\gamma_{\:\:\nu} -\bar{\nabla}^\gamma h_{\mu\nu} \right){\biggl]}. \label{O2:TLLaux}
\end{eqnarray}

Although we start with a seed that has a single harmonic, \eqref{O2:TLL} excites many more harmonics at second order. Moreover,  harmonics in both sectors of perturbations are typically excited. It follows that the next task is to decompose the source $T^{(2)}_{\mu\nu}(h^{(1)})$ as a sum of the fundamental scalar blocks $\mathcal{T}^{(k)}_{({\rm \bf s})\ell_{\rm \bf s},m_{\rm \bf s}}$ and the fundamental vector blocks $\mathcal{T}^{(k)}_{({\rm \bf v})\ell_{\rm \bf v},m_{\rm \bf v}}$, as described in \eqref{decomposeT}. There is no systematic way to identify {\it \`a priori} what scalar and vector harmonics are excited.  The number of excited harmonics is nevertheless finite. Indeed, the trignometric addition formulas and the quadratic nature of the source in $h^{(1)}$ described by  \eqref{initialdata} guarantee that  quantum numbers of the excited harmonics are sums and differences of the quantum numbers of the two initial modes and bounded above by $ \{ \ell,m, \omega\}_{k=2}\leq 2\times \{ \ell,m, \bar{\omega}\}_{k=1}$. 

With these guides we identify precisely the excited harmonics by a direct inspection of $T^{(2)}_{\mu\nu}(h^{(1)})$. Concretely, for each normal mode seed identified in the first column, the harmonics that are excited at second order are displayed in the second column of of Tables  \ref{Table:geonsS1}, \ref{Table:geonsS2} and \ref{Table:geonsV}. 

In this process, and for each scalar building block  $\mathcal{T}^{(2)}_{({\rm \bf s})\ell_{\rm \bf s},m_{\rm \bf s}}$, we find the associated seven components $ \{ \mathcal{T}_{ab}^{(2)},\mathcal{T}^{(2)}_{a},\mathcal{T}^{(2)}_L,\mathcal{T}^{(2)}_T \}_{\ell_{\rm \bf s},m_{\rm \bf s}}$: see \eqref{KI:scalar} and Appendix \ref{appendix:KIscalar}. Similarly, for each vector harmonic block $\mathcal{T}^{(2)}_{({\rm \bf v})\ell_{\rm \bf v},m_{\rm \bf v}}$ we identify the associated three components $\{ \mathcal{T}_{a},\mathcal{T}_T \}_{\ell_{\rm \bf v},m_{\rm \bf v}}$: see \eqref{KI:vector} and Appendix \ref{appendix:KIvector}. 

Following a Kodama-Ishibashi  procedure \cite{Kodama:2003kk}, next we find the master equation \eqref{eq:master} with potential \eqref{eq:masterPotV} that the second order master variables $ \Phi^{(2)}_{({\rm \bf j})\ell_{\rm \bf j},m_{\rm \bf j}}$ must satisfy. That is, for each excited harmonic listed in \eqref{O2:excitedS}-\eqref{O2:excitedV},  we find the Kodama-Ishibashi source term $\widetilde{\mathcal{T}}^{(2)}_{({\rm \bf j})\ell_{\rm \bf j},m_{\rm \bf j}}$ as a function of the original $\mathcal{T}^{(2)}_{({\rm \bf j})\ell_{\rm \bf j},m_{\rm \bf j}}$ that we had identified in our previous step $-$ see \eqref{sourceKI}.

For each excited harmonic we are then able to solve the inhomogeneous master ODEs \eqref{eq:master} and find the associated $\Phi^{(2)}_{({\rm \bf j})\ell_{\rm \bf j},m_{\rm \bf j}}$.
The differential map \eqref{KI:metricS} reconstructs $h^{(2)}_{({\rm \bf j})\ell_{\rm \bf j},m_{\rm \bf j}}$ from $\Phi^{(2)}_{({\rm \bf j})\ell_{\rm \bf j},m_{\rm \bf j}}$ in the gauge \eqref{KI:gaugeS}.
Next, we impose the appropriate boundary conditions following the procedure described in detail in \eqref{KIasymp}-\eqref{KI:BCorigin}, adding gauge transformations to our  $h^{(2)}_{({\rm \bf j})\ell_{\rm \bf j},m_{\rm \bf j}}$ when necessary.  Namely, we require each $h^{(2)}_{({\rm \bf j})\ell_{\rm \bf j},m_{\rm \bf j}}$ to have a regular centre and, at asymptotic infinity, we require it is asymptotically globally AdS. 

Here, it is important to observe that each one of the harmonics excited by the source has a frequency $\omega$ that is given by a sum or difference of the frequency $\bar{\omega}$ of our linear seed, \emph{i.e.} $\omega \in \{0,2\bar{\omega}\}$. However, in the excited spectrum listed in the second column of Tables \ref{Table:geonsS1}$-$\ref{Table:geonsV} we do not find a single source combination of quantum numbers $\{\ell, p,\omega\}$ that coincides with one of the elements $\{\ell_{\rm \bf j},p_{\rm j},\bar{\omega}_{\rm \bf j} \}$ of the linear frequency spectra \eqref{spectrumS} or \eqref{spectrumV}. That is, for any case one has  $\omega \neq 1+\ell_{\rm \bf s}+2p$ and $\omega \neq 2+\ell_{\rm \bf v}+2p$. In equivalent words, there are no resonances and the solution can be made straightforwardly regular at the origin with a choice of integrations constants of the homogeneous solution of $\Delta_L h_{\mu\nu}^{(2)} = T^{(2)}_{\mu\nu}$.

We conclude that {\it each one} of the harmonic building blocks that is excited at second order can be made regular at the origin and manifestly asymptotically global AdS (after a gauge transformation). Consequently, the total metric \eqref{gexpansion} up to second order, $g^{(2)} = \bar{g}+ \varepsilon^1 h^{(1)}+\varepsilon^2 h^{(2)}$, is regular everywhere and asymptotically global AdS. As a mandatory check of our computation, we do confirm that this total metric satisfies the Einstein equation \eqref{eq:action} up to $\mathcal{O}(\varepsilon^2)$. 

The second order analysis is still not complete. At leading order, the frequency of the seed $h^{(1)}$ is given by a normal mode frequency $\bar{\omega}$  of global AdS. But as we climb the perturbation ladder, the background is changing and there is no reason why the fundamental frequencies of the system should keep unchanged. It follows that we should allow the frequencies themselves to be  corrected at each perturbation order:
 \begin{equation}\label{O2:wexpansion}
 \bar{\omega} \to \omega = \bar{\omega}+ \varepsilon^2 \omega^{(2)}.
\end{equation}
For completeness we observe that the structure of the perturbed equations indicates that these frequencies receive corrections only at even orders, $\omega= \bar{\omega}+\sum_{j=1} \varepsilon^{2j}  \omega^{(2j)}$.

\subsection{Third order analysis. Irremovable secular resonances or geons.  \label{sec:Grav3geon}}

Moving to third order, $k=3$, we first compute the energy-momentum tensor $T^{(3)}_{\mu\nu}$,
\begin{equation}
T^{(3)}_{\mu\nu}(h^{(1)},h^{(2)}) =R_{\mu\nu}(g^{(2)})+\frac{3}{L^2}\,g^{(2)}_{\mu\nu}.
\label{O3:TLL}
\end{equation}
It is a $\mathcal{O}(\varepsilon^3)$ quantity constructed from the total second order metric and its derivatives. To compute it we use
 \begin{eqnarray}\label{O3:TLLaux}
&& g^{(2)}_{\mu\nu}= \bar{g}_{\mu\nu}+ \varepsilon^1 h^{(1)}_{\mu\nu}+\varepsilon^2 h^{(2)}_{\mu\nu}\,, \nonumber\\
&& g_{(2)}^{\mu\nu}= \bar{g}^{\mu\nu}+ \varepsilon^1 H_{(1)}^{\mu\nu}+\varepsilon^2 H_{(2)}^{\mu\nu}+\varepsilon^3 H_{(3)}^{\mu\nu} \,, 
\end{eqnarray}
where
 \begin{eqnarray}\label{O3:TLLaux2}
&& H_{(1)}^{\mu\nu}\equiv \bar{g}^{\mu\alpha} \bar{g}^{\nu\beta} h_{\alpha\beta}^{(1)},
\qquad 
 H_{(2)}^{\mu\nu} \equiv h_{(2)}^{\mu\nu}-\bar{g}^{\mu\alpha} h_{(1)}^{\nu\beta} h_{\alpha\beta}^{(1)},
\qquad
  H_{(3)}^{\mu\nu} \equiv -\bar{g}^{\mu\alpha}\left( H_{(1)}^{\nu\beta} h_{\alpha\beta}^{(2)} + H_{(2)}^{\nu\beta} h_{\alpha\beta}^{(1)}\right) \nonumber\\
&&
\end{eqnarray}
are such that $g^{(2)}_{\mu\alpha} g_{(2)}^{\alpha\nu}=\delta_\mu^\nu$ up to $\mathcal{O}(\varepsilon^3)$. 
The $\varepsilon^3$ terms in \eqref{O3:TLL} source the third order metric perturbation $h^{(3)}$ that we now want to find solving the linearized Einstein equation \eqref{eq:perturb} at order $k=3$. 

As described in \eqref{decomposeT}, we have to decompose the source $T^{(3)}$ as a sum of fundamental scalar, $\mathcal{T}^{(3)}_{({\rm \bf s})\ell_{\rm \bf s},m_{\rm \bf s}}$, and vector, $\mathcal{T}^{(3)}_{({\rm \bf v})\ell_{\rm \bf v},m_{\rm \bf v}}$, blocks. 
So we need to identify the quantum numbers $\{\ell_{\rm \bf s}, m_{\rm \bf s}, \omega \}$ and the seven functions $ \{ \mathcal{T}_{ab}^{(3)},\mathcal{T}^{(3)}_{a},\mathcal{T}^{(3)}_L,\mathcal{T}^{(3)}_T \}_{\ell_{\rm \bf s},m_{\rm \bf s}}$ of each excited scalar harmonic: see \eqref{KI:scalar} and Appendix \ref{appendix:KIscalar}. Similarly, for each  excited vector harmonic we must find the quantum numbers $\{\ell_{\rm \bf v}, m_{\rm \bf v}, \omega \}$ and the three functions $ \{ \mathcal{T}^{(3)}_{a},\mathcal{T}^{(3)}_T \}_{\ell_{\rm \bf v},m_{\rm \bf v}}$. 

Using a Clebsch-Gordan analysis as a guide and a direct inspection of \eqref{O3:TLL} we find that, for each normal mode identified in the first column, the harmonics that are excited at third order are those given in the third column of Tables  \ref{Table:geonsS1}, \ref{Table:geonsS2} and \ref{Table:geonsV}. The excited modes with highest frequency have $\omega=3\,\bar{\omega}$ and those  with lower frequency have $\omega=\bar{\omega}$.

Out of the excited  harmonics  there are always a few that stand out. The reason is two-folded.
One and only one of them singles out because its quantum numbers $\{ \ell_{\rm j}, m_{\rm j}, p_{\rm j},\omega_{\rm j}\}$ exactly match the quantum numbers of the normal mode we started with. This special mode is shown in the fourth column of Tables  \ref{Table:geonsS1}, \ref{Table:geonsS2} and \ref{Table:geonsV}. 
The other(s) single out for, although {\it not} present in the initial seed, their quantum numbers are such that their frequency matches one of the normal mode frequencies of the quantised linear spectrum \eqref{spectrumS} or  \eqref{spectrumV}. Such special mode(s) are identified in the last column of Tables  \ref{Table:geonsS1}, \ref{Table:geonsS2} and \ref{Table:geonsV}. 

In both these special cases, we have  {\it resonances} since the source quantum numbers $\{ \ell_{\rm \bf j}, m_{\rm \bf j}, \omega_{\rm \bf j} \}$ match the normal mode quantum numbers $\{ \ell_{\rm \bf j}, m_{\rm \bf j}, \bar{\omega}_{\rm \bf j} \}$ of the global AdS background. The far-reaching consequences of these observations become clear when we try to impose regularity of the solutions at the origin of the spacetime. This is precisely our next task.

For each excited harmonic listed in the third column of Tables \ref{Table:geonsS1}-\ref{Table:geonsV}, alike in the second order analysis, we use \eqref{sourceKI} to find the Kodama-Ishibashi source term $\widetilde{\mathcal{T}}^{(3)}_{({\rm \bf j})\ell_{\rm \bf j},m_{\rm \bf j}}$ as a function of the original $\mathcal{T}^{(3)}_{({\rm \bf j})\ell_{\rm \bf j},m_{\rm \bf j}}$ that we  identified in our previous step. This sources the Kodama-Ishibashi master equation \eqref{eq:master} with potential \eqref{eq:masterPotV} that the third order master variables $ \Phi^{(3)}_{({\rm \bf j})\ell_{\rm \bf j},m_{\rm \bf j}}$ must obey.
It is then possible to solve this inhomogeneous master ODE \eqref{eq:master} for $\Phi^{(3)}_{({\rm \bf j})\ell_{\rm \bf j},m_{\rm \bf j}}$. It has a particular contribution associated to the source $\mathcal{T}^{(3)}$ but also a homogeneous solution with two associated integration constants that we henceforth denote by $\alpha^{(3)}_{\ell_{\rm \bf j},m_{\rm \bf j}}$ and $\beta^{(3)}_{\ell_{\rm \bf j},m_{\rm \bf j}}$.
We can now apply the Kodama-Ishibashi differential map \eqref{KI:metricS} to reconstruct $h^{(3)}_{({\rm \bf j})\ell_{\rm \bf j},m_{\rm \bf j}}$ from $\Phi^{(3)}_{({\rm \bf j})\ell_{\rm \bf j},m_{\rm \bf j}}$ in the gauge \eqref{KI:gaugeS}.   

At this stage, a generic $h^{(3)}_{({\rm \bf j})\ell_{\rm \bf j},m_{\rm \bf j}}$ (\emph{i.e.} for all the modes displayed in the third column of the Tables) is a function of the two third order Kodama-Ishibashi amplitudes $\alpha^{(3)}_{\ell_{\rm \bf j},m_{\rm \bf j}}$ and $\beta^{(3)}_{\ell_{\rm \bf j},m_{\rm \bf j}}$. The metric perturbation of the special resonant harmonic in the fourth column of Tables \ref{Table:geonsS1}-\ref{Table:geonsV}, and only this, also depends on the frequency correction $\omega^{(2)}$ introduced  as part of the perturbation scheme in \eqref{O2:wexpansion}. The several $\{ \alpha^{(3)}_{\ell_{\rm \bf j},m_{\rm \bf j}}, \beta^{(3)}_{\ell_{\rm \bf j},m_{\rm \bf j}} \}$ and the frequency correction $\omega^{(2)}$ are to be fixed by the boundary conditions.

We  impose the boundary conditions as described in \eqref{KIasymp}-\eqref{KI:BCorigin}. Start with the non-resonant vector modes. A judicious choice of the Kodama-Ishibashi amplitudes $\alpha^{(3)}_{\ell_{\rm \bf v},m_{\rm \bf v}}$ and $\beta^{(3)}_{\ell_{\rm \bf v},m_{\rm \bf v}}$ allows to make each of these vector harmonic perturbations asymptotically global AdS and regular at the origin. Consider next the non-resonant scalar modes. A choice of the Kodama-Ishibashi amplitudes $\alpha^{(3)}_{\ell_{\rm \bf s},m_{\rm \bf s}}$ and $\beta^{(3)}_{\ell_{\rm \bf s},m_{\rm \bf s}}$, together with the addition of a gauge transformation generated by \eqref{KI:gaugeS0} and \eqref{gaugetransf}, allows to explicitly get a regular two-tensor $(h^{(3)}_{({\rm \bf j})\ell_{\rm \bf j},m_{\rm \bf j}})_{\mu\nu}dx^\mu dx^\nu$ at the origin that is moreover manifestly asymptotically global AdS.

Take now the special resonant harmonics listed in the fourth and fifth columns of Tables \ref{Table:geonsS1}-\ref{Table:geonsV}. Recall that these have quantum numbers that are found  in the normal mode spectrum \eqref{spectrumS} or \eqref{spectrumV} of the global AdS background.  We are thus in a resonant situation where the source $\widetilde{\mathcal{T}}(r) \cos\left( \omega t - m \phi \right)$ of our differential equation has the {\it same} $\{t,\phi\}$ functional dependence  $-$ because $\omega\equiv \bar{\omega}$ $-$ as its homogeneous solution $\Phi_H(r)\cos\left( \bar{\omega} t - m \phi \right)$. In such cases, standard PDE theory instructs us that the most general solution of the inhomogeneous differential equation is a sum of the {\it homogeneous} solution and a {\it particular} solution that is of the form $ \Phi_{P,s}(r) \,t \,\sin\left( \bar{\omega} t - m \phi \right)+\Phi_{P,c}(r) \,t \, \cos\left( \bar{\omega} t - m \phi \right)$. Such a particular solution is {\it secular} since it grows polynomially in time. We find that the equations of motion require that $\Phi_{P,c}(r)=0$.
We are left with the sum of the homogeneous $\Phi^{(3)}_H$ and of the particular $\Phi_{P}^{(3)}\equiv\Phi_{P,s}$ solutions:\footnote{In the main text we are oversimplifying the discussion to keep it clear. In the actual computation for the resonant cases we use the trignometric addition formula $\cos\left( \bar{\omega}_{\ell_{\rm \bf s}} t - m_{\rm \bf s} \phi \right)=\cos\left( \bar{\omega}_{\ell_{\rm \bf s}} t \right) \cos\left( m_{\rm \bf s} \phi \right)-\sin\left( \bar{\omega}_{\ell_{\rm \bf s}} t \right) \sin\left( m_{\rm \bf s} \phi \right)$ and similarly for the sine to decompose the source tensor into $\mathcal{T}^{(3)}_{({\rm \bf s})\ell_{\rm \bf s},m_{\rm \bf s}}=\mathcal{T}_c(t,r,\theta) \cos(m_{\rm \bf s}\phi)+\mathcal{T}_s(t,r,\theta) \sin(m_{\rm \bf s}\phi)$. Then, for each resonant mode, there is not one but  a pair of Kodama-Ishibashi master variables. One is associated to the cosine source contribution $\widetilde{\mathcal{T}}_c(t,r)$ and the other to the sine contributions $\widetilde{\mathcal{T}}_s(t,r)$. The first one is essentially described in the end of the day by \eqref{secular}, while the second has a similar description after replacing $\cos \leftrightarrow \sin$ in  \eqref{secular}. 
} 
\begin{eqnarray}\label{secular}
\Phi^{(3)}_{({\rm \bf j})\ell_{\rm \bf j},m_{\rm \bf j}}(t,r,\phi)&=&\Phi^{(3)}_H\left(r;\alpha^{(3)}_{\ell_{\rm \bf j},m_{\rm \bf j}},\beta^{(3)}_{\ell_{\rm \bf j},m_{\rm \bf j}}\right)\cos\left( \bar{\omega}_{\ell_{\rm \bf j},p_{\rm \bf j}} t - m_{\rm \bf j} \phi \right) \\
&& +\Phi^{(3)}_P\left(r;\widetilde{\alpha}^{(3)}_{\ell_{\rm \bf j},m_{\rm \bf j}},\widetilde{\beta}^{(3)}_{\ell_{\rm \bf j},m_{\rm \bf j}}\right)\,t\,\sin\left( \bar{\omega}_{\ell_{\rm \bf j},p_{\rm \bf j}} t - m_{\rm \bf j} \phi \right) .\nonumber 
\end{eqnarray}
The particular solution $\Phi^{(3)}_P(r)$ obeys the Kodama-Ishibashi master equation \eqref{eq:master} {\it without} a source. Once we solve it, we can find $\Phi^{(3)}_H(r)$ which satisfies the Kodama-Ishibashi master equation \eqref{eq:master}  with an inhomogeneous term that is sourced by $\Phi^{(3)}_P(r)$. 
Note that $\Phi^{(3)}_P(r)$ depends on the two arbitrary integration constants $\widetilde{\alpha}^{(3)}_{\ell_{\rm \bf s},m_{\rm \bf s}}$ and $\widetilde{\beta}^{(3)}_{\ell_{\rm \bf s},m_{\rm \bf s}}$. Similarly $\Phi^{(3)}_H(r)$ is a linear combination of two independent solutions with arbitrary amplitudes $\alpha^{(3)}_{\ell_{\rm \bf s},m_{\rm \bf s}}$ and $\beta^{(3)}_{\ell_{\rm \bf s},m_{\rm \bf s}}$. These are to be fixed by the boundary conditions that the total solution must obey. 

Imposing the asymptotic global AdS boundary condition on these resonant solutions follows a strategy similar to the one for generic non-resonant modes. The only difference is that this time we have a homogeneous contribution $\cos\left( \bar{\omega}_{\ell_{\rm \bf s}} t - m_{\rm \bf s} \phi \right)$ and a secular contribution $t\,\sin\left( \bar{\omega}_{\ell_{\rm \bf s}} t - m_{\rm \bf s} \phi \right)$ with the extra pair of integration constants associated to this particular solution. A judicious choice of one of the homogeneous and particular integration constants, say  $\beta^{(3)}_{\ell_{\rm \bf s},m_{\rm \bf s}}$ and $\widetilde{\beta}^{(3)}_{\ell_{\rm \bf s},m_{\rm \bf s}}$, together with a gauge transformation \eqref{KI:gaugeS0} with \eqref{gaugetransf} yields a manifestly asymptotically global AdS perturbation. Crucially, as we will find out below, it turns out that the unique choice for the secular integration constant is $\widetilde{\beta}^{(3)}_{\ell_{\rm \bf s},m_{\rm \bf s}}=0$.

We still need to impose regularity of  the total solution $\Phi^{(3)}_H+\Phi^{(3)}_P$ at the origin. The solution can be made regular at the origin with a {\it non-vanishing} selection of the {\it secular} integration constant $\widetilde{\alpha}^{(3)}_{\ell_{\rm \bf s},m_{\rm \bf s}}$. Thus, we have a regular solution but at the cost of having a growing secular resonant mode. 
In this process, the homogeneous integration constant $\alpha^{(3)}_{\ell_{\rm \bf s},m_{\rm \bf s}}$, that was not fixed in this process, is left undetermined at order $k=3$. 

Yet, we might still ask wether we can remove such a resonance with a {\it resumation} or a {change of basis} procedure. To analyse this possibility we have to further distinguish the  class of resonant modes in the fourth column of the Tables \ref{Table:geonsS1}-\ref{Table:geonsV} from those in the last column (if present). Indeed, the resonant modes in the fourth column  are further special because their quantum numbers  coincide with those present in our seed (first column). Consequently, their Kodama-Ishibashi equation and solution \eqref{secular} depends  also on the frequency correction $\omega^{(2)}$ introduced in \eqref{O2:wexpansion}. This dependence is such that the aforementioned choice of $\widetilde{\alpha}^{(3)}_{\ell_{\rm \bf s},m_{\rm \bf s}}$ (that makes the resonant solution regular at the origin) turns out not to be unique but is a function of $\omega^{(2)}$. We can then use this freedom to chose uniquely the frequency correction $\omega^{(2)}$ to make the solution regular at the origin and at the same time annihilate the secular integration constant, $\widetilde{\alpha}^{(3)}_{\ell_{\rm \bf s},m_{\rm \bf s}}=0$. The frequency correction $(-L\,\omega^{(2)})$ choice that does this job is in between brackets in the fourth column of Tables \ref{Table:geonsS1}-\ref{Table:geonsV}. \footnote{\label{foot:poincare}The resummation procedure \eqref{O2:wexpansion} whereby one takes into account the dependence of the frequency on the expansion parameter $\varepsilon$ is a standard procedure within the Poincar\'e-Lindstedt method to find periodic solutions. Essentially, it follows from the fact that $\cos\left[ \left( \bar{\omega}+\varepsilon^2 \omega^{(2)} \right) t \right] =\cos(\bar{\omega} t)+\varepsilon^2\,\omega^{(2)}\, t \, \sin(\bar{\omega} t) +\mathcal{O}(\varepsilon^4)$ for  $\varepsilon^2  \omega^{(2)} \ll \bar{\omega}$. See e.g. \cite{Hunter:2004}.} 
Recall that the asymptotic boundary condition requires that the other secular integration constant $\widetilde{\beta}^{(3)}_{\ell_{\rm \bf s},m_{\rm \bf s}}$ vanishes. We thus conclude that the introduction of the frequency correction in \eqref{O2:wexpansion} is not only natural from a physical and perturbation scheme perspective, but it also amounts to a resumation procedure that removes the resonances  associated to the $\mathcal{O}(\varepsilon^3)$ modes with quantum numbers that match the seed mode.

However, no such resummation procedure is available in the case of the resonant modes ({\it if present}) displayed in the last column of Tables \ref{Table:geonsS1}-\ref{Table:geonsV}. More specifically, we are not aware of a resummation procedure that removes these resonances without {\it i)} promoting the mode amplitudes to be functions of time (which is not periodic) or {\it ii)} adding new modes to the linear seed as done in  \cite{Rostworowski:2016isb,Rostworowski:2017tcx,Martinon:2017uyo}. In particular, their quantum numbers do not match those of the seed and thus we do not have frequency corrections to avoid the secular terms. These modes source a solution that is asymptotically global AdS and regular everywhere  but contains {\it irremovable} resonances with a secular growing amplitude.
Consequently, perturbation theory with expansion parameter $\varepsilon$ breaks down at third order  with secular growing modes linear in time, $\varepsilon^3 t$. 

It is instructive to analyse the necessary and sufficient conditions for the existence of resonances at $\mathcal{O}(\varepsilon^3)$. Such resonances can occur only when 
\begin{equation}
\bar{\omega}_1+\bar{\omega}_2-\bar{\omega}_3=\omega\,,\qquad m_1+m_2-m_3=m\,,\qquad \ell\geq \hbox{min}\{2,|m|\},
\label{cond:resonance1}
\end{equation}
where $\bar{\omega}_i$ are one of the frequencies of the quantized linear spectrum \eqref{spectrumS} or \eqref{spectrumV} that is/are present in the linear seed, and $\{\ell,m,p,\omega\}$ are the quantum numbers characterizing an excited mode at order $\mathcal{O}(\varepsilon^3)$. 
For a linear seed that contains a single scalar (${\rm j}={\rm s}$) or vector mode (${\rm j}={\rm v}$) mode, the three sets of $\{\ell,m,p,\bar{\omega}\}_{1,2,3}$ are the same, $\{\ell_{\rm j},m_{\rm j},p_{\rm j},\bar{\omega}_{\rm j}\}$, and the condition \eqref{cond:resonance1} is simply 
\begin{equation}
\bar{\omega}_{\rm j}=\omega\,,\qquad m_{\rm j}=m\,,\qquad \ell\geq \hbox{min}\{2,|m|\}. 
\label{cond:resonance2}
\end{equation} 
If the excited mode at $\mathcal{O}(3)$ happens to be a scalar mode, it follows from \eqref{spectrumS} that  \eqref{cond:resonance2} can be rewritten as
\begin{equation}
c_{\rm j}+\ell_{\rm j}+2p_{\rm j}=1+\ell+2p\quad \Leftrightarrow \quad \ell=\ell_{\rm s}+ c_{\rm j}-1+2\left( p_{\rm j}-p \right)\,,\quad \ell\geq \hbox{min}\{2,|m|\}.
\label{cond:resonanceS}
\end{equation}
where $c_{\rm j}=1$ ($c_{\rm j}=2$) for a linear seed with a scalar (vector) mode.   
Alternatively, the excited mode at $\mathcal{O}(3)$ can be a vector mode in which case it follows from \eqref{spectrumV} that  \eqref{cond:resonance2} can be rewritten as
\begin{equation}
c_{\rm j}+\ell_{\rm j}+2p_{\rm j}=2+\ell+2p\quad \Leftrightarrow \quad \ell=\ell_{\rm s}+c_{\rm j}-2+2\left( p_{\rm j}-p \right)\,,\quad \ell\geq \hbox{min}\{2,|m|\}. 
\label{cond:resonanceV}
\end{equation} 
Given a linear seed with a single scalar or vector mode, we cannot predict wether the nonlinear Einstein equation does excite at third order a mode obeying the resonant conditions \eqref{cond:resonanceS} or/and \eqref{cond:resonanceV}. We need to do the actual computation to find the answer.  
This analysis summarized in Tables \ref{Table:geonsS1} and \ref{Table:geonsS2} concludes that, when we start with a seed that has a single scalar mode such that i) $\ell_{\rm s}=|m_{\rm s}|\geq 2$ and $p_{\rm s}=0$ or ii) $\ell_{\rm s}= 2$, $m_{\rm s}=0,1$ and $p_{\rm s}=0$, the only excited mode at $\mathcal{O}(3)$ that obeys \eqref{cond:resonanceS} has $p=p_{\rm s}=0$ and $\ell=\ell_{\rm s}$. Moreover, there are no excited modes satisfying \eqref{cond:resonanceV}. 
In both these families of linear seeds, we thus find that a resonance appears in a mode that coincides with the mode present in the linear seed and this resonance can be removed with a shift on the frequency as explained above. Similarly, the analysis summarized in Table \ref{Table:geonsV} also finds that when the linear seed contains a single vector mode with $\ell_{\rm v}= 2$, $m_{\rm v}=0$ and $p_{\rm v}=0$, the only excited mode at $\mathcal{O}(3)$ that obeys \eqref{cond:resonanceV} has $p=p_{\rm v}=0$ and $\ell=\ell_{\rm v}=0$. In addition, there are no excited modes satisfying \eqref{cond:resonanceS}. Again, this is a removable resonance with a frequency correction. 

All other linear seeds with a single scalar or with a single vector mode develop, in addition to the removable resonance, irremovable resonances that obey \eqref{cond:resonanceS} or \eqref{cond:resonanceV} at order  $\mathcal{O}(\varepsilon^3)$ with $\{\ell,m,p,\omega\}\neq \{\ell_{\rm j},m_{\rm j},p_{\rm j},\bar{\omega}_{\rm j}\}$. Again, these irremovable resonances are listed in the last column of Tables \ref{Table:geonsS1}-\ref{Table:geonsV}.

Before ending this section, it is worth emphasizing that we had to solve several coupled systems of ODEs to arrive to our conclusions. To guarantee that we had not missed something in the process, for each case we analysed we have explicitly checked that the total gravitational field (\emph{i.e.} built from the sum of all excited harmonics), $g^{(3)} = \bar{g}+ \varepsilon^1 h^{(1)}+\varepsilon^2 h^{(2)}+\varepsilon^3 h^{(3)}$, does solve the Einstein equation \eqref{eq:action} up to $\mathcal{O}(\varepsilon^3)$. 

\subsection{Geons that are the back-reaction of a single normal mode of AdS  \label{sec:Grav4geon}}

Tables \ref{Table:geonsS1}-\ref{Table:geonsV} show that there are a few individual gravitational normal modes of AdS$_4$ that do not develop irremovable secular resonances when back-reacted at $\mathcal{O}(\varepsilon^3)$. These are the modes with the word ``None'' in the fourth column. One of the main results of the present study is that back-reactions of individual normal modes that do not generate irremovable resonances at third order are to be seen as truly exceptional cases. We will argue below that most of these actually can be back-reacted at any order to make a time-periodic smooth soliton. When this nonlinear extension exists we say that we have a  {\it geon}, \emph{i.e.} an asymptotically AdS smooth gravitational solution without a horizon that is the back-reaction up to any order of {\it special} cases of AdS normal modes. 
Only a few individual gravitational normal modes of AdS$_4$ have a nonlinear geonic. Although they are exceptional cases, there is a countable infinite number of geons in these conditions. The educated examples that we have analysed give strong evidence to the following conjecture.

{\it The only individual gravitational normal modes of AdS$_4$ that can be back-reacted up to third order to yield a time-periodic horizonless solution of the Einstein equation are:}
\begin{eqnarray}
&&\hspace{-4cm} \hbox{$\bullet \,$ Scalar modes with $\ell_{\rm s}=|m_{\rm s}|\geq 2$ and $p_{\rm s}=0$, or}\nonumber \\ 
&&\hspace{-4cm}  \hbox{$\bullet \,$ Scalar modes with $\ell_{\rm s}= 2$, $m_{\rm s}=0,1$ and $p_{\rm s}=0$, or}\nonumber \\ 
&&\hspace{-4cm}  \hbox{$\bullet \,$ Vector modes with $\ell_{\rm v}= 2$, $m_{\rm v}=0$ and $p_{\rm v}=0$.} 
\label{geons}
\end{eqnarray}
In one of these cases, namely for the back-reaction of a scalar mode with $\{\ell_{\rm s},m_{\rm s},p_{\rm s},\bar{\omega}_{\rm s}\}=\{2,2,0,3/L\}$, we have further extended the computation up to fifth order. Irremovable resonances are also absent at $\mathcal{O}(\varepsilon^5)$ \cite{Dias:2011ss}. The same is true for the scalar mode with $\{\ell_{\rm s},m_{\rm s},p_{\rm s},\bar{\omega}_{\rm s}\}=\{4,4,0,5/L\}$ \cite{Martinon:2017uyo}. 
Actually, these cases can be back-reacted up to any order in perturbation theory since they have a full nonlinear extension constructed numerically in \cite{Horowitz:2014hja,Martinon:2017uyo}. Such case describes a geon. 

These studies for the $\{\ell_{\rm s},m_{\rm s},p_{\rm s},\bar{\omega}_{\rm s}\}=\{2,2,0,3/L\}$ case, and the observation that the structure of the problem is the same for all $\ell_{\rm s}=|m_{\rm s}|\geq 2$ and $p_{\rm s}=0$ modes, allows to conjecture that the first case in our list \eqref{geons} can be nonlinearly extended to yield a geon.  We do not have a similar argument for the two last cases in list \eqref{geons}. A perturbative extension to higher order, $\geq 5$, or a full nonlinear study is required and left for the future. Accordingly, in the \ref{Table:geonsS1}-\ref{Table:geonsV} we leave these three cases with a question mark: ``Geon?''.

There are reasons to consider the conclusions summarized in the list \eqref{geons} as an unexpected result. Indeed, instead of the gravitational sector, consider for a moment a spherically symmetric real or a complex scalar field in AdS$_4$. A linear perturbation analysis yields a quantized spectrum of frequencies given by $\bar{\omega}_p \,L= 3+2 p$ where $p\in\{0,1,2\ldots \}$ is again the radial overtone. The back-reaction of {\it all} these scalar normal modes yields a smooth asymptotically AdS oscillon (for a real spherically symmetric scalar field) or boson star (for a spherically symmetric complex scalar field) {\it independently} of the radial overtone $p$. On the other hand, in the gravitational sector only those individual normal modes with $p=0$ can be back-reacted to yield a geon and they must further have $\ell_{\rm s}=m_{\rm s}\geq 2$ (or be one of the three cases in the two last items of list \eqref{geons}). This is a sharp difference between the behaviour of spherically symmetric scalar fields and gravitons in global AdS. Ultimately, this difference follows from the fact that the gravitational linear spectrum is degenerate; see \eqref{spectrumS} and \eqref{spectrumV}. 

This fundamental difference has further consequences. As described before, the appearance of irremovable resonances in the perturbative analysis is interpreted as evidence that a black hole forms in a time evolution, no matter how small the amplitude of the initial data is. This is also known as the nonlinear instability of AdS. In short, irremovable resonances in the perturbative analysis correspond to gravitational collapse in the time evolution problem. Several numerical studies available in the literature indeed confirm that there is a set of initial data with non-zero measure that indeed leads to black hole formation for arbitrarily small amplitude of the spherical scalar field shell. 
{\it \`A priori} there is a good argument to expect the nonlinear instability and associated gravitational collapse to a black hole to be shut-down once we break spherical symmetry ($\ell\neq 0$). Indeed, breaking this symmetry amounts to give angular momentum (when $m\neq 0$) to the system and it would be natural to expect the centrifugal effects to balance the gravitational collapse and halt the black formation. Quite remarkably, our perturbative analysis indicates that this intuition is too naive and incorrect. Even more remarkably, the addition of angular momentum further contributes to the nonlinear instability and associated black hole formation. For even initial data with a single non-spherical normal mode of AdS can generically develop resonances at third order, unlike the spherical symmetric scalar field case.   
This observation is one of the main conclusions of the present study.

In view of these arguments, geons that are the back-reaction of a single normal mode are thus a very special class of solutions. It is thus important to study their properties in some detail. Geons are a 1-parameter family of solutions that we can take to be parametrized by their energy. To find their angular momentum $J$ as a function of their energy $E$, we compute the energy-momentum tensor $T^{(4)}_{\mu\nu}$ that sources the fourth-order solution,
\begin{equation}
T^{(4)}_{\mu\nu}(h^{(1)},h^{(2)},h^{(3)}) =R_{\mu\nu}(g^{(3)})+\frac{3}{L^2}\,g^{(3)}_{\mu\nu},
\label{O4:TLL}
\end{equation}
where
 \begin{eqnarray}\label{O4:TLLaux}
&& g^{(3)}_{\mu\nu}= \bar{g}_{\mu\nu}+ \varepsilon^1 h^{(1)}_{\mu\nu}+\varepsilon^2 h^{(2)}_{\mu\nu}+\varepsilon^3 h^{(3)}_{\mu\nu}\,, \nonumber\\
&& g_{(4)}^{\mu\nu}= \bar{g}^{\mu\nu}+ \varepsilon^1 H_{(1)}^{\mu\nu}+\varepsilon^2 H_{(2)}^{\mu\nu}+\varepsilon^3 H_{(3)}^{\mu\nu}+\varepsilon^4 H_{(4)}^{\mu\nu} \,, 
\end{eqnarray}
$H_{(1)},H_{(2)},H_{(3)}$  are defined in \eqref{O3:TLLaux2} and
 \begin{eqnarray}\label{O3:TLLaux2}
 H_{(4)}^{\mu\nu} \equiv -\bar{g}^{\mu\alpha}\left( H_{(3)}^{\nu\beta} h_{\alpha\beta}^{(1)} + H_{(2)}^{\nu\beta} h_{\alpha\beta}^{(2)}+H_{(1)}^{\nu\beta} h_{\alpha\beta}^{(3)}\right).
\end{eqnarray}
One has $g^{(3)}_{\mu\alpha} g_{(3)}^{\alpha\nu}=\delta_\mu^\nu$ up to $\mathcal{O}(\varepsilon^4)$. 

We can now pull-back the total energy-momentum tensor $T_{\mu\nu}=\varepsilon^2 T^{(2)}_{\mu\nu}+\varepsilon^3 T^{(3)}_{\mu\nu}+\varepsilon^4 T^{(4)}_{\mu\nu}$ to a 3-dimensional spatial hypersurface  $\Sigma_t$, with unit normal $n$ and induced metric $\sigma^{\mu\nu}=g_{(3)}^{\mu\nu}+n^{\mu}n^{\nu}$, and project it along the normal direction. We can contract this with the Killing vector $\xi=\partial_t$ (of AdS$_4$) that generates time translations, or with the axisymmetric Killing vector generator $\chi=\partial_\phi$. Finally, integrating the result over the volume of the hypersurface we get the energy or angular momentum of the solution, respectively:
\begin{eqnarray} \label{geonEJ}
E&=&-\int_{\Sigma_t}\sqrt{\sigma} \,T_A^{\:B}  \xi^A n_B\,,\nonumber\\
J&=& \int_{\Sigma_t}\sqrt{\sigma} \,T_A^{\:B}  \chi^A n_B\,.
\end{eqnarray}

For the geon that is the back-reaction of the scalar normal mode $\{ \ell_{\rm \bf s},m_{\rm \bf s}, p_{\rm \bf s},\bar{\omega}_{\ell_{\rm \bf s}}\}=\{4,4,0,5/L\}$ the energy and angular momentum are:
\begin{eqnarray} \label{geonEJ4}
E&=&\varepsilon^2 \,\frac{\pi  L}{G_N} \,\frac{2480625}{16}\left[1-\varepsilon^2
\left( \frac{57578367303705}{6845696} +\frac{6780375 \pi ^2}{16} -2 \alpha^{(3)}_{4,4}  \right)\right],\nonumber\\
J&=& \varepsilon^2 \,\frac{\pi  L^2}{G_N} \,\frac{496125}{4}\left[1-\varepsilon^2
\left( \frac{4424359561185}{526592}+\frac{6780375 \pi ^2}{16} -2 \alpha^{(3)}_{4,4}  \right)\right],
\end{eqnarray}
where we have already replaced the frequency correction $\omega^{(2)}$ by the value $\omega_{4,4,0}^{(2)} = -\frac{7010569125}{77792}\, \frac{1}{L}$ listed in Table \ref{Table:geonsS2}
and $\alpha^{(3)}_{4,4}$ is the integration constant that is left undetermined at order $k=3$ (after requiring the solution to be regular everywhere and asymptotically global AdS).

As a non-trivial check of our computations, we can verify that \eqref{geonEJ4} and the frequency correction  \eqref{O3:wCorrections}  are such that the first law for the geon, $dE=\frac{\omega}{m}\,dJ$, is obeyed up to $\mathcal{O}(\varepsilon^5)$,
\begin{eqnarray} \label{geonEJ}
\partial_\varepsilon E- \frac{\bar{\omega}_{4,4,0}+\varepsilon^2 \omega_{4,4,0}^{(2)}}{4}\,\partial_\varepsilon J=\mathcal{O}(\varepsilon^5).
\end{eqnarray}

Insofar we have not discussed the  physical meaning of the expansion parameter $\varepsilon$.
Yet, we can define it without ambiguity. With no loss of generality, fix the arbitrary integration constant $\alpha^{(3)}_{4,4,0}$ to be such that it annihilates the $\varepsilon^4$ contribution in $E$ in \eqref{geonEJ4}. With this choice for $\alpha^{(3)}_{4,4,0}$  we can now use the expression for $E$ in \eqref{geonEJ4} as a defining condition for the expansion parameter, 
\begin{equation} \label{def:ep}
\varepsilon{\bigl |}_{4,4,0} \equiv \frac{4\sqrt{G_N}}{1575\sqrt{\pi L}}\,\sqrt{E},
\end{equation}
{\it i.e.}, the expansion parameter is proportional to the square root of the energy of the geon. Therefore, a geon with energy $E$ that at linear order is described by a scalar normal mode $\{ \ell_{\rm \bf s},m_{\rm \bf s}, p_{\rm \bf s},\bar{\omega}_{\ell_{\rm \bf s}}\}=\{4,4,0,5/L\}$ of global AdS has angular momentum and frequency
\begin{eqnarray} \label{geonEJ4final}
J {\bigl |}_{4,4,0}&=& \frac{4}{5}\,E L\left( 1+ \frac{890231}{15315300}\,\frac{G_N \,E}{\pi  L}  \right)+\mathcal{O}\left(E^3\right), \nonumber\\
\omega {\bigl |}_{4,4,0}&=& \frac{5}{L}\left( 1- \frac{890231}{7657650}\, \frac{G_N\,E}{\pi  L} \right)+\mathcal{O}\left(E^2\right).
\end{eqnarray}
In these expressions, perturbation theory is valid for values of $E$ such that the next-to-leading term is much smaller than the leading term.
 
We can repeat a similar analysis to get the properties of the geon that at linear order is described by a scalar normal mode $\{ \ell_{\rm \bf s},m_{\rm \bf s}, p_{\rm \bf s}, \bar{\omega}_{\ell_{\rm \bf s}}\}=\{6,6,0,7/L\}$ of global AdS. We use the frequency correction $\omega_{6,6,0}^{(2)}= - \frac{8231910851500875}{3090464}\,\frac{1}{L}$  listed in Table \ref{Table:geonsS2} and we choose an integration constant $\alpha^{(3)}_{6,6,0}$ (that is left undetermined at order $k=3$) to be such that the $\varepsilon^4$ contribution to $E$ vanishes. Then $\varepsilon$ is simply proportional to $\sqrt{E}$, 
\begin{equation} \label{def:ep660}
\varepsilon{\bigl |}_{6,6,0} \equiv \frac{2}{72765} \sqrt{\frac{2 G_N}{15 \pi  L}}\sqrt{E}.
\end{equation}
It follows that this geon is described by 
\begin{eqnarray} \label{geonEJ6final}
J {\bigl |}_{6,6,0}&=& \frac{6}{7}\,E L\left( 1+\frac{502800391}{26235721512}\,\frac{G_N\,E}{\pi  L} \right)+\mathcal{O}\left(E^3\right), \nonumber\\
\omega {\bigl |}_{6,6,0}&=& \frac{7}{L}\left( 1- \frac{502800391}{13117860756} \,\frac{G_N\,E}{\pi  L} \right)+\mathcal{O}\left(E^2\right).
\end{eqnarray}

Repeating the very same procedure for the geon that at linear order is described by the AdS scalar normal mode $\{ \ell_{\rm \bf s},m_{\rm \bf s}, p_{\rm \bf s}, \bar{\omega}_{\ell_{\rm \bf s}}\}=\{3,3,0,4/L\}$, we find that its angular momentum and frequency can be written in terms of its energy as 
\begin{eqnarray} \label{geonEJ3final}
J {\bigl |}_{3,3,0}&=& \frac{3}{4}\,E L\left( 1+\frac{74351}{576576}\,\frac{G_N\,E}{\pi  L} \right)+\mathcal{O}\left(E^3\right), \nonumber\\
\omega {\bigl |}_{3,3,0}&=& \frac{4}{L}\left( 1- \frac{74351}{288288} \,\frac{G_N\,E}{\pi  L} \right)+\mathcal{O}\left(E^2\right),
\end{eqnarray}
and the expansion parameter is nothing but $\varepsilon \equiv \frac{2}{15} \sqrt{\frac{G_N}{15 \pi  L}}\sqrt{E}$.

The angular momentum  and frequency of a geon with energy $E$ that at linear order is described by a scalar normal mode $\{ \ell_{\rm \bf s},m_{\rm \bf s},p_{\rm \bf s}, \bar{\omega}_{\ell_{\rm \bf s}}\}=\{2,2,0,3/L\}$ is \cite{Dias:2011ss}:
\begin{eqnarray} \label{geonEJ2final}
J {\bigl |}_{2,2,0}&=& \frac{2}{3}\,E L\left( 1+\frac{4901}{11340}\,\frac{G_N\,E}{\pi  L} \right)+\mathcal{O}\left(J^3\right), \nonumber\\
\omega {\bigl |}_{2,2,0}&=& \frac{3}{L}\left( 1-\frac{4901}{5670} \,\frac{G_N\,E}{\pi  L} \right)+\mathcal{O}\left(E^2\right),
\end{eqnarray}
with $\varepsilon=\frac{16}{9} \sqrt{\frac{G_N}{\pi  L}}\sqrt{E}$.

We can also back-react the normal mode $\{ \ell_{\rm \bf s},m_{\rm \bf s},p_{\rm \bf s}, \bar{\omega}_{\ell_{\rm \bf s}}\}=\{2,1,0,3/L\}$  up to $\mathcal{O}(3)$. The angular momentum and frequency of this third-order solution are,
\begin{eqnarray} \label{geonEJ1final}
J {\bigl |}_{2,1,0}&=& \frac{1}{3}\,E L\left( 1+\frac{41}{162}\,\frac{G_N\,E}{\pi  L} \right)+\mathcal{O}\left(J^3\right), \nonumber\\
\omega {\bigl |}_{2,1,0}&=& \frac{3}{L}\left( 1-\frac{41}{81} \,\frac{G_N\,E}{\pi  L} \right)+\mathcal{O}\left(E^2\right),
\end{eqnarray}
with $\varepsilon=\frac{8}{9} \sqrt{\frac{G_N}{\pi  L}}\sqrt{E}$.

All geons have the remarkable property that they are invariant {\it only} under the helical Killing vector field,
\begin{equation} \label{helicalKVF}
K\equiv\partial_t+\frac{\omega}{m}\partial_\phi.
\end{equation}
Therefore, these solitonic solutions  are not time symmetric neither axisymmetric. Instead, they are  time-periodic \cite{Dias:2011ss}.

\begin{figure}[ht]
\centerline{
\includegraphics[width=0.55\textwidth]{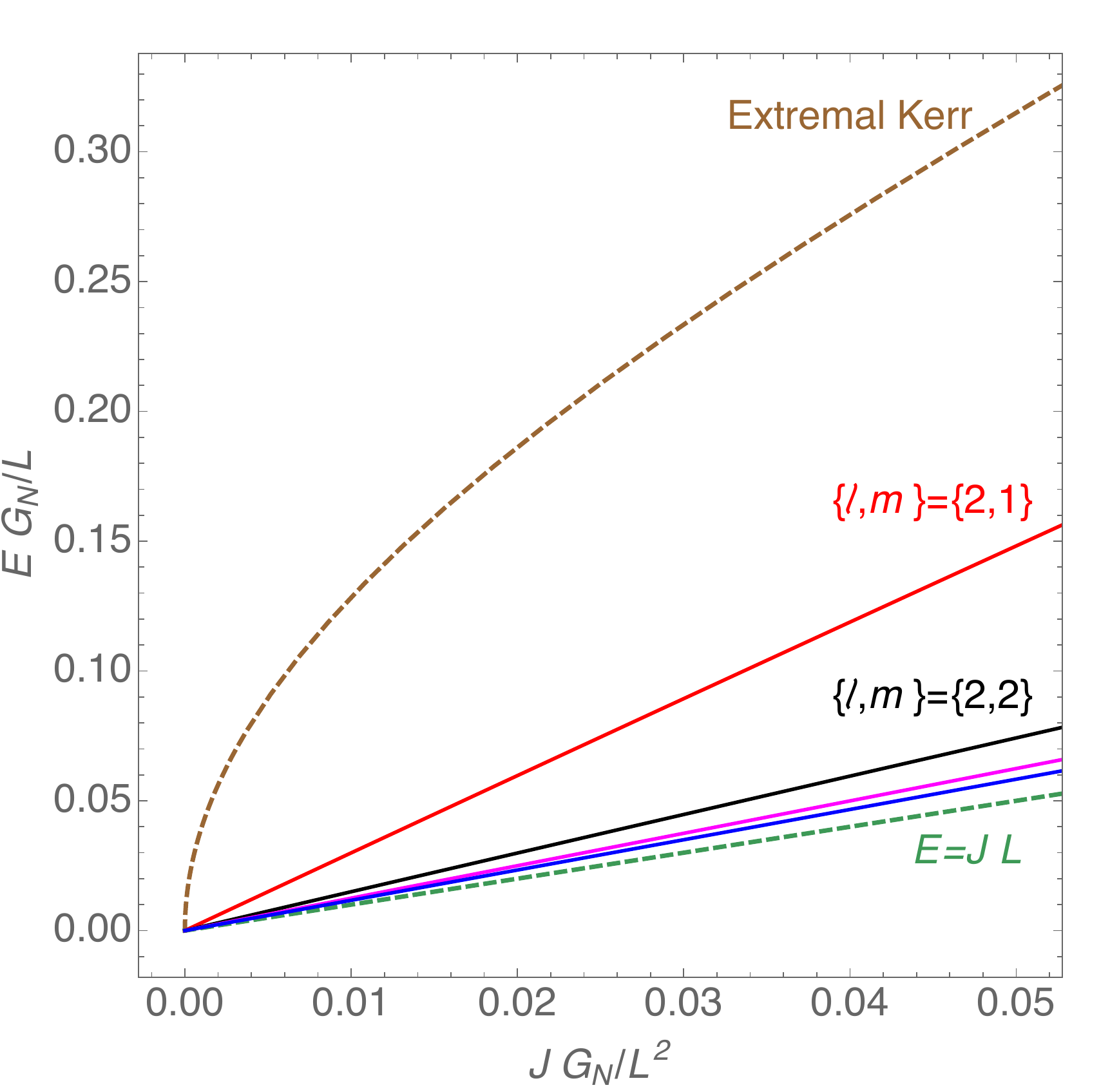}
}
\caption{Phase diagram of global AdS stationary solutions, for small $E/L$ and $J/L^2$. Non-extremal Kerr-AdS BHs exist only above the extremal black line (dashed brown region).   The dashed green curve in the bottom is the $E=J L$ line. From the top to the bottom, the lines describe  geons with $\{\ell_{\rm_s},m_{\rm s},p_{\rm s}\}$ given by $\{2,1,0\}$, $\{2,2,0\}$, $\{4,4,0\}$ and $\{6,6,0\}$, respectively. This diagram does not display the black resonators of \cite{Dias:2015rxy}.}
\label{fig:phasediag}
\end{figure}

We can now consider the special cases with $m_{\rm \bf j}=0$, \emph{i.e.} the back-reaction of normal modes that have no angular momentum.
The frequency of the third-order solution that at linear order is described by the scalar normal mode $\{ \ell_{\rm \bf s},m_{\rm \bf s},p_{\rm \bf s}, \bar{\omega}_{\ell_{\rm \bf s}}\}=\{2,0,0,3/L\}$ is:\footnote{The energy has a $\mathcal{O}(\varepsilon^4)$ contribution that depends on an integration constant that is left undetermined after imposing the boundary condition at third order. We choose this integration constant to eliminate the $\mathcal{O}(\varepsilon^4)$ correction to the energy. The energy is then just a function of $\varepsilon^2$ which we use to express the expansion parameter as the energy of the geon. We replace this in the expression \eqref{O2:wexpansion} with the frequency correction in Table \ref{Table:geonsS1} to get \eqref{geonEsL2m0final}.}
\begin{eqnarray} \label{geonEsL2m0final}
\omega {\bigl |}_{2,0,0}&=& \frac{3}{L}\left( 1-\frac{407}{630} \,\frac{G_N\,E}{\pi  L} \right)+\mathcal{O}\left(E^2\right),
\end{eqnarray}
and the expansion parameter is interpreted as the energy of the solution, $\varepsilon \equiv \frac{8}{3} \sqrt{\frac{2G_N}{3 \pi L}}\sqrt{E}$. 

Finally, consider the only solution in the vector sector that yields no irremovable resonances at $\mathcal{O}(3)$. It can be constructed back-reacting the vector normal mode $\{ \ell_{\rm \bf v},m_{\rm \bf v},p_{\rm \bf v}, \bar{\omega}_{\ell_{\rm \bf v}}\}=\{2,0,0,4/L\}$. It breaks spherical symmetry since $\ell_{\rm \bf v} \neq 0$ but is axisymmetric and thus has no angular momentum. Its frequency is 
\begin{eqnarray} \label{geonEvL2m0final}
\omega_{\rm v} {\bigl |}_{2,0,0}&=& \frac{4}{L}\left( 1-\frac{1469}{1260} \,\frac{G_N\,E}{\pi  L} \right)+\mathcal{O}\left(E^2\right),
\end{eqnarray}
and the expansion parameter is just the energy, $\varepsilon \equiv  \frac{16}{\sqrt{3 \pi L}}\sqrt{G_N E}$.

The third-order solutions described by \eqref{geonEsL2m0final} and \eqref{geonEvL2m0final} are   axisymmetric (since $m_{\rm j}=0$) and time-dependent. That is, $\partial_\phi$ is a Killing vector field but  $\partial_t$ is not. Moreover, these solutions  break the SO(3) spherical symmetry (since $\ell_{\rm j}=2$).

It is instructive to display the geons in a phase diagram of solutions that asymptote to global AdS$_4$. In particular, it is important to locate the 1-parameter family of geons with $\ell_{\rm s}=m_{\rm s}$ and $p_{\rm s}=0$ relatively to the extremal Kerr solution and to the BPS relation $E= J L$. We do this Fig. \ref{fig:phasediag}, which displays a phase diagram of energy versus the angular momentum. We find that, for a given angular momentum, (non-axisymmetric) geons always have less energy than the extremal (and thus any other) Kerr black hole. Moreover, again for fixed $J$, as we increase  $\ell_{\rm s}=m_{\rm s}$, the energy of the geons decreases. In the limit where $\ell_{\rm s}=m_{\rm s}\to \infty$ the geon curves approach progressively the BPS line $E= J L$. Another family of solutions that asymptotes to global AdS$_4$, but is not represented in the phase diagram of Fig. \ref{fig:phasediag} are the black resonators \cite{Dias:2015rxy}. Black resonators  are time-periodic black holes with a single Killing vector field (that generates their horizon) and whose zero-size limit is a geon. We ask the reader to see \cite{Dias:2015rxy} for further details.

\begin{figure}[ht]
\centerline{
\includegraphics[width=1.0\textwidth]{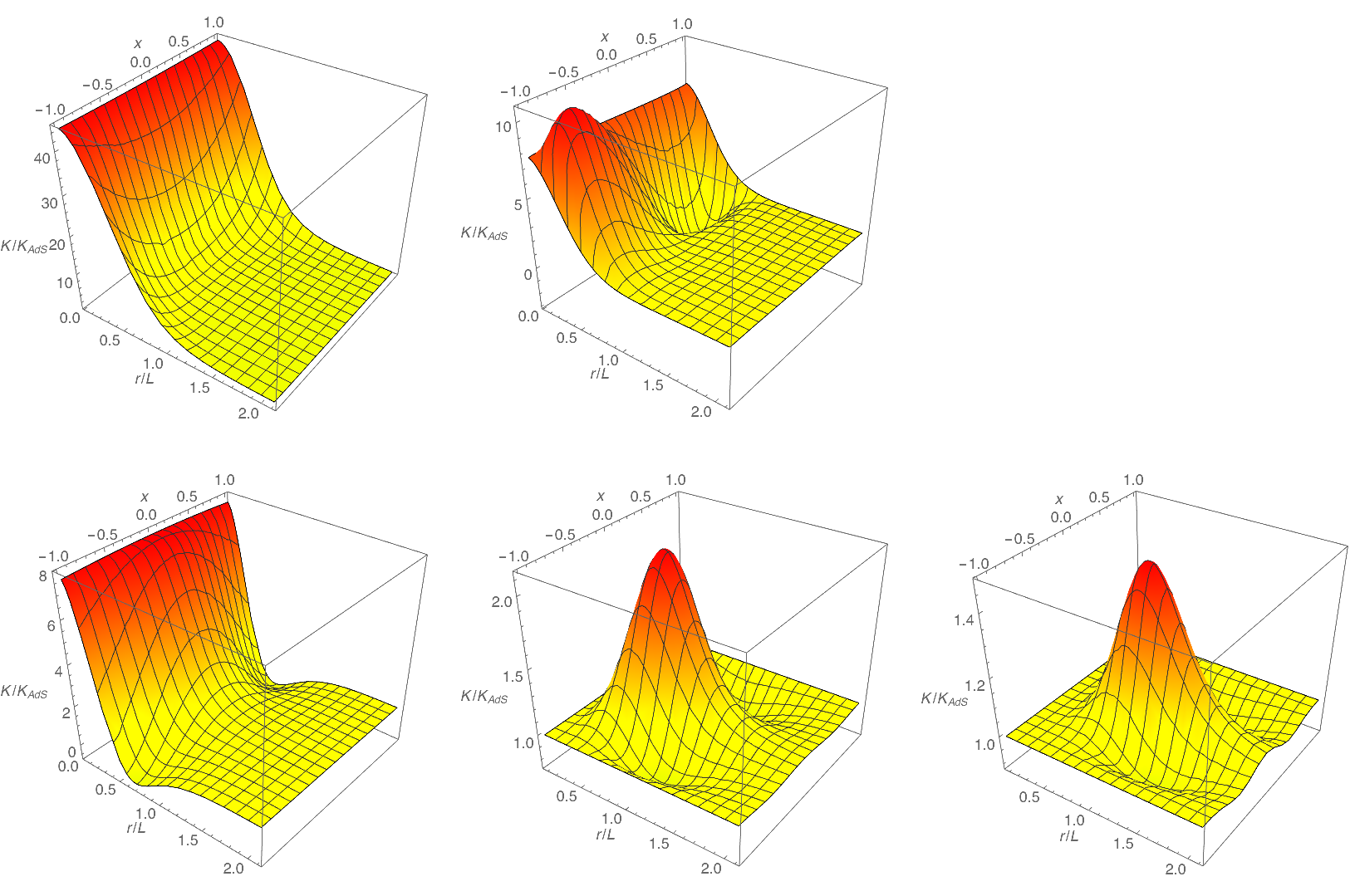}
}
\caption{Kretschman curvature, normalized with respect to the AdS curvature $K_{AdS}$, in the co-rotating frame of the geons as a function of the radial direction in AdS units, $r/L$, and of the polar direction $x=\cos\theta$. As we move from left to right and top down, the plots describe the geons with $\{\ell_{\rm s},m_{\rm s}\}=\{2,0\}$, $\{\ell_{\rm s},m_{\rm s}\}=\{2,1\}$, $\{\ell_{\rm s},m_{\rm s}\}=\{2,2\}$, $\{\ell_{\rm s},m_{\rm s}\}=\{4,4\}$, $\{\ell_{\rm s},m_{\rm s}\}=\{6,6\}$. All the geons have the same energy $E G_N/L=0.5$.}
\label{fig:curvature}
\end{figure}

Another  quantity that is of interest is the curvature of the geon as a function of the spatial directions. A good gauge invariant measure of the curvature is the Kretschman scalar, $K=R_{\mu\nu\alpha\beta}R^{\mu\nu\alpha\beta}$. For a geon, this quantity depends on all spacetime coordinates $\{t,r,\theta,\phi\}$. However, we can move to a frame that co-rotates with the geon, as dictated by it helical Killing vector field \eqref{helicalKVF}. This is achieved if the observer co-rotates  with angular velocity $\omega/m$ \emph{i.e.} if we set $\phi=\omega t$. It is further convenient to consider the curvature of the geon normalized to the curvature of the global AdS background, $K_{AdS}=24/L^4$. To compare appropriately the spatial pattern of curvature of the different geons we take all the geons to have the same energy. The curvatures of  several geons computed in these conditions are plotted in Fig. \ref{fig:curvature}. We see that the highest local curvature is reached in the axisymmetric case with $\{\ell_{\rm s},m_{\rm s}\}=\{2,0\}$ in which case the maximum curvature is localized at the origin of the radial coordinate. However, as $\ell_{\rm s}$ increases we find that the maximum of the curvature progressively moves away from the origin of the radial space and its maximum value decreases (again, for the same total energy). This result is consistent with the following observations. Geons with $\ell_{\rm s}=m_{\rm s}$ and $p_{\rm s}=0$ rotate with angular velocity $\frac{\omega_{\ell_{\rm s},p_{\rm s}}L}{m_{\rm s}}=\frac{1+\ell_{\rm s}}{\ell_{\rm s}}$, which decreases and approaches $1$ as $\ell_{\rm s}$ increases. On the other hand, the angular momentum increases as $\ell_{\rm s}$ grows. This means that momentum of inertia of a geon, for fixed energy, must increase as $\ell_{\rm s}$ grows, \emph{i.e.} its energy must localize at larger radial distances. This is what Fig. \ref{fig:curvature} concludes. These results have further consequences that we postpone until the end of this section.

\begin{figure}[ht]
\centerline{
\includegraphics[width=.53\textwidth]{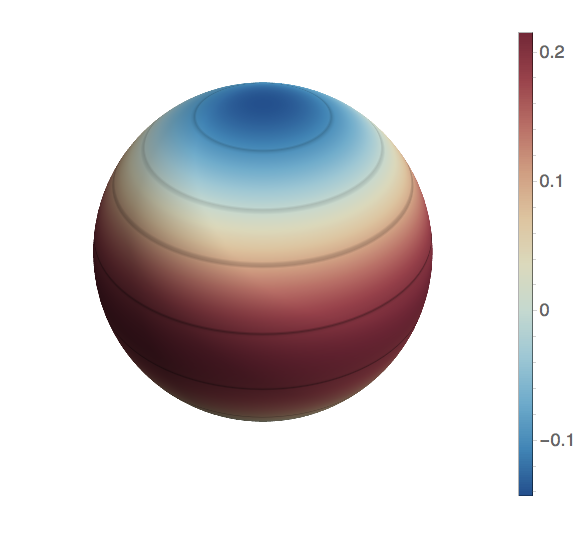}
\includegraphics[width=.5\textwidth]{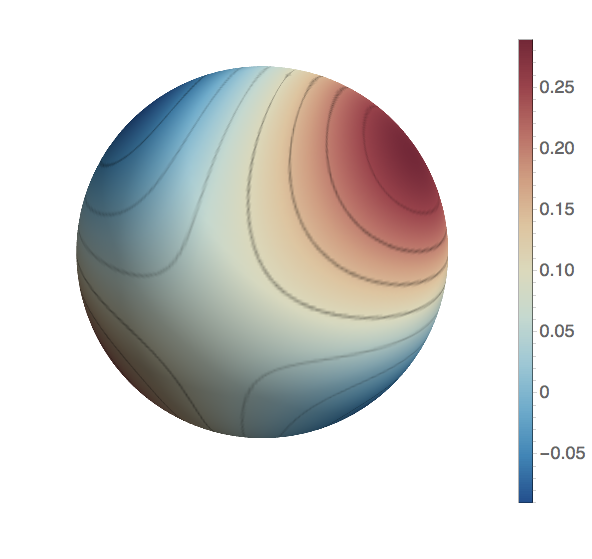}
}
\caption{Energy density $\rho$ on the holographic $S^2$ boundary. {\it Left panel:} $\{\ell_{\rm s},m_{\rm s}\}=\{2,0\}$.  {\it Right panel:}  $\{\ell_{\rm s},m_{\rm s}\}=\{2,1\}$. In these figures (and in Figs. \ref{fig:holoE2} and \ref{fig:holoE3}) the vertical axis defines the North and South poles and the total energy is the same for all solutions. When $m_{\rm s}\neq 0$, the solution is rotating around the vertical axis with angular velocity $\omega_{\ell_{\rm s},0}/m_{\rm s}$.}
\label{fig:holoE1}
\end{figure}
  
\begin{figure}[ht]
\centerline{
\includegraphics[width=.5\textwidth]{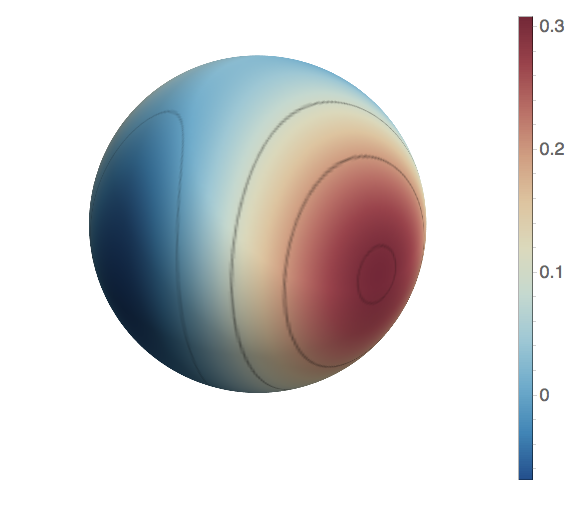}
\includegraphics[width=.5\textwidth]{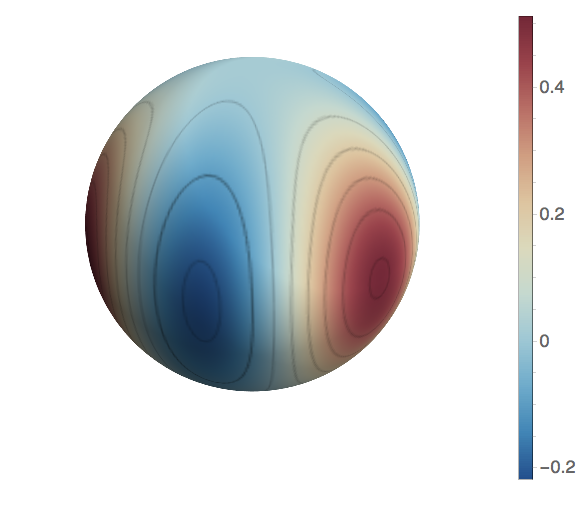}
}
\caption{Energy density $\rho$ on the holographic $S^2$ boundary. {\it Left panel:} $\{\ell_{\rm s},m_{\rm s}\}=\{2,2\}$.  {\it Right panel:}  $\{\ell_{\rm s},m_{\rm s}\}=\{3,3\}$. See caption of Fig. \ref{fig:holoE1} for further information.}
\label{fig:holoE2}
\end{figure}
  
\begin{figure}[ht]
\centerline{
\includegraphics[width=.5\textwidth]{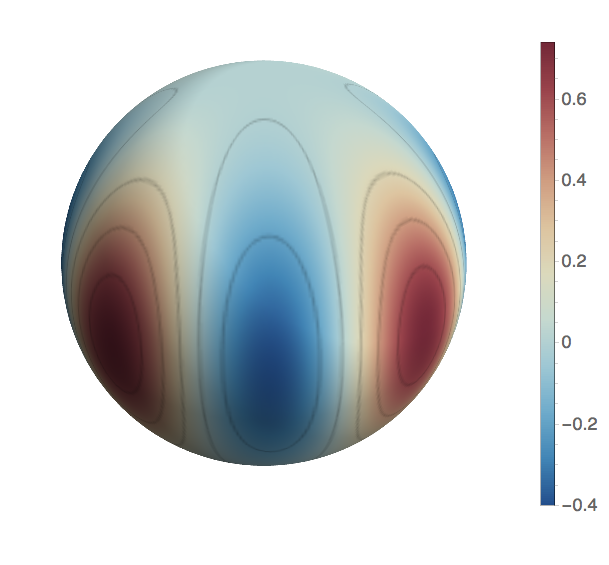}
\includegraphics[width=.58\textwidth]{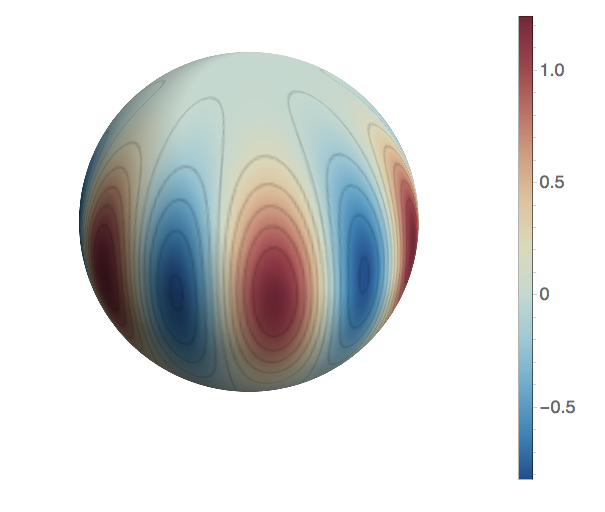}
}
\caption{Energy density $\rho$ on the holographic $S^2$ boundary. {\it Left panel:} $\{\ell_{\rm s},m_{\rm s}\}=\{4,4\}$.  {\it Right panel:}  $\{\ell_{\rm s},m_{\rm s}\}=\{6,6\}$. See caption of Fig. \ref{fig:holoE1} for further information.}
\label{fig:holoE3}
\end{figure}

The energy and angular momentum densities of the geons measured in the holographic boundary of global AdS, $r\to \infty$, are also interesting to analyse. Firstly, because they give the densities of the CFT  states that are dual to the geons in the context of the AdS$_4$/CFT$_3$ correspondence. Second, because geons typically break spherical symmetry and axisymmetry and thus their energy and angular momentum densities have a non-trivial profile as we move along the coordinates $\theta$ and $\phi$ of the $S^2$ boundary. To determine these densities we need to compute the expectation value of the holographic stress tensor $\langle \tau_{ik} \rangle$, with $i,k=\{t,\theta,\phi\}$. This is done using either the holographic renormalization procedure of \cite{Balasubramanian:1999re} or  \cite{deHaro:2000vlm}. We can then pull-back $\langle \tau_{ik} \rangle$ into a 3-dimensional spatial hypersurface  $\Sigma_t$, with unit normal $n$ and induced metric $\sigma^{ik}=g^{ik}+n^{i}n^{k}$. Contracting this with the vector $\xi=\partial_t$ that generates time translations yields the energy density $\rho$, while contracting it with the vector $\chi=\partial_\phi$ that generates azimuthal translations yields the angular momentum density $j$: 
\begin{eqnarray} \label{HoloDensities}
\rho&=&- \langle T_i^{\,k} \rangle n_k \xi^i,\nonumber\\
j&=& \langle T_i^{\,k} \rangle n_k\chi^i.
\end{eqnarray}
We plot the energy density $\rho$ on the holographic $S^2$ boundary in Figs. \ref{fig:holoE1}-\ref{fig:holoE3} for $\{\ell_{\rm s},m_{\rm s}\}=\{2,0\}$, $\{\ell_{\rm s},m_{\rm s}\}=\{2,1\}$, $\{\ell_{\rm s},m_{\rm s}\}=\{2,2\}$, $\{\ell_{\rm s},m_{\rm s}\}=\{3,3\}$, $\{\ell_{\rm s},m_{\rm s}\}=\{4,4\}$ and $\{\ell_{\rm s},m_{\rm s}\}=\{6.6\}$. All these figures represent an instant in time. Except for the case $\{\ell_{\rm s},m_{\rm s}\}=\{2,0\}$, the reader should imagine them as rotating with angular velocity $\frac{\omega_{\ell_{\rm s},0}L}{m_{\rm s}}=\frac{1+\ell_{\rm s}}{m_{\rm s}}$ (\emph{i.e.} in a time-periodic way).  Not surprisingly, the profile of the energy density is that of the spherical harmonic $Y_{\ell_{\rm s}}^{m_{\rm s}}(\theta,\phi)$ that describes the linear seed of the geon. All the solutions have the same total energy. We see that the maximum of the energy density is increasing as $\ell_{\rm s}$ (and $m_{\rm s}$) grows, \emph{i.e.} the energy of the system is localizing in a progressively smaller area. This suggests that in the limit where $\ell_{\rm_s}=m_{\rm s}\to\infty$ the maximum of the energy density will grow unbounded in an infinitesimal area. Note that this seems to suggest that a geon with unit angular velocity (in AdS units) must necessarily be singular, since this would need $\ell\to+\infty$. This is in agreement with the analysis of \cite{Niehoff:2015oga}.
   
\subsection{Individual normal modes without a nonlinear extension  \label{sec:Grav5geon}}
 
Consider now the back-reaction of individual gravitational normal modes that are not described by \eqref{geons}. We may say that these are the majority of the modes given the restrictive conditions imposed in \eqref{geons}. At third order in perturbation theory, seeds with such a {\it single} normal mode generate one or more irremovable secular resonances. This is the perturbative signature that AdS is nonlinearly unstable (even in the absence of matter). 

We can attempt to extrapolate the consequences of this breakdown of perturbation theory to a time evolution scenario that starts with initial data consisting of a single geon that is not in the conditions of  \eqref{geons}. The expansion parameter is $\epsilon$ and the irremovable resonances have amplitude $\epsilon^3 t$. Therefore, the appearance of resonances and associated breakdown of the perturbative analysis should signal a violent event at a timescale $t \geq 1/\varepsilon^2$. It is natural to expect this event to be the formation of a black hole. Indeed, this is what happens in time evolution simulations of a spherically symmetric scalar field collapse in AdS. 

It is however important to emphasize that this conclusion is in stark contrast with what happens within spherical symmetry. Indeed, for the latter to trigger the nonlinear instability it is fundamental to have initial data with at least two normal modes. On the other hand, our prediction is that in the non-spherical sector this occurs even for initial data with a {\it single} normal mode.

In this sense we might say that the non-spherical gravitational sector is more prone to secular resonances and nonlinear instabilities than the spherical scalar field sector. The reason for this difference ultimately lies on the fact that the linear spectrum of frequencies of the gravitational normal modes is degenerate $-$ see \eqref{spectrumS} and  \eqref{spectrumV} $-$ while the frequency spectrum of a spherically symmetric scalar field in AdS is not. That is, the gravitational spectrum is described by {\it two} quantum numbers, namely the angular quantum number $\ell$ and the radial overtone $p$ (and has two sectors), while the frequency spectrum of a spherically symmetric scalar field is described by just the single quantum number $p$.

\section{Direct and inverse turbulent gravitational cascades \label{sec:Grav}}

In this section we show that the weakly perturbative turbulent mechanism predicts/explains the existence of direct but also inverse frequency cascades to be observed in (future)  full time evolution simulations that break spherical symmetry.  We want to do this in one of the simplest possible ways. Although seeds with a single gravitational normal mode can already trigger irremovable secular resonances (vide the last section), these are always such that their frequency is the same as the normal mode frequency we start with. 
In the weakly perturbative turbulent formalism we work with, the smoking gun of a frequency cascade is  given by the appearance of irremovable resonances at third order that have frequencies different from the seed. This necessarily requires that we start with a seed that is the superposition of at least two normal modes, \emph{i.e.} it requires a collision of two or more normal modes. 
Much of the systematic computational machinery required do analyse this collision is exactly the same as that described in the previous section where we considered simpler seeds with just a single mode.

\subsection{Leading order analysis. Two-mode seed.   \label{sec:Grav1}}

 In \cite{Dias:2011ss} we have chosen to start with a linear combination of the two simplest scalar modes, namely $\{ \ell_{\rm \bf s},m_{\rm \bf s},p_{\rm \bf s},\bar{\omega}_{\ell_{\rm \bf s}}\}=\{2,2,0,3/L\}$  and  $\{ \ell_{\rm \bf s},m_{\rm \bf s},p_{\rm \bf s},\bar{\omega}_{\ell_{\rm \bf s}}\}=\{4,4,0,5/L\}$. We found that this leads to a direct cascade, but no hints of an inverse cascade was  observed. This is essentially because one of the frequencies we start with, $\bar{\omega}_{\ell_{\rm \bf s}}=3/L$, is already the lowest frequency that can be found in the characteristic oscillation spectrum of the system: see \eqref{spectrumS} and \eqref{spectrumV}. 

Our aim in this section is to show that the weakly perturbative turbulent analysis can predict the existence of direct but also inverse frequency cascades. For that, we start with one of the simplest seeds that develops frequency cascades in both directions.
Namely, we consider a linear combination of two scalar modes each one of which only has the $\cos\left( \bar{\omega}_{\ell_{\rm \bf s}} t - m_{\rm \bf s} \phi \right) $ contribution $-$ see \eqref{hmetricS} $-$ and specified by the harmonic quantum numbers and amplitudes:
 \begin{eqnarray}
\hbox{\bf Seed:} && \{ \ell_{\rm \bf s},m_{\rm \bf s},p_{\rm \bf s},\bar{\omega}_{\ell_{\rm \bf s}}\}=\{4,4,0,5/L\} \quad  \hbox{with amplitude}\:\:\mathcal{A}^{(1)}_{({\rm \bf s}) 4} \varepsilon,\quad \hbox{and} \nonumber\\ 
&&  \{ \ell_{\rm \bf s},m_{\rm \bf s},p_{\rm \bf s}, \bar{\omega}_{\ell_{\rm \bf s}}\}=\{6,6,0,7/L\} \quad  \hbox{with amplitude}\:\:\mathcal{A}^{(1)}_{({\rm \bf s}) 6}\varepsilon\,,
\label{initialdata}
\end{eqnarray}
where it is understood that $\mathcal{A}^{(1)}_{({\rm \bf s}) 4}$ and $\mathcal{A}^{(1)}_{({\rm \bf s}) 6}$ are $\mathcal{O}(1)$ quantities. We will use the short-hand notation $\omega_4 \equiv\omega_{4,4,0}$ and $\omega_6 \equiv\omega_{6,6,0}$.

\subsection{Second order analysis. Frequency corrections.  \label{sec:Grav2}}

The second order metric $ h^{(2)}$ satisfies the linearized Einstein equation \eqref{eq:perturb}, $\Delta_L h_{\mu\nu}^{(2)} = T^{(2)}_{\mu\nu}$, with $\Delta_L$ given by \eqref{linearGab} and source tensor $T^{(2)}$ given by \eqref{O2:TLL}-\eqref{O2:TLLaux}.
Although we start with the seed \eqref{initialdata} that only has two scalar harmonics, \eqref{O2:TLL} excites many more harmonics at second order and even vector harmonics. 
Like in the single mode seed case we decompose the source $T^{(2)}_{\mu\nu}(h^{(1)})$ as a sum of the fundamental scalar blocks $\mathcal{T}^{(k)}_{({\rm \bf s})\ell_{\rm \bf s},m_{\rm \bf s}}$ and the fundamental vector blocks $\mathcal{T}^{(k)}_{({\rm \bf v})\ell_{\rm \bf v},m_{\rm \bf v}}$, as described in \eqref{decomposeT}. The trignometric addition formulas and the quadratic nature of the source in $h^{(1)}$ guarantee that  quantum numbers of the excited harmonics are sums and differences of the quantum numbers of the two initial modes   \eqref{initialdata}  and bounded above by $ \{ \ell_{\rm \bf s},m_{\rm \bf s}, \omega_{\ell_{\rm \bf s},p_{\rm \bf s}}\}\leq 2\times\{6,6,7/L\}$. With these guides we identify precisely the excited harmonics by a direct inspection of $T^{(2)}_{\mu\nu}(h^{(1)})$.

We then find that at second order, the seed \eqref{initialdata} excites 15 scalar harmonics and 10 vector harmonics. The 15 scalar harmonics are described by the quantum numbers: 
 \begin{eqnarray}
&& \{\ell_{\rm \bf s},m_{\rm \bf s}, \omega \}=\{\ell_{\rm \bf s},0,0\} \quad  \hbox{with}\:\: \ell_{\rm \bf s}=0,2,4,6,8,10,12; \nonumber\\ 
&& \{\ell_{\rm \bf s},m_{\rm \bf s}, \omega\}=\{\ell_{\rm \bf s},2,2/L\} \quad  \hbox{with}\:\: \ell_{\rm \bf s}=2,4,6,8,10;
\nonumber\\ 
&& \{\ell_{\rm \bf s},m_{\rm \bf s}, \omega\}={\bigl\{ } \{8,8,10/L\}, \{10,10,12/L\}, \{12,12,14/L\} 
{\bigr\} },
\label{O2:excitedS}
\end{eqnarray}
while the 10 vector harmonics have quantum numbers:
 \begin{eqnarray}
&& \{\ell_{\rm \bf v},m_{\rm \bf v}, \omega\}=\{\ell_{\rm \bf v},0,0\} \quad  \hbox{with}\:\: \ell_{\rm \bf v}=1,3,5,7,9,11; \nonumber\\ 
&& \{\ell_{\rm \bf v},m_{\rm \bf v}, \omega\}=\{\ell_{\rm \bf v},2,2/L\} \quad  \hbox{with}\:\: \ell_{\rm \bf v}=3,5,7,9.
\label{O2:excitedV}
\end{eqnarray}

Following a systematic procedure akin the Kodama-Ishibashi analysis \cite{Kodama:2003kk} described in the previous sections, next we find the master equation \eqref{eq:master} with potential \eqref{eq:masterPotV} that the second order master variables $ \Phi^{(2)}_{({\rm \bf j})\ell_{\rm \bf j},m_{\rm \bf j}}$ must satisfy. That is, for each excited harmonic listed in \eqref{O2:excitedS}-\eqref{O2:excitedV},  we find the Kodama-Ishibashi source term $\widetilde{\mathcal{T}}^{(2)}_{({\rm \bf j})\ell_{\rm \bf j},m_{\rm \bf j}}$ as a function of the original $\mathcal{T}^{(2)}_{({\rm \bf j})\ell_{\rm \bf j},m_{\rm \bf j}}$ that we had identified in our previous step: see \eqref{sourceKI}.

Like in the single mode seed of the previous section, we use the Kodama-Ishibashi formalism
to solve the 25 inhomogeneous master ODEs \eqref{eq:master} to find $\Phi^{(2)}_{({\rm \bf j})\ell_{\rm \bf j},m_{\rm \bf j}}$. We then use the differential map in Appendix \ref{appendix:KI}  to reconstruct the metric $h^{(2)}_{({\rm \bf j})\ell_{\rm \bf j},m_{\rm \bf j}}$ from $\Phi^{(2)}_{({\rm \bf j})\ell_{\rm \bf j},m_{\rm \bf j}}$ in a given gauge. Finally, we impose the appropriate boundary conditions as described in detail in \eqref{KIasymp}-\eqref{KI:BCorigin}, adding gauge transformations to our  $h^{(2)}_{({\rm \bf j})\ell_{\rm \bf j},m_{\rm \bf j}}$ when necessary.  Namely, we require each $h^{(2)}_{({\rm \bf j})\ell_{\rm \bf j},m_{\rm \bf j}}$ to have a regular centre, and at asymptotic infinity, we require it is asymptotically globally AdS. 
 
Here, it is important to observe that each one of the harmonics excited by the source has a frequency 
$\omega$ that is  given by a sum or difference of the frequencies $\bar{\omega}_4, \bar{\omega}_6$ of our seed \eqref{initialdata}, \emph{i.e.} $\omega \in \{0, \bar{\omega}_6\pm \bar{\omega}_4, \bar{\omega}_4+\bar{\omega}_4,\bar{\omega}_6+\bar{\omega}_6\}$. However, in the excited spectrum \eqref{O2:excitedS} and \eqref{O2:excitedS} we do not find a single source combination of quantum numbers $\{\ell, \omega\}$ that coincides  with one of the elements $\{\ell_{\rm \bf j},\bar{\omega}_{\rm \bf j} \}$ of the linear frequency spectra \eqref{spectrumS} or \eqref{spectrumV}, that is $\omega \neq 1+\ell_{\rm \bf s}+2p$ in \eqref{O2:excitedS} and $\omega \neq 2+\ell_{\rm \bf v}+2p$ in \eqref{O2:excitedV}. That is to say, there are no resonances and the solution can be made straightforwardly regular at the origin with a choice of integrations constants of the homogeneous solution of $\Delta_L h_{\mu\nu}^{(2)} = T^{(2)}_{\mu\nu}$. 
It follows that the total metric \eqref{gexpansion} up to second order, $g^{(2)} = \bar{g}+ \varepsilon^1 h^{(1)}+\varepsilon^2 h^{(2)}$, is regular everywhere and asymptotically global AdS. As a check of our computation, we can confirm that this total metric satisfies the Einstein equation \eqref{eq:action} up to $\mathcal{O}(\varepsilon^2)$. 

Similarly to the single mode seed case, next we consider a Poincar\'e-Lindstedt frequency correction (see \eqref{O2:wexpansion} and footnote \eqref{foot:poincare}) for the two seed frequencies,
 \begin{eqnarray}\label{O2:wexpansion46}
&& \omega_{4} = \bar{\omega}_4+ \varepsilon^2 \omega_4^{(2)}= \frac{5}{L}+ \varepsilon^2 \omega_4^{(2)}, \nonumber\\
&& \omega_{6} = \bar{\omega}_6+ \varepsilon^2 \omega_6^{(2)}= \frac{7}{L}+ \varepsilon^2 \omega_6^{(2)}.
\end{eqnarray}
The structure of the perturbed equations indicates that these frequencies receive corrections only at even orders $k$, $\omega_{4,6}= \bar{\omega}_{4,6}+\sum_{j=1} \varepsilon^{2j}  \omega_{4,6}^{(2j)}$.

\subsection{Third order analysis. Irremovable secular resonances.  \label{sec:Grav3}}

At third order, $k=3$, we start by computing the energy-momentum tensor $T^{(3)}_{\mu\nu}$ that sources \eqref{eq:perturb}. This is given by \eqref{O3:TLL}-\eqref{O3:TLLaux2}.

As described in \eqref{decomposeT}, we decompose the source $T^{(3)}$ as a sum of fundamental scalar blocks $\mathcal{T}^{(3)}_{({\rm \bf s})\ell_{\rm \bf s},m_{\rm \bf s}}$ and vector blocks $\mathcal{T}^{(3)}_{({\rm \bf v})\ell_{\rm \bf v},m_{\rm \bf v}}$.  This procedure identifies the quantum numbers $\{\ell_{\rm \bf s}, m_{\rm \bf s}, \omega \}$ and the seven functions $ \{ \mathcal{T}_{ab}^{(3)},\mathcal{T}^{(3)}_{a},\mathcal{T}^{(3)}_L,\mathcal{T}^{(3)}_T \}_{\ell_{\rm \bf s},m_{\rm \bf s}}$ of each excited scalar harmonic: see \eqref{KI:scalar} and Appendix \ref{appendix:KIscalar}. Similarly, for each  excited vector harmonic we find the quantum numbers $\{\ell_{\rm \bf v}, m_{\rm \bf v}, \omega \}$ and the three functions $ \{ \mathcal{T}^{(3)}_{a},\mathcal{T}^{(3)}_T \}_{\ell_{\rm \bf v},m_{\rm \bf v}}$. From a Clebsch-Gordan analysis we know that for scalar (vector) modes the minimum and maximum $\ell_{\rm \bf s}$ ($\ell_{\rm \bf v}$) that can be excited are (let $\ell_2=6$, $\ell_1=4$):
\begin{eqnarray}
&& \ell_{\rm \bf s}{\bigl |}_{min}=\hbox{min}\{ \ell_1, |\ell_2-2\ell_1|\},\qquad \ell_{\rm \bf s}{\bigl |}_{max}=3\ell_2; \nonumber\\ 
&& \ell_{\rm \bf v}{\bigl |}_{min}=\hbox{min}\{ \ell_1+1, |\ell_2-2\ell_1|+1\},\qquad \ell_{\rm \bf v}{\bigl |}_{max}=3\ell_2-1;
\label{O3:rangeModes}
\end{eqnarray}
Using this guide and a direct inspection of \eqref{O3:TLL}, we find that at third order the seed \eqref{initialdata} and associated second order metric excites 30 scalar harmonics and 22 vector harmonics. The 30 scalar harmonics have the quantum numbers: 
 \begin{eqnarray}
&& \{\ell_{\rm \bf s},m_{\rm \bf s}, \omega \}=\{\ell_{\rm \bf s},2,3/L\} \quad  \hbox{with}\:\: \ell_{\rm \bf s}=2,4,6,8,10,12,14; \nonumber\\ 
&& \{\ell_{\rm \bf s},m_{\rm \bf s}, \omega\}=\{\ell_{\rm \bf s},4,5/L\} \quad  \hbox{with}\:\: \ell_{\rm \bf s}=4,6,8,10,12,14,16;
\nonumber\\ 
&& \{\ell_{\rm \bf s},m_{\rm \bf s}, \omega\}=\{\ell_{\rm \bf s},6,7/L\} \quad  \hbox{with}\:\: \ell_{\rm \bf s}=6,8,10,12,14,16,18;   \\ 
&& \{\ell_{\rm \bf s},m_{\rm \bf s}, \omega\}=\{\ell_{\rm \bf s},8,9/L\} \quad  \hbox{with}\:\: \ell_{\rm \bf s}=8,10,12,14,16;
\nonumber\\ 
&& \{\ell_{\rm \bf s},m_{\rm \bf s}, \omega\}=
{\bigl\{ } \{12,12,15/L\}, \{14,14,17/L\}, \{16,16,19/L\}, \{18,18,21/L\} 
{\bigr\} },\nonumber
\label{O3:excitedS}
\end{eqnarray}
(note that the excited mode with highest frequency has $\omega=3\,\bar{\omega}_6$ as expected),
while the 22 vector harmonics have quantum numbers:
 \begin{eqnarray}
&& \{\ell_{\rm \bf v},m_{\rm \bf v}, \omega \}=\{\ell_{\rm \bf v},2,3/L\} \quad  \hbox{with}\:\: \ell_{\rm \bf v}=3,5,7,9,11,13; \nonumber\\ 
&& \{\ell_{\rm \bf v},m_{\rm \bf v}, \omega\}=\{\ell_{\rm \bf v},4,5/L\} \quad  \hbox{with}\:\: \ell_{\rm \bf v}=5,7,9,11,13,15;
\nonumber\\ 
&& \{\ell_{\rm \bf v},m_{\rm \bf v}, \omega\}=\{\ell_{\rm \bf v},6,7/L \} \quad  \hbox{with}\:\: \ell_{\rm \bf v}=7,9,11,13,15,17; \\ 
&& \{\ell_{\rm \bf v},m_{\rm \bf v}, \omega\}=\{\ell_{\rm \bf v},8,9/L\} \quad  \hbox{with}\:\: \ell_{\rm \bf v}=9,11,13,15. \nonumber
\label{O3:excitedV}
\end{eqnarray}

Out of the 30 scalar harmonics \eqref{O3:excitedS}  there are four that stand out as special because they have a quantum numbers $\{ \ell_{\rm \bf s}, m_{\rm \bf s}, p_{\rm \bf s}, \omega \}$ that are  already present in the linear frequency spectrum \eqref{spectrumS}.
These are  
\begin{equation}
\{ \ell_{\rm \bf s}, m_{\rm \bf s}, p_{\rm \bf s}, \omega \}= \{4,4,0,5/L \} \quad\hbox{and}\quad  \{ \ell_{\rm \bf s}, m_{\rm \bf s}, p_{\rm \bf s}, \omega \}=\{6,6,0,7/L \},
\label{O3:removableR}
\end{equation}
and
\begin{equation}
\{ \ell_{\rm \bf s}, m_{\rm \bf s}, p_{\rm \bf s}, \omega \}=\{2,2,0,3/L \} \quad\hbox{and}\quad  \{ \ell_{\rm \bf s}, m_{\rm \bf s}, p_{\rm \bf s}, \omega \}=\{8,8,0,9/L \}.
\label{O3:irremovableR}
\end{equation}
In these four cases, since the source quantum numbers $\{ \ell_{\rm \bf s}, m_{\rm \bf s}, p_{\rm \bf s}, \omega \}$ match the normal mode quantum numbers $\{ \ell_{\rm \bf s}, m_{\rm \bf s}, p_{\rm \bf s}, \bar{\omega}_{\rm \bf s} \}$ of the global AdS background, we say that we have resonances, as discussed in the previous section. The two resonances \eqref{O3:removableR} are further special because they have quantum numbers that match those that were present in our seed \eqref{initialdata}. Because of the specific seed we started with, no vector mode in the list \eqref{O3:excitedV} matches the vector normal mode spectrum \eqref{spectrumV}. 

For each excited harmonic listed in \eqref{O3:excitedS}-\eqref{O3:excitedV}, we now proceed exactly like in the single mode seed case analysis described in detail in section \ref{sec:Grav3geon}. That is, we use the Kodama-Ishibashi formalism to find the most general inhomogeneous solution $h^{(3)}_{({\rm \bf j})\ell_{\rm \bf j},m_{\rm \bf j}}$. It depends on the seed amplitudes $\mathcal{A}^{(1)}_{({\rm \bf s}) 4}$ and $\mathcal{A}^{(1)}_{({\rm \bf s}) 6}$  introduced in \eqref{initialdata}, and on the  third order Kodama-Ishibashi amplitudes $\alpha^{(3)}_{\ell_{\rm \bf j},m_{\rm \bf j}}$ and $\beta^{(3)}_{\ell_{\rm \bf j},m_{\rm \bf j}}$. The metric perturbation of each of the two special resonant harmonics listed in \eqref{O3:removableR}, and only these, also depends on the frequency correction $\omega_4^{(2)}$ {\it or} $\omega_6^{(2)}$ introduced  as part of the perturbation scheme in \eqref{O2:wexpansion46}. The several $\{ \alpha^{(3)}_{\ell_{\rm \bf j},m_{\rm \bf j}}, \beta^{(3)}_{\ell_{\rm \bf j},m_{\rm \bf j}} \}$ and the two frequency corrections $\omega_4^{(2)}$ and $\omega_6^{(2)}$ are to be fixed by the boundary conditions. Exactly  like in the single mode seed case, we require the solution to be asymptotically global AdS and regular everywhere, including the origin. We ask the reader to revisit section \ref{sec:Grav3geon} for the details.

Typically, we conclude that a judicious choice of the Kodama-Ishibashi amplitudes $\alpha^{(3)}_{\ell_{\rm \bf v},m_{\rm \bf v}}$ and $\beta^{(3)}_{\ell_{\rm \bf v},m_{\rm \bf v}}$ allows to make each of the 22 vector harmonic perturbations listed in \eqref{O3:excitedV} asymptotically global AdS and regular at the origin. Consider next the scalar modes listed in \eqref{O3:excitedS} {\it except} the four singled out in  \eqref{O3:removableR} and \eqref{O3:irremovableR}. For these 26 scalar modes, a selection of the Kodama-Ishibashi amplitudes $\alpha^{(3)}_{\ell_{\rm \bf s},m_{\rm \bf s}}$ and $\beta^{(3)}_{\ell_{\rm \bf s},m_{\rm \bf s}}$, this time complemented  with a gauge transformation generated by \eqref{KI:gaugeS0} and \eqref{gaugetransf}, allows to explicitly get a regular two-tensor 
$(h^{(3)}_{({\rm \bf j})\ell_{\rm \bf j},m_{\rm \bf j}})_{\mu\nu}dx^\mu dx^\nu$ at the origin. Moreover, it is manifestly asymptotically global AdS. It should be mentioned that typically all the integration constants are fixed by the boundary conditions. However, in a few cases, some are left undetermined at this third order and would be fixed only at higher order (as occurs sometimes in perturbation problems).

Take now the two resonant modes \eqref{O3:removableR}  whose quantum numbers  coincide with those present in our seed \eqref{initialdata}. It follows that their gravitational field solution depends not only on the integration constants $\alpha^{(3)}_{\ell_{\rm \bf s},m_{\rm \bf s}}, \beta^{(3)}_{\ell_{\rm \bf s},m_{\rm \bf s}}, \widetilde{\alpha}^{(3)}_{\ell_{\rm \bf s},m_{\rm \bf s}},  \widetilde{\beta}^{(3)}_{\ell_{\rm \bf s},m_{\rm \bf s}}$ but also on the frequency correction $\omega_4^{(2)}$ (in the first case of \eqref{O3:removableR}) or $\omega_6^{(2)}$ (in the second case of \eqref{O3:removableR}).  A selection of one of the integration constants together with a gauge transformation yields an asymptotically global AdS solution. We can then choose the frequency corrections $\omega_4^{(2)}$, in the first case of  \eqref{O3:removableR}, or $\omega_6^{(2)}$, in the second case of \eqref{O3:removableR}, to make the solution regular at the origin, 
\begin{eqnarray}\label{O3:wCorrections}
&&  \omega_4^{(2)} = -\frac{7010569125}{77792}\, \frac{1}{L}\,,\nonumber\\
&&  \omega_6^{(2)} = - \frac{8231910851500875}{3090464}\,\frac{1}{L}\,.
\end{eqnarray} 
Naturally, these values agree with those listed in Table \ref{Table:geonsS2} where we considered each of the two seed modes \eqref{initialdata} in isolation. Again note that the frequency correction \eqref{O2:wexpansion46} and  \eqref{O3:wCorrections} is effectively a Poincar\'e-Lindstedt resummation procedure that avoids introducing a periodic secular contribution to make the solution regular at the origin. (Further note that in this process one of the integration constants is left undetermined: it is to be fixed at higher order). 

Finally let us discuss the two resonant harmonics listed in  \eqref{O3:irremovableR}. These have quantum numbers that can be found already in the normal mode spectrum \eqref{spectrumS} of the global AdS background, but they do not coincide with the linear seed \eqref{initialdata} quantum numbers. These describe irremovable resonances since we cannot eliminate a particular solution that is secular if we require the asymptotically global AdS solution to be regular at the origin.  
As pointed out in Section \ref{sec:Grav2geon}, we are not aware of any resummation procedure for these resonant modes \eqref{O3:irremovableR}. Their quantum numbers do not match the seed \eqref{initialdata} and thus we do not have a Poincar\'e-Lindstedt frequency correction to avoid the associated   secular contributions. 
The modes \eqref{O3:irremovableR} source a solution that is asymptotically global AdS and regular everywhere  but contains {\it irremovable} resonances with a secular growing amplitude.

Consequently, perturbation theory with expansion parameter $\varepsilon$ breaks down at third order  with secular growing modes linear in time, $\varepsilon^3 t$. Extrapolating the consequences of this perturbative result to a time evolution scenario that starts with the initial data \eqref{initialdata} with amplitudes of order $\varepsilon$, we take this to be strong evidence for expecting black hole formation at a timescale $t \sim 1/\varepsilon^2$.

The first mode in  \eqref{O3:irremovableR} leading to an irremovable resonance has $\omega=\frac{3}{L}\equiv\bar{\omega}_{\ell_{\rm \bf s}=2, p_{\rm \bf s}=0}$. From \eqref{spectrumS} this is a normal mode frequency {\it below} the two seed frequencies \eqref{initialdata}. On the other hand, the second mode in  \eqref{O3:irremovableR} has $\omega=\frac{9}{L}\equiv\bar{\omega}_{\ell_{\rm \bf s}=8, p_{\rm \bf s}=0}$ which is a normal mode frequency {\it above} the two seed frequencies.
Thus, the seed \eqref{initialdata} demonstrates that the weakly perturbative turbulent mechanism predicts that a given seed with a certain spectrum of normal mode frequencies of AdS excites at third order irremovable resonances that have {\it both} higher and lower frequency than those present in the seed.

In an attempt to understand which of the cascades is likely to dominate faster the time-evolution, we have compared the coefficient of the direct and inverse cascades. In order to do this in a gauge invariant way, we computed the boundary holographic stress energy tensor \cite{Balasubramanian:1999re,deHaro:2000vlm} and its concomitant energy density \eqref{HoloDensities}. The amplitudes of the two secular contributions in the energy density are
\begin{eqnarray} 
&& A {\bigl |}_{\omega=3/L}= \varepsilon^3 t \,\left(\mathcal{A}^{(1)}_{({\rm \bf s}) 4}\right)^2 \mathcal{A}^{(1)}_{({\rm \bf s}) 6} \frac{15601838844375}{7072 \pi }\frac{1}{G_N L^4}\cos^2\theta,\nonumber \\
&& A{\bigl |}_{\omega=9/L}= \varepsilon^3 t \,\mathcal{A}^{(1)}_{({\rm \bf s}) 4} \left(\mathcal{A}^{(1)}_{({\rm \bf s}) 6}\right)^2 \frac{1181585584929605625}{162656 \pi }\frac{1}{G_N L^4} \cos^8\theta \,.
\label{secularAmp}
\end{eqnarray}
As discussed in detail in Section \ref{sec:Grav4geon}, the expansion parameters $\mathcal{A}^{(1)}_{({\rm \bf s}) 4} \varepsilon$ and $\mathcal{A}^{(1)}_{({\rm \bf s}) 6} \varepsilon$ introduced in \eqref{initialdata} have a clear physical interpretation: their are related to the energy $E$ we deposit in each of the normal modes of the initial data.  We assume that each of the modes in the initial seed carries {\it equal energy} $E$ and  replace $\mathcal{A}^{(1)}_{({\rm \bf s}) 4} \varepsilon$ by \eqref{def:ep} and  $\mathcal{A}^{(1)}_{({\rm \bf s}) 6} \varepsilon$ by \eqref{def:ep660}.

In these conditions we can now compare the ratio (normalized with the appropriate angular factor) between the two secular terms \eqref{secularAmp}. We find that\footnote{For reference, the ratio between the removable secular amplitudes in the energy density, $A {\bigl |}_{\omega=5/L}$ and $A {\bigl |}_{\omega=7/L}$, is of order 1.} 
\begin{equation} 
\frac{A{\bigl |}_{\omega=9/L}}{A {\bigl |}_{\omega=3/L}}\frac{1}{\cos^6\theta}\sim 13.
\label{ratiosecular}
\end{equation}
That is, if we assume that each of the modes in the seed carries {\it equal energy}, the direct cascade is a factor of 10 larger than the inverse cascade. Extrapolating these perturbative results to a time evolution simulation that starts with initial data \eqref{initialdata}, we conjecture that \eqref{ratiosecular} signals that the direct cascade should dominate over the inverse cascade and black hole formation is likely to occur at late times.

\section{Gravitational sector is more populated by secular resonances  \label{sec:Grav4}}

In this final section we want to highlight the fact that in a sense AdS has a much stronger nonlinear instability in the non-spherical gravitational sector than in the spherically symmetric scalar field sector. On one hand this is because  seeds with a single mode already develop irremovable resonances as discussed in Section \ref{sec:geons}. This is not the case in the spherically symmetric scalar field sector. Naturally, these irremovable resonances then add to those that unavoidably appear from the nonlinear interactions between such modes in seeds consisting of two or more of these modes. On the other hand, there seems to be an enhancement on the number of irremovable resonances when we start with two generic modes in these conditions. That is, modes that in isolation develop resonances tend, when combined with others with different frequency, to increase the number of irremovable resonances that are proportional to the seed amplitude of two or more modes. 

As a simple example that illustrates this point consider the seed that collides a scalar normal mode with a vector normal mode: 
 \begin{eqnarray}
\hbox{\bf Seed:} && \{ \ell_{\rm \bf s},m_{\rm \bf s},p_{\rm \bf s},\bar{\omega}_{\ell_{\rm \bf s}}\}=\{4,4,0,5/L\} \quad  \hbox{with amplitude}\:\:\mathcal{A}^{(1)}_{({\rm \bf s}) 4} \varepsilon,\quad \hbox{and} \nonumber\\ 
&&  \{ \ell_{\rm \bf v},m_{\rm \bf v},p_{\rm \bf v}, \bar{\omega}_{\ell_{\rm \bf v}}\}=\{7,6,0,9/L\} \quad  \hbox{with amplitude}\:\:\mathcal{A}^{(1)}_{({\rm \bf v}) 7}\varepsilon\,,
\label{initialdata2}
\end{eqnarray}
The first is a scalar mode that in isolation does not develop irremovable resonances, while the second is a vector mode that  does so: see Tables \ref{Table:geonsS2} and \ref{Table:geonsV}.     
(We will use the short-hand notation $\omega_4 \equiv\omega_{4,4,0}$ and $\omega_7 \equiv\omega_{7,6,0}$).

At second order, 16 scalar harmonics are excited,
 \begin{eqnarray}
&& \{\ell_{\rm \bf s}, m_{\rm \bf s}, \omega \}=\{\ell_{\rm \bf s},0,0\} \quad  \hbox{with}\:\: \ell_{\rm \bf s}=0,2,4,6,8,10,12,14; \nonumber\\ 
&& \{\ell_{\rm \bf s}, m_{\rm \bf s}, \omega\}=\{\ell_{\rm \bf s},2,4/L\} \quad  \hbox{with}\:\: \ell_{\rm \bf s}=4,6,8,10;
\nonumber\\ 
&& \{\ell_{\rm \bf s}, m_{\rm \bf s}, \omega\}={\bigl\{ } \{8,8,10/L\}, \{10,10,14/L\}, \{12,12,18/L\},\{12,14,18/L\} 
{\bigr\} },
\label{O2:excitedS2}
\end{eqnarray}
in addition to 13 vector harmonics, 
 \begin{eqnarray}
&& \{\ell_{\rm \bf v}, m_{\rm \bf v}, \omega\}=\{\ell_{\rm \bf v},0,0\} \quad  \hbox{with}\:\: \ell_{\rm \bf v}=1,3,5,7,9,11,13; \nonumber\\ 
&& \{\ell_{\rm \bf v}, m_{\rm \bf v}, \omega\}=\{\ell_{\rm \bf v},2,2/L\} \quad  \hbox{with}\:\: \ell_{\rm \bf v}=3,5,7,9,11;\nonumber\\
&& \{\ell_{\rm \bf v}, m_{\rm \bf v}, \omega\}= \{10,11,14/L\}.
\label{O2:excitedV2}
\end{eqnarray}
Naturally the frequencies that can be excited at second order are $\omega \in \{0, \bar{\omega}_7\pm \bar{\omega}_4, \bar{\omega}_4+\bar{\omega}_4,\bar{\omega}_7+\bar{\omega}_7\}$.
All these modes are asymptotically global AdS and regular at the origin without resonances.

The third order excites 34 scalar harmonics,
 \begin{eqnarray}
&& \{\ell_{\rm \bf s}, m_{\rm \bf s}, \omega \}=\{2,\ell_{\rm \bf s},1/L\} \quad  \hbox{with}\:\: \ell_{\rm \bf s}=2,4,6,8,10,12,14; \nonumber\\ 
&& \{\ell_{\rm \bf s}, m_{\rm \bf s}, \omega\}=\{4,\ell_{\rm \bf s},5/L\} \quad  \hbox{with}\:\: \ell_{\rm \bf s}=4,6,8,10,12,14,16,18;
\nonumber\\ 
&& \{\ell_{\rm \bf s}, m_{\rm \bf s}, \omega\}=\{6,\ell_{\rm \bf s},9/L\} \quad  \hbox{with}\:\: \ell_{\rm \bf s}=6,8,10,12,14,16,18,20;   \\ 
&& \{\ell_{\rm \bf s}, m_{\rm \bf s}, \omega\}=\{8,\ell_{\rm \bf s},13/L\} \quad  \hbox{with}\:\: \ell_{\rm \bf s}=8,10,12,14,16,18;
\nonumber\\ 
&& \{\ell_{\rm \bf s}, m_{\rm \bf s}, \omega\}=
{\bigl\{ } \{12,12,15/L\}, \{14,14,19/L\}, \{16,16,23/L\}, \{16,18,23/L\}, \{18,18,27/L\} 
{\bigr\} },\nonumber
\label{O3:excitedS2}
\end{eqnarray}
and 31 vector harmonics,
 \begin{eqnarray}
&& \{\ell_{\rm \bf v}, m_{\rm \bf v}, \omega \}=\{\ell_{\rm \bf v},2,1/L\} \quad  \hbox{with}\:\: \ell_{\rm \bf v}=3,5,7,9,11,13,15; \nonumber\\ 
&& \{\ell_{\rm \bf v}, m_{\rm \bf v}, \omega\}=\{\ell_{\rm \bf v},4,5/L\} \quad  \hbox{with}\:\: \ell_{\rm \bf v}=5,7,9,11,13,15,17;
\nonumber\\ 
&& \{\ell_{\rm \bf v}, m_{\rm \bf v}, \omega\}=\{\ell_{\rm \bf v},6,9/L \} \quad  \hbox{with}\:\: \ell_{\rm \bf v}=7,9,11,13,15,17,19,21; \\ 
&& \{\ell_{\rm \bf v}, m_{\rm \bf v}, \omega\}=\{\ell_{\rm \bf v},8,13/L\} \quad  \hbox{with}\:\: \ell_{\rm \bf v}=9,11,13,15,17, \nonumber \\
&& \{\ell_{\rm \bf v}, m_{\rm \bf v}, \omega\}= {\bigl\{ } \{15,14,19/L\}, \{17,16,23/L\}, \{19,18,27/L\}, \{21,18,27/L\} {\bigr \} }. \nonumber
\label{O3:excitedV2}
\end{eqnarray}
Thus, the excited frequencies at third order are $\omega \in \{2\bar{\omega}_4\pm \bar{\omega}_7, 2\bar{\omega}_4\pm\bar{\omega}_4, 2\bar{\omega}_7\pm\bar{\omega}_7, 2\bar{\omega}_7\pm \bar{\omega}_4 \}$.

Note that the modes that  {\it can} (but do not necessarily do) generate resonances are those whose quantum numbers match one of the two conditions,
 \begin{eqnarray}  \label{ResonantCondition}
&& \omega_{\rm \bf s} L= 1+\ell_{\rm \bf s} +2 p_{\rm \bf s}, \quad \hbox{for} \quad \ell_{\rm s}\geq m_{\rm s}, \:\: \:\: p_{\rm \bf s}\in\{0,1,2\ldots \}, \nonumber\\
&&\omega_{\rm \bf v} L= 2+\ell_{\rm \bf v} +2 p_{\rm \bf v}, \quad \hbox{for} \quad \ell_{\rm v}\geq m_{\rm v}, \:\: \:\: p_{\rm \bf v}\in\{0,1,2\ldots \},
  \end{eqnarray}
for some values of the pair $\{ \ell_{\rm j}\geq2, p_{\rm j} \}$.

Out of these excited modes there are two resonances that, in addition to satisfy \eqref{ResonantCondition}, do have the same quantum numbers as the seed \eqref{initialdata2}. One is a scalar mode and the other a vector mode, namely:
\begin{eqnarray}\label{O3:removableR2a} 
&&\{\ell_{\rm \bf s}, m_{\rm \bf s}, p_{\rm \bf s}, \omega\}=\{4,4,0,5/L\}, \nonumber\\
&&\{\ell_{\rm \bf v}, m_{\rm \bf v}, p_{\rm \bf v}, \omega\}=\{7,6,0,9/L\}.
\end{eqnarray}
These are removable resonances since we can avoid them if we introduce the Poincar\'e-Lindstedt frequency corrections,
\begin{eqnarray}
&&  \omega_4^{(2)} = - \frac{1}{L}\left( \frac{7010569125}{77792}+\frac{1860284480041845}{7607296}\frac{\lp\mathcal{A}^{(1)}_{({\rm \bf v}) 7}\rp^2}{\lp\mathcal{A}^{(1)}_{({\rm \bf s}) 4}\rp^2} \right)\,, \label{O3:wCorrections1}\\
&&  \omega_7^{(2)} = - \frac{1}{L}\left( \frac{8548214990390361}{19124224}+\frac{21681637296525}{271960832} \frac{\lp\mathcal{A}^{(1)}_{({\rm \bf s}) 4}\rp^2}{\lp\mathcal{A}^{(1)}_{({\rm \bf v}) 7}\rp^2} \right)\,.  \label{O3:wCorrections2}
\end{eqnarray} 
Here, we notice that the first contribution to these corrections agrees with those listed in Tables \ref{Table:geonsS2} and \ref{Table:geonsV} for the case where we started with a seed that had the normal modes \eqref{initialdata2} in isolation. In equivalent words, it is a shift that removes a resonance of the type $2\bar{\omega}_4-\bar{\omega}_4=5/L$ in \eqref{O3:wCorrections1} and  $2\bar{\omega}_7-\bar{\omega}_7=9/L$ in \eqref{O3:wCorrections2}.
However, the second contribution is new. It appears due to the interaction of the two initial modes in \eqref{initialdata2} and it is proportional to the relative seed amplitudes of the two modes. That is, it is a correction that removes a resonance of the type $\bar{\omega}_4+\bar{\omega}_7-\bar{\omega}_7=5/L$ in \eqref{O3:wCorrections1} and $\bar{\omega}_7+\bar{\omega}_4-\bar{\omega}_4=9/L$  in \eqref{O3:wCorrections2}.
Interestingly, these second contributions do not appear when we start with a two-mode seed where each of the modes in isolation can be back-reacted to yield a geon. In particular, note that there is no analogue of these contributions in the spherically symmetric scalar field sector because {\it all} individual spherically symmetric scalar field normal modes can be back-reacted to yield a oscillon or a boson star.

Besides these removable resonances, there are also several irremovable resonances, namely:    
\begin{eqnarray}
&& \{ \ell_{\rm \bf s}, m_{\rm \bf s}, p_{\rm \bf s}, \omega \}= {\bigl\{ }  \{6,6,1,9/L \}, \{8,6,0,9/L \}{\bigr \} };  \nonumber\\
&& \{ \ell_{\rm \bf s}, m_{\rm \bf s}, p_{\rm \bf s}, \omega \}=  {\bigl\{ } \{8,8,2,13/L \}, \{10,8,1,13/L \}, \{12,8,0,13/L \}  {\bigr \} };  \nonumber\\
&& \{ \ell_{\rm \bf v}, m_{\rm \bf v}, p_{\rm \bf v}, \omega \}=  {\bigl\{ }  \{9,8,1,13/L \}, \{11,8,0,13/L \}  {\bigr \} }. \label{O3:irremovableR2}
\end{eqnarray}
The first line resonances in this list are of the type $2\bar{\omega}_7-\bar{\omega}_7=9/L$ and thus coincide with those already listed in the last column of Table  \ref{Table:geonsV}  when we start the linear order with the single normal mode  $\{ \ell_{\rm \bf v},m_{\rm \bf v},p_{\rm \bf v}, \bar{\omega}_{\ell_{\rm \bf v}}\}=\{7,6,0,9/L\}$. Again note that the these two resonances have the same frequency but are generated by different combinations of $\ell$ and $p$. On the other hand, the second and third line resonances are of the type $2\bar{\omega}_7-\bar{\omega}_4=13/L$ and thus a consequence of the two-mode collision \eqref{initialdata2}.

The simple seed \eqref{initialdata2}, \emph{i.e.} a linear combination of a scalar mode that can be back-reacted to give a geon plus a vector mode that in isolation is already secular resonant, illustrates one of the main results of our study.
Indeed this combination generates much more irremovable resonances at third order than if we start with two modes that can be both back-reacted to give a geon. In the latter case we get just two irremovable resonances: see \eqref{initialdata} and \eqref{O3:irremovableR} in Sections \ref{sec:Grav1}-\ref{sec:Grav3}. This is very much what happens in the spherically symmetric scalar field sector where any mode of the seed can always be back-reacted to give a soliton (oscillon or boson star). But we found in Section \ref{sec:geons} that the gravitational sector is much different from the spherically symmetric scalar field sector for: 1) we have now two sectors of normal modes that have a degenerate frequency spectrum that depends not only on $p$ but also $\ell$ and different combinations of these can give the same frequency, 2) unlike the spherically symmetric scalar field counterparts, most of the individual gravitational normal modes cannot be back-reacted to give a soliton (geon), namely all that do not satisfy the conditions listed in \eqref{geons}. 

As a consequence of these properties, when we collide two or more normal modes that individually do not have geon nonlinear extensions, the system typically generates much more than just a pair of irremovable resonances\footnote{However, if we start with a linear combination of all normal modes that have the same frequency, we get regular geons \cite{Rostworowski:2016isb,Rostworowski:2017tcx,Martinon:2017uyo}.}. This is clearly illustrated by \eqref{initialdata2} and \eqref{O3:irremovableR2}. Had we replaced the first mode in  \eqref{initialdata2} by a normal mode that also does not have a geon nonlinear extension and we would generate even more irremovable resonances. In addition, if we start with not two but several modes with different frequencies each one of each cannot be indovidually back-reacted to give a geon, then a cascading generation of irremovable resonances occurs. Such a dramatic generation of irremovable resonances is not observed at all in the weakly turbulent perturbative analysis of the spherically symmetric scalar field. We take this as evidence suggesting that the time evolution of the gravitational nonlinear instability of AdS might differ from its spherically symmetric scalar field counterpart. In a sense it might be more dramatic, although our perturbation theory analysis still breaks down only at $\mathcal{O}(3)$ and thus for timescales of order $t\geq 1/\varepsilon^2$. 

Finally, note that with the seed \eqref{initialdata2} we just have a direct cascade, \emph{i.e.} irremovable resonances with lower frequency that the ones we start with in \eqref{initialdata2} are not generated. For these two modes are such that the lowest frequency that can be excited $2\bar{\omega}_4- \bar{\omega}_7=1/L$ is not in the normal mode spectra \eqref{spectrumS} and \eqref{spectrumV}. 

\vskip .5cm
\centerline{\bf Acknowledgements}
\vskip .2cm

O.J.C.D. is supported by the STFC Ernest Rutherford grants ST/K005391/1 and ST/M004147/1. This research received funding from the European Research Council under the European Community's 7th Framework Programme (FP7/2007-2013)/ERC grant agreement no. [247252]. 
\begin{appendix}
\section{Kodama-Ishibashi gauge invariant formalism \label{appendix:KI}}

In the Kodama-Ishibashi (KI) formalism  \cite{Kodama:2003jz,Kodama:2003kk} (which is an extension of the  Regge-Wheeler$-$Zerilli decomposition \cite{Regge:1957td,Zerilli:1970se} to backgrounds with a cosmological constant; see footonote \ref{footKI}), the most general perturbation $h$ $-$ actually any regular two-tensor $\mathcal{T}$ $-$ of global AdS is decomposed  into a superposition of two classes of modes:  scalar and vector. Scalar and vector modes are expanded in terms of the scalar $\scalar(x,\phi)$  and vector $\vector_i(x,\phi)$  harmonics that we review next. In this discussion, we use the polar coordinate $x=\cos\theta$. To ease the notation, in this appendix we omit the subscripts $\{{\rm \bf s},{\rm \bf v}\}$ in the quantum number $m$, and we omit the quantum numbers $\ell_{\rm \bf j}$ and $m_{\rm \bf j}$ in the master variable $\Phi_{({\rm \bf j})}$ and in the perturbations $h^{({\rm \bf j})}$.

\subsection{Scalar modes \label{appendix:KIscalar}}
Scalar modes are given by   \cite{Kodama:2003jz}
\begin{eqnarray}\label{KI:scalar}
  h^{\rm \bf s}_{ab}= f_{ab} \scalar, \qquad  
  h^{\rm \bf s}_{ai}= r f_a \scalar_i  , \qquad 
  h^{\rm \bf s}_{ij}= 2r^2 \left( H_L\gamma_{ij}\scalar + H_T \scalar_{ij} \right),
\end{eqnarray}
where $(a,b)$ are components in the orbit spacetime parametrized by $\{t,r\}$, $(i,j)$ are legs on the sphere,  $\{f_{ab},f_a,H_T,H_L\}$ are functions of  $(t,r)$, and $\scalar$ is the KI scalar harmonic, 
$ \scalar_i = -\lambda_{\rm \bf s}^{-1/2} D_i \scalar$,   $ \scalar_{ij}  = \lambda_{\rm \bf s}^{-1} D_i D_j \scalar + \frac{1}{2}\gamma_{ij} \scalar$,
$\gamma_{jk}$ is the unit radius metric on $S^2$  and $D_j$ is the associated covariant derivative.
Assuming the ansatz $S(x,\phi)=e^{i m \phi} Y_{\ell_{\rm \bf s}}^{m }(x)$, the scalar harmonic equation $\left (\triangle_{S^2} + \lambda_{\rm \bf s} \right)\scalar=0$  ($\triangle_{S^2} = \gamma^{jk}  D_j D_k$) reduces to
\begin{equation}\label{KI:scalarHeq}
\partial _x\left[\left(1-x^2\right)\partial _x Y_{\ell_{\rm \bf s}}^{m }(x)\right]+\left(\lambda_{\rm \bf s} -\frac{m^2}{1-x^2}\right)Y_{\ell_{\rm \bf s}}^{m }(x)=0\,.   
\end{equation}
Its regular solutions, with normalization $\int _0^{2\pi }d\phi\int _{-1}^1 \mathrm{d}x\,\left | \scalar \right|^2 =1$, are
\begin{eqnarray}\label{KI:scalarH}
&& \scalar(x,\phi)=\sqrt{\frac{2\ell_{\rm \bf s} +1}{4\pi }\frac{(\ell_{\rm \bf s} -m)!}{(\ell_{\rm \bf s} +m)!}}\, P_{\ell_{\rm \bf s}}^m(x)\,e^{i m \phi}\equiv  Y_{\ell_{\rm \bf s}}^{m }(x,\phi)\,, \\ 
&& \hbox{with} \quad \lambda_{\rm \bf s}=\ell_{\rm \bf s}\left(\ell_{\rm \bf s}+1\right),\qquad  \ell_{\rm \bf s}=0,1,2,\cdots,\quad |m|\leq  \ell_{\rm \bf s}.\nonumber
\end{eqnarray}
where $P_{\ell}^m(x)$ is the associated Legendre polynomial. Hence, the KI scalar harmonic $ \scalar(x,\phi)$ is the standard scalar spherical harmonic $Y_{\ell_{\rm \bf s}}^{m }(x,\phi)$.

An infinitesimal gauge vector $\xi$ can also be decomposed in terms of  scalar harmonics as
\begin{equation}\label{KI:gaugeS0}
 \xi_{a}=P_a(t,r) \scalar \,, \qquad 
 \xi_i = r L(t,r)  \scalar_{i} \,.
\end{equation}

Scalar modes are fully described by the KI scalar variable $\Phi_{({\rm \bf s})}\equiv \Phi^{(k)}_{({\rm \bf s})\ell_{\rm \bf s},m_{\rm \bf s}}(t,r)$  which obeys the KI master equation for scalar modes \eqref{eq:master}. Once we solve this master equation, there is a differential map that allows to  reconstruct the scalar metric perturbations \eqref{KI:scalar} from the scalar gauge invariant master variable $\Phi_{({\rm \bf s})}$. 
To get it first observe that, for $\ell_{\rm \bf s}\neq 0$, we can use \eqref{KI:gaugeS0} to choose the gauge
\begin{equation}\label{KI:gaugeS}
  f_t=f_r=H_T=0 \,.
\end{equation} 
In this gauge, the other metric components can be expressed in terms of the scalar master variable $\Phi_{({\rm \bf s})}(t,r)$, of the source components $\mathcal{T}(t,r)$, and their first derivatives as
\begin{eqnarray}\label{KI:metricS}
&& \hspace{-0.5cm} f_{tt}=  -\left( r\partial_t^2 +f\partial_r +\frac{\ell_{\rm s}(\ell_{\rm s}+1)}{2r}\right)\Phi\nonumber \\ 
&& \hspace{0.5cm}+ \frac{r}{\ell_{\rm s}^2+\ell_{\rm s} -2}\left[r \mathcal{T}_{tt} +f^2 \left(r \mathcal{T}_{rr} -\frac{4 r}{\sqrt{\ell_{\rm s}  (\ell_{\rm s} +1)}}\mathcal{T}_{r}+2 \int \mathcal{T}_{tr} \, \mathrm{d}t\right)\right]\nonumber \\ 
&& \hspace{0.5cm} -  f \left[4 r \mathcal{T}_{L}+\frac{4}{\sqrt{\ell_{\rm s}  (\ell_{\rm s} +1)}}\int \mathcal{T}_{t} \, \mathrm{d}t+\frac{r}{\ell_{\rm s}  (\ell_{\rm s} +1)} \left(\frac{2 r^2}{f}\partial_t^2 +2 r \partial_r+\left(\ell_{\rm s}^2+\ell_{\rm s} +6\right) \right) \mathcal{T}_{T}\right], \nonumber \\ 
&& \hspace{-0.5cm}  f_{tr}= -\left(r \partial_r+ \frac{1}{f}\right)   \partial_t\Phi  \nonumber \\ 
&& \hspace{0.5cm}  + \frac{2r^2}{\ell_{\rm s}^2+\ell_{\rm s} -2} \left[ \mathcal{T}_{tr}  -\frac{2}{\sqrt{\ell_{\rm s}  (\ell_{\rm s} +1)}}\,\mathcal{T}_{t} -\frac{r^2}{\ell_{\rm s} (\ell_{\rm s}  +1)} \left( \partial _r + \frac{3 f+1}{r f} \right) \partial_t  \mathcal{T}_{T} \right] \nonumber \\ 
&& \hspace{-0.5cm} f_{rr}=  -\frac{1}{f} \left( \frac{r}{f}\partial_t^2 +\partial_r +\frac{\ell_{\rm s}(\ell_{\rm s}+1)}{2r}\right)\Phi\nonumber \\ 
&& \hspace{0.5cm}+ \frac{r}{\ell_{\rm s}^2+\ell_{\rm s} -2}\left[\frac{r}{f^2}\,\mathcal{T}_{tt} +r\left(\mathcal{T}_{rr} -\frac{4}{\sqrt{\ell_{\rm s}  (\ell_{\rm s} +1)}}\mathcal{T}_{r}+2 \int \mathcal{T}_{tr} \, \mathrm{d}t\right)\right]\nonumber \\ 
&& \hspace{0.5cm} -\frac{1}{f} \left[ 4r \mathcal{T}_{L}+\frac{4}{\sqrt{\ell_{\rm s}  (\ell_{\rm s} +1)}}\int \mathcal{T}_{t} \, \mathrm{d}t
+\frac{r}{\ell_{\rm s}  (\ell_{\rm s} +1)} \left( \frac{2r^2}{f}\partial_t^2 +2 r \partial_r+\left(5\ell_{\rm s}^2+5\ell_{\rm s} -2\right) \right) \mathcal{T}_{T} \right], \nonumber \\ 
&& \hspace{-0.5cm} 
H_L=  -\frac{f}{2} \left(\partial_r +\frac{\ell_{\rm s}(\ell_{\rm s}+1)}{2r f}\right)\Phi+\frac{r}{\ell_{\rm s}^2+\ell_{\rm s} -2}{\biggl [} f \int \mathcal{T}_{tr} \, \mathrm{d}t -\frac{2f}{\sqrt{\ell_{\rm s}  (\ell_{\rm s} +1)}} \int \mathcal{T}_{t} \, \mathrm{d}t \nonumber \\ 
&& \hspace{0.5cm} -\frac{r}{2\ell_{\rm s} (\ell_{\rm s}  +1)}\left( 2r f \partial_r +(\ell_{\rm s}^2+\ell_{\rm s} +6f)\right)  \mathcal{T}_{T}  {\biggr ]}.
\end{eqnarray}

We want the perturbations (and thus the associated sources) to keep the $t-\phi$ symmetry of the background, {\it i.e.} the invariance of $h_{\mu\nu} dx^\mu dx^\nu$ under $\{t,\phi\} \to \{-t,-\phi\}$. Therefore we must do the replacements $\{ f_{tr} \to -i\, f_{tr},f_t \to -i\, f_t \}$ in \eqref{KI:scalar}. 
Further note that the global AdS background has the Killing isometries generated by $\partial_t$ and $\partial_\phi$. This allows a Fourier decomposition $e^{i m \phi}$ of any tensor along $\phi$ as we did above, but it also permits to Fourier expand $\{f_{ab},f_a,H_T,H_L\}$ along the time direction as $e^{-i \omega t}$ (where it is to be understood that $m\equiv m_{\rm \bf s}$ and $\omega\equiv \omega_{\ell_{\rm \bf s},p_{\rm \bf s}}$).
In addition, because we will go beyond the linear order $k=1$ in perturbation theory, we need to use a real representation for the spherical harmonics. This means that we have to decompose each component of the perturbation as a sum of a $\cos\left( \omega t - m\phi \right)= {\rm Re}[e^{-i \omega_{\ell_{\rm \bf s},p_{\rm \bf s}}}e^{i m \phi}]$ contribution and a $\sin\left( \omega t - m\phi \right)= -{\rm Im}[e^{-i \omega_{\ell_{\rm \bf s},p_{\rm \bf s}}}e^{i m \phi}]$ contribution.

\subsection{Vector modes \label{appendix:KIvector}}
On the other hand, the KI  vector modes are given by  \cite{Kodama:2003jz}
\begin{equation}\label{KI:vector}
   h^{\rm \bf v}_{ab}=0 \,, \qquad 
  h^{\rm \bf v}_{ai}=r h_a \vector_i ,\qquad 
  h^{\rm \bf v}_{ij} = -2 \lambda_{\rm \bf v}^{-1} r^2 h_T  D_{(i}\vector_{j)} \,, 
\end{equation} 
where $\{h_a, h_T\}$ are functions of $(t,r)$ and the KI  vector harmonics $\vector_j$ are the  solutions of 
\begin{equation}\label{KI:vectorHeq}
\left (\triangle_{S^2} + \lambda_{\rm \bf v} \right)\vector_j=0 \,, \qquad D_j \vector^j=0 \quad 
   \hbox{(Transverse condition)}.
\end{equation} 
The regular vector harmonics can be written in terms of the spherical harmonic $Y_{\ell_{\rm \bf v}}^{m }(x,\phi)$ as
\begin{eqnarray}\label{KI:vectorH}
&& \vector_j\,dx^j=-\frac{i \,m}{1-x^2}Y_{\ell_{\rm \bf v}}^{m }(x,\phi )\,dx+\left(1-x^2\right)\partial_x Y_{\ell_{\rm \bf v}}^{m }(x,\phi )\,d\phi\,,  \\
&& \hbox{with}    \qquad \lambda_{\rm \bf v}=\ell_{\rm \bf v}\left(\ell_{\rm \bf v}+1\right)-1,\qquad  \ell_{\rm \bf v}=1,2,\cdots,\quad |m|\leq  \ell_{\rm \bf v}.\nonumber
\end{eqnarray}

As for the scalar modes, we want the perturbations (and thus the associated sources) to keep the $t-\phi$ symmetry of the background. Therefore we need to do the replacements $\{ h_{r} \to -i\, h_{r},h_T \to -i\, h_T \}$ in \eqref{KI:vector}. 
Vector modes are fully described by the KI scalar variable $\Phi_{({\rm \bf v})}\equiv \Phi^{(k)}_{({\rm \bf v})\ell_{\rm \bf v},m_{\rm \bf v}}(t,r)$  which obeys the KI master equation for scalar modes \eqref{eq:master}. 

An infinitesimal gauge vector $\xi$ also has a vectorial harmonic decomposition, 
\begin{equation}\label{KI:gaugeV0}
 \xi_{A}=0 \,, \qquad 
 \xi_i = r L(t,r) \vector_{i} \,.
\end{equation} 
Modes with $m=0$ have $D_{(i}\vector_{j)}=0$ and we can choose  the gauge $h_r=0$.
On the other hand, for modes with $m\neq 0$ we can choose the gauge $h_T=0$ and  write straightforwardly $h_a=h_a(\Phi_{({\rm \bf v})})$ \cite{Kodama:2003jz}.

\end{appendix}

\bibliography{refsTurb}{}
\bibliographystyle{JHEP}

\end{document}